\documentclass[a4paper]{article}
\pdfoutput=1 
\usepackage{ae,aecompl}
\usepackage{jheppub} 

\usepackage[T1]{fontenc}
\usepackage[none]{hyphenat}
\usepackage[italian,english]{babel}
\usepackage{hyperref}
\usepackage{changepage}
\usepackage{ifpdf}
\usepackage{subfigure}
\usepackage{amssymb}
\usepackage{upgreek}
\usepackage{amsfonts}
\usepackage{epsf}
\usepackage{rotating}
\usepackage{graphicx}
\usepackage{amsmath}
\usepackage{fancyhdr}
\usepackage{lineno}
\usepackage{dutchcal}
\usepackage{babel}
\usepackage{graphics}
\usepackage{pstricks}
\usepackage{color}
\usepackage{multirow}
\usepackage{slashed}
\usepackage{physics}
\usepackage{caption}
\usepackage{cancel}
\usepackage{ytableau}
\usepackage{array,booktabs, makecell}
\usepackage[toc,page]{appendix}
\usepackage{pifont}
\usepackage{float}

\usepackage{tikz}
\usepackage{tikz-feynman}
\tikzfeynmanset{compat=1.0.0}

\newcommand{\lsim}{\mathrel{\mathop{\kern 0pt \rlap
			{\raise.2ex\hbox{$<$}}}
		\lower.9ex\hbox{\kern-.190em $\sim$}}}
\newcommand{\gsim}{\mathrel{\mathop{\kern 0pt \rlap
			{\raise.2ex\hbox{$>$}}}
		\lower.9ex\hbox{\kern-.190em $\sim$}}}

\newcommand{\be}{\begin{equation}}
	\newcommand{\ee}{\end{equation}}
\newcommand{\bea}{\begin{eqnarray}}
	\newcommand{\eea}{\end{eqnarray}}

\newcommand{\half}{\frac{1}{2}}

\newcommand{\sarah}{\texttt{SARAH} 4.15.1 }

\newcommand{\relic}{\Omega h^2}
\newcommand{\relicobs}{\Omega h^2_{\text{obs}}}

\author[a]{Pram Milan P. Robin,}
\author[b]{Priyotosh Bandyopadhyay}
\affiliation[a, b]{Indian Institute of Technology Hyderabad, Kandi,  Sangareddy-502285, Telangana, India}

\emailAdd{ph22resch01001@iith.ac.in}
\emailAdd{bpriyo@phy.iith.ac.in}

\title{\boldmath Perspective of inert quartet in the context of perturbativity and dark matter phenomenology}

\preprint{IITH-PH-0001/26}

\abstract{In this article we consider a $\mathbb{Z}_2$-odd $SU(2)$ quartet with hypercharge $Y = +\frac{1}{2}$ as an extension of the Standard Model whose scalar potential which introduces three additional Higgs portal and two self-couplings. We first investigate the possibility of having Landau poles (LPs) in one-loop and Fixed Points (FPs) in two-loop $\beta$-functions of the Higgs quartic couplings. The role of portal and self-couplings with and without residual phases is extensively investigated in obtaining the Fixed Point at two-loop. The model also can provide us with $\mathbb{Z}_2$-odd neutral scalar as the possible dark matter. However not always the lightest state corresponds to the neutral states, and we look into one-loop mass correction for an enhanced dark matter parameter space. This also gives rise to interesting phenomenology of the next-to-lightest particle which can be singly charged, doubly charged or neutral scalar.  We performed a detailed study of dark matter relic calculation with one-loop masses and with direct detection bounds, and found out that, unlike the minimal inert extensions of $SU(2)$ multiplets, here the dark matter mass can go beyond 15 TeV without crossing the observed relic. Finally,  we summarized with a few benchmark points for future studies.}

\begin{document} 
\maketitle
\flushbottom

\section{Introduction}\label{sec:intro}
Finding of the Standard Model (SM) Higgs boson around 125 GeV has been a direct confirmation of the existence of at least one scalar in the electroweak symmetry breaking (EWSB) \cite{ATLAS:2012yve, CMS:2012qbp}. Addition of scalar generally enhances vacuum stability \cite{Jangid:2020qgo,Bandyopadhyay:2024plc,Bandyopadhyay:2020djh, Bandyopadhyay:2020otm,Chiang:2020rcv,Chakrabarty:2015yia,Bandyopadhyay:2021ipw,Jangid:2020dqh,Khan:2016sxm}, however, often these models are constrained by the perturbativity \cite{Kanemura:2023wap,Gonderinger:2009jp,Athron:2018ipf}. The minimal extensions like SM gauge singlet, $SU(2)$ doublet, triplets are well studied in the literature \cite{Jangid:2020qgo,Bandyopadhyay:2023joz,Benincasa:2022elt,Banerjee:2019luv,Branco:2011iw,Ferreira:2004yd,Kanemura:2023wap,Datta:2016nfz,Parashar:2024cqq,Kanemura:2025xkf,Jangid:2023jya,Chowdhury:2015yja,Bandyopadhyay:2025jlg,Bandyopadhyay:2024gyg}. The $SU(2)$ quartet representation and its phenomenology is studied in \cite{Zeng:2019tlw,AbdusSalam:2013eya,Chowdhury:2023uyd,Jurciukonis:2024bzx,Chakraborty:2025kcl}, however it lacks on restricting the model on perturbativity and one-loop mass splitting essential  for certain parameter space which do not have a viable  dark matter (DM) at the tree-level.  Similarly,  extension of colour groups, i.e. $SU(3)$ are also studied with the advent of Leptoquarks \cite{Bandyopadhyay:2016oif,Bandyopadhyay:2017tlq,Bandyopadhyay:2021kue}. In our previous article we showed that these scenarios in the very minimal form of inert case (i.e., with $\mathbb{Z}_2$-odd extension) show a theoretically interesting feature of having Fixed Points (FPs) at some UV scales \cite{Bandyopadhyay:2025ilx}. Above some critical values of the portal and self-couplings the theory has similar behaviour of $O(N)$-symmetric $\Phi^4$ theory \cite{Shrock:2016hqn, Shrock:2023xcu} at least at the two-loop level. To be precise these inert models showed Landau pole (LP) at one-loop $\beta$-function and Fixed Point at two-loop $\beta$-function. We found that existence of the residual phases among the different components of the Higgs multiplets causes destructive interference in attaining those FPs. We first encounter such residual phases in the context of Inert Doublet Model (IDM) \cite{Bandyopadhyay:2025ilx} which act like a spoiler. Whereas, models like Inert Singlet Model (ISM) and Inert Triplet Model (ITM) do not have such residual phases, thus they tend to have FP more easily as we go above some critical values of the portal couplings   \cite{Bandyopadhyay:2025ilx}.

In this article we consider a $\mathbb{Z}_2$-odd $SU(2)$ quartet . The minimal setup with a possible charge neutral dark matter demands the hypercharge must be $\pm \frac{1}{2}$, and for this study we consider that to be as $+\frac{1}{2}$, which provides us a doubly charged scalar, two singly charged  scalars and two neutral scalars, the lightest of the neutral being the possible dark matter candidate. Appearance of two singly charged, one doubly charged Higgs makes the model phenomenologically more interesting. This extension provides us with more such residual phases along with more degrees of freedom. In the first part of the paper, we investigate such interplay in the context of FPs . Compared to inert singlet, doublet, and triplet as shown in \cite{Bandyopadhyay:2025ilx}, the FPs can appear much earlier in this case due to larger degrees of freedom with the two potential  FP enhancer couplings, as we discuss them later.

 As we consider the case of real dark matter in this set up, we found that the mass hierarchy between the lightest stable particle (LSP) and next-to-lightest stable particle (NLSP) can be reversed if we consider one-loop mass corrections. This enables us to open up some parameter space for real dark matter only at one-loop \cite{Chun:2009mh, Cirelli:2005uq, Bandyopadhyay:2010wp}.
 We investigate such parameter spaces along with the total parameter space with an eligible dark matter. The corresponding  dark matter relic density and direct detection bounds are considered. The theoretical bounds from perturbativity, tree-level stability and the existence of FPs are also pointed out while we fix a few benchmark points. This includes points with correct DM relic, underabundant points as well a point with relatively larger magnetite of individual Higgs-portal couplings suitable for the first order phase transitions \cite{Bandyopadhyay:2021ipw, Bandyopadhyay:2025jlg,Ahriche:2007jp,Espinosa:2011ax}.

The article is arranged as follows. In \autoref{model} we describe the model and the terms in the Lagrangian. Then in \autoref{RGEs} we discuss the renormalization group equations (RGEs) or $\beta$-functions of the Higgs self and Higgs portal quartic couplings at both one- and two-loop level, where we also discuss about the possibility of Landau poles at one-loop and Fixed Points at two-loop level. In \autoref{massspectrum}, we consider the one-loop mass corrections to the components of the quartet and the parameter spaces that have viable dark matter candidates. Then, we discuss the phenomenology of dark matter in \autoref{DMpheno} including bounds from the relic density and direct detection emphasizing the influence of annihilation and co-annihilation processes that are possible in this model. Finally, in \autoref{conclusion} we summarize the important conclusions that we detailed in this study.

\section{Inert quartet extension of the Standard Model}\label{model}
In this model, we extend the Standard Model with an $SU(2)$ scalar quartet (also called quadruplet or 4-plet) $X$ which is in a  $\mathbf{4}$-representation of $SU(2)$ and has a weak isospin $\frac{3}{2}$. In this paper, we will choose the hypercharge of $X$ to be $+\half$ \footnote{We adopt the convention for the electroweak charge relation as $Q = T_3 + Y$ for the electroweak charge  relation where Q denotes the electric charge, $T_3$  represents the third component of the $SU(2)_L$ weak isospin,and $Y$ signifies the weak hypercharge of the multiplet components.}.  In addition, to have a dark matter candidate, we also require $X$ to be odd under a global $\mathbb{Z}_2$ symmetry, while the SM particles are even. Like in any other inert scalar models, this $\mathbb{Z}_2$ odd nature prohibits $X$ to participate in the  EWSB. Thus we will also mention this extension of the Standard Model as the Inert 4-plet Model (I4M).

 The column vector form of the 4-plet is given as,
\begin{equation}
	X = \begin{pmatrix}
		X^{++} \\  X^{+}\\  X^0 \\ X^{-}
	\end{pmatrix},\label{eq:4plet-definition}
\end{equation}
where the components $X^{++}$, $X^{+}$, $X^{0}$,  $X^{-}$ has $T_3$ charges $\frac{3}{2}$, $\half$, $-\half$, and $-\frac{3}{2}$ respectively. The neutral complex component $X^0 = \frac{1}{\sqrt{2}}(X_R +i\, X_I)$, where $X_R$ corresponds to the CP-even real scalar  and $X_I$ is the CP-odd real scalar.   We can also represent the 4-plet as an $SU(2)$ tensor of rank 3, denoted by $X_{ijk}$, $i,j,k=1,2$. The tensor components are identified with the components of column vector in \autoref{eq:4plet-definition} as follows:
\begin{equation}\label{eq:4plet-tensor-components}
	\begin{split}
		X_{111} &= X^{++}, \ \ \ \ \ \ \ \ 
		X_{112} = X_{121} = X_{211} = \frac{X^{+}}{\sqrt{3}} , \\
		X_{122} &= X_{212} = X_{221}  = \frac{X^{0}}{\sqrt{3}} , \ \ \ \ \ \ \ \ 
		X_{222} = X^{-}.
	\end{split}
\end{equation}
With this identification of components, the tensor representation is equivalent description to the column vector form. The scalar potential is given by,
\begin{equation}
	V = V_{\text{Higgs}} + V_{\text{4-plet}} + V_{\text{portal}},
\end{equation}
where,
\begin{equation}
	V_{\text{Higgs}} = - \mu_H^2 \Phi^\dagger \Phi + \lambda_H (\Phi^\dagger \Phi )^2, \label{eq:Higgs-potential}
\end{equation}
where $\Phi = \begin{pmatrix}
	\phi^+ & \phi^0
\end{pmatrix}^T$ is the SM Higgs doublet,
\begin{equation}
	\begin{split}
		V_{\text{4-plet}} = -\mu_X^2 \sum_{\text{indices}} &X^*_{i j k}\ X_{i j k} +  \lambda_{Q1} \sum_{\text{indices}} X^*_{i j k}\ X_{i j k}  \ X^*_{l m n}\  X_{l m n}\ \\ &+\  \lambda_{Q2}
		\sum_{\text{indices}} X^*_{i j k}\  X_{k l m}\ X^*_{l m n}\ X_{n i j}, \label{eq:4plet-self-potential}
	\end{split}
\end{equation} 
and
\begin{equation}
\begin{split}
V_{\text{portal}} = \lambda_{P1} \sum_{\text{indices}}^{} \Phi^*_{l} \Phi_{l}\ &X^*_{ijk}\  X_{ijk} + \lambda_{P2} \sum_{\text{indices}}^{} \Phi_l \  X^*_{lij}\ X_{ijk}\  \Phi^*_{k} \\ &+ \left(\lambda_{P3} \sum_{\text{indices}}^{} X_{imn}\, X_{jpq}\, \Phi^*_{i}\, \Phi^*_{j}\, \epsilon_{mp}\, \epsilon_{nq}  + \text{h.c.}\right) . \label{eq:4plet-portal-potential}
\end{split}
\end{equation}
All the summations of the indices in the above equations run from 1 to 2. The parameters $\lambda_{Q1}$, and $\lambda_{Q2}$ are quartic self-couplings of the 4-plet and $\lambda_{P1}$, $\lambda_{P2}$, and $\lambda_{P3}$ are the Higgs portal couplings. The term with coefficient $\lambda_{P3}$ is complex and we have to take the hermitian conjugate to make it real. In our analysis we choose the $\lambda_{P3}$ to be real. Later in this article we will explain the role of these couplings along with their residual phases. 

The electroweak symmetry breaking (EWSB) conditions are given by,
\begin{equation}
	\begin{split}
		\phi^0 = \begin{pmatrix}
			0\\
			\frac{1}{\sqrt{2}}(v_h + h )
		\end{pmatrix} ,
	\end{split}
\end{equation}
where $v_h =$ 246 $GeV$ is the vacuum expectation value (vev) of the SM Higgs. Since the $X$ is $\mathbb{Z}_2$-odd, it does not  get any vev and it does not mix with the SM neutral scalars components.  After the EWSB, the masses of the neutral scalar fields are given by,
\begin{equation}
\begin{split}
		M_h^2 &=  {2\lambda_H} v^2_h, \\
	    M^2_{X_R} &= {-\mu_X^2 + \frac{v_h^2 (3 \lambda_{P1} + 2 \lambda_{P2} - 4 \lambda_{P3})}{6} }, \\ 
	    M^2_{X_I} &= {-\mu_X^2 + \frac{v_h^2 (3 \lambda_{P1} + 2 \lambda_{P2} + 4 \lambda_{P3})}{6} }. \label{eq:4plet-neutral-masses}
\end{split}
\end{equation}
For the positive definite choices of all the portal couplings $X_R$ becomes the lightest $\mathbb{Z}_2$-odd particle, which can be a probable dark matter candidate as we discuss about it later.  The mass of the doubly charged scalar is given by
\begin{equation}
\begin{split}
		M^2_{X^{++}} &= -\mu_X^2 + \frac{v_h^2 \lambda_{P1}}{2}. \label{eq:4plet-doubly-charged-masses}
\end{split}
\end{equation}
The singly-charged scalars $X^+$, $(X^-)^*$  mixes with each other via $\lambda_{P3}\,v^2_h$, while they get diagonal mass contributions via $\lambda_{P1}v^2_h, \, \lambda_{P2}v^2_h$ and these are expressed via the mass matrix given below
\begin{equation}
	\begin{split}
		M^2_{\text{charged.}} = \begin{pmatrix}
			-\mu_X^2 + \frac{v_h^2}{6} (3 \lambda_{P1} + \lambda_{P2})  & \frac{\lambda_{P3} v_h^2}{\sqrt{3}} \\
			\frac{\lambda_{P3} v_h^2}{\sqrt{3}} & -\mu_X^2 + \frac{v_h^2}{6} (\lambda_{P1} + \lambda_{P2})
		\end{pmatrix}
	\end{split}
\end{equation}
The singly-charged mass eigenstates, discussed in \autoref{Appdx:singly-charged-mixing}, $X_1^\pm$, $X_2^\pm$ have masses given by,
\begin{equation}
\begin{split}
M^2_{X_1^\pm} &=  -\mu_X^2 + \frac{v_h^2}{6} \left(3 \lambda_{P1} + 2 \lambda_{P2} -  \sqrt{\lambda_{P2}^2 + 12 \lambda_{P3}^2}\right), \\
M^2_{X_2^\pm} &=  -\mu_X^2 + \frac{v_h^2}{6} \left(3 \lambda_{P1} + 2 \lambda_{P2} +  \sqrt{\lambda_{P2}^2 + 12 \lambda_{P3}^2}\right).
\label{eq:4plet-singly-charged-masses}
\end{split} .
\end{equation}

Notice that if the portal coupling $\lambda_{P3}$ vanishes, then the $X_R$ and $X_I$ become mass degenerate. Similarly, if the  parameters $\lambda_{P2}$ and $\lambda_{P3}$ both are zero, then singly charged and doubly charged Higgs bosons will be mass degenerate. For this requirement the real $X_R$ and imaginary $X_I$ part of the neutral scalar also becomes mass degenerate as can be seen from \autoref{eq:4plet-neutral-masses} giving rise to a complex dark matter.  

In conclusion, after EWSB, there are five BSM scalars namely a real scalar $X_R$, a pseudo-scalar $X_I$, two singly charged scalars $X^\pm_{1,2}$ and a doubly charged scalar $X^{\pm\pm}$.\footnote{We define $X^-_{1,2} = (X^+_{1,2})^*$ and $X^{--} = (X^{++})^*$.} The free parameters in the theory are the quartic self-couplings of the 4-plet, $\lambda_{Q1}$ and $\lambda_{Q2}$, the quartic portal couplings $\lambda_{P1}$, $\lambda_{P2}$ and $\lambda_{P3}$, and the bare mass coupling of 4-plet $\mu_X^2$. The derivation of the requirements for the potential to be bounded from below (BFB) are quite involved for higher multiplets of $SU(2)$, in particular for the quartet because of the added complexity in the potential due to its specific hypercharge $Y = \half$. We use the results in \cite{Kannike:2023bfh} to evaluate necessary and sufficient conditions for the tree-level stability for the potential described in the \autoref{eq:4plet-portal-potential}, \autoref{eq:4plet-self-potential}, and \autoref{eq:Higgs-potential}.
\begin{equation}\label{BFB}
	\begin{split}
		\lambda_H > 0, \ \ & \ \ \lambda_{Q1} > 0,\ \ \  \ \lambda_{Q1} + \lambda_{Q2} > 0, \\
		\lambda_{P1} \ge & - 2 \sqrt{\lambda_{H} (\lambda_{Q1} + \lambda_{Q2})}, \\
		\lambda_{P1} + \lambda_{P2} + 2 |\lambda_{P3}|\ \ge & - 2 \sqrt{\lambda_{H} (\lambda_{Q1} + \half \lambda_{Q2})}.
	\end{split}
\end{equation}

\section{RG evolution and Fixed Points}\label{RGEs}
In this section we shall investigate the evolution of the running Higgs quartic couplings with the renormalization scale $\mu$. 
We are interested in the RG evolution of the quartic couplings. The two-loop $\beta$-functions were calculated by implementing the model into  \sarah \cite{Staub:2013tta} and are given in \autoref{appendix:beta-functions-4plet}. The one-loop $\beta$-function for the SM-like Higgs quartic coupling $\lambda_H$ is given in the \autoref{eq:4plet-beta-1lp-lambdaH}. 
\begin{equation}
	\begin{split}
\beta^{\text{1-loop}}_{\lambda_H} &=	\frac{1}{16\pi^2}\Big(\frac{27}{200}g_1^4 + \frac{9}{20} g_1^2 g_2^2 + \frac{9}{8} g_2^4 + 24 \lambda_H^2 + \frac{13}{4} \lambda_{P1}^2 + \frac{7}{2} \lambda_{P1} \lambda_{P2} \\ &\ \ \ \ + \frac{53}{36} \lambda_{P2}^2 + \frac{28}{9} \lambda_{P3}^2 + 12 \lambda_H Y_t^2 - \frac{9}{5} g_1^2 \lambda_H - 
9 g_2^2 \lambda_H - 6 Y_t^4\Big). \label{eq:4plet-beta-1lp-lambdaH}
	\end{split}
\end{equation}
As expected there are positive contributions proportional to $\lambda^2_{P1, P2, P3}$ and $\lambda^2_{H}$ and negative contribution proportional to $Y^4_t$, which again reiterates the fact that the addition of scalars bring stability to the Higgs potential \cite{Jangid:2020dqh,Jangid:2020qgo, Bandyopadhyay:2016oif,Bandyopadhyay:2021kue,Bandyopadhyay:2021ipw}.  However, the two-loop $\beta$-function which are given in \autoref{appendix:beta-functions-4plet} are used for the numerical evolution of the couplings and presented in the plots along with the one-loop ones.

Keeping the spirit of finding the Fixed Point in inert BSM models in \cite{Bandyopadhyay:2025ilx}, we focus on the same for inert $SU(2)$ quartet model in this article. Since the 4-plet has hypercharge same as the inert doublet in IDM and the scalar potentials of both models contain a complex term and its hermitian conjugate, it is more sensible to compare these two models. Here by Fixed Point we mean that only $\beta$-functions corresponding to the scalar couplings are approximately vanishing. Also, as we detailed in our previous work that some couplings may cross perturbative limit while attaining the constant value. However, we are interested in knowing how the structure of potential that affect the occurrence of these Fixed Points \cite{Bandyopadhyay:2025ilx}.

\begin{figure}[h]
	\begin{center}
		\mbox{\hspace*{-0.3cm}
			\subfigure[]{\includegraphics[height=0.228\textheight, width=0.42\linewidth,angle=0]{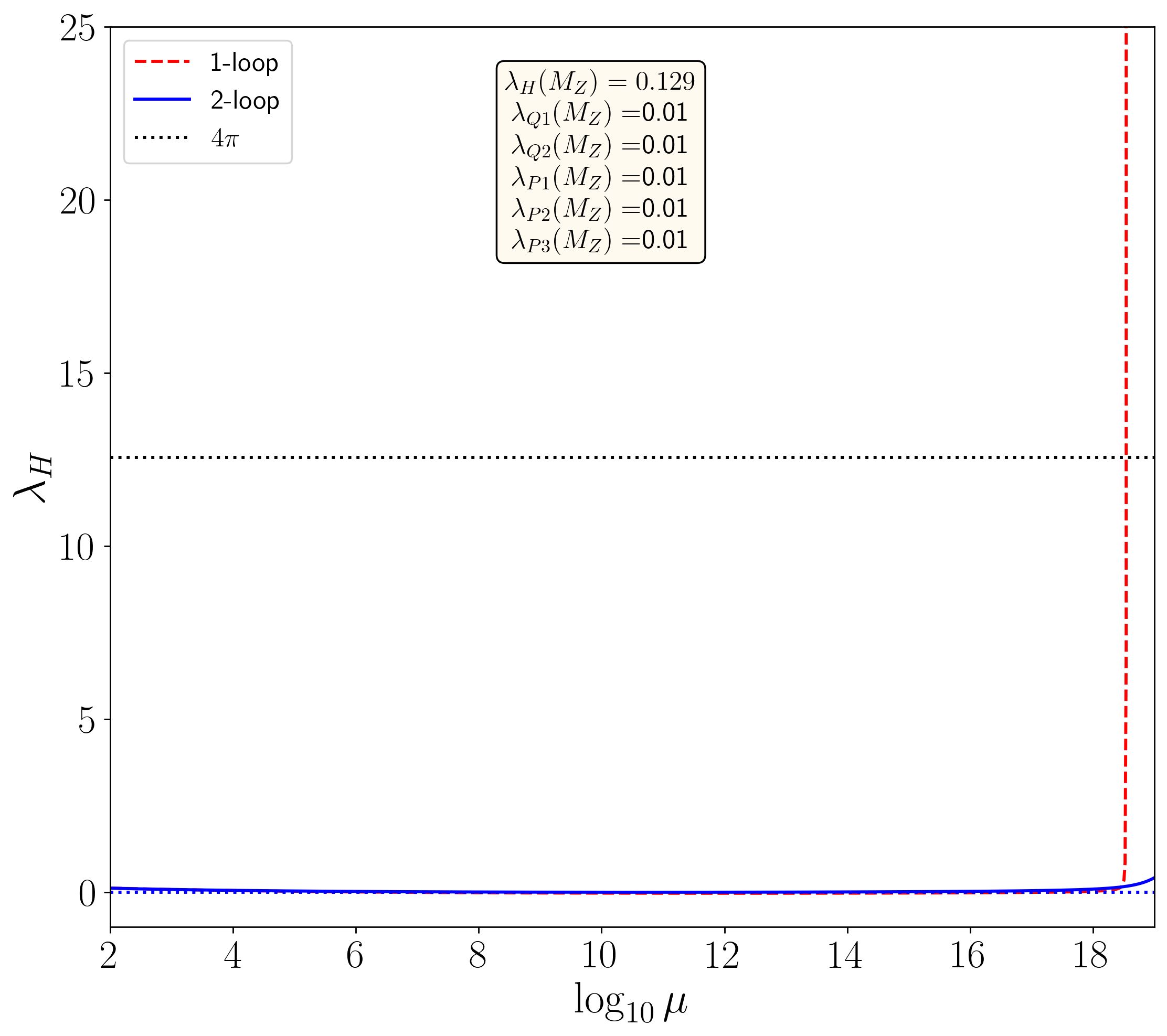}}	
			\hspace*{0.25cm}
			\subfigure[]{\includegraphics[height=0.23\textheight,width=0.4\linewidth,angle=0]{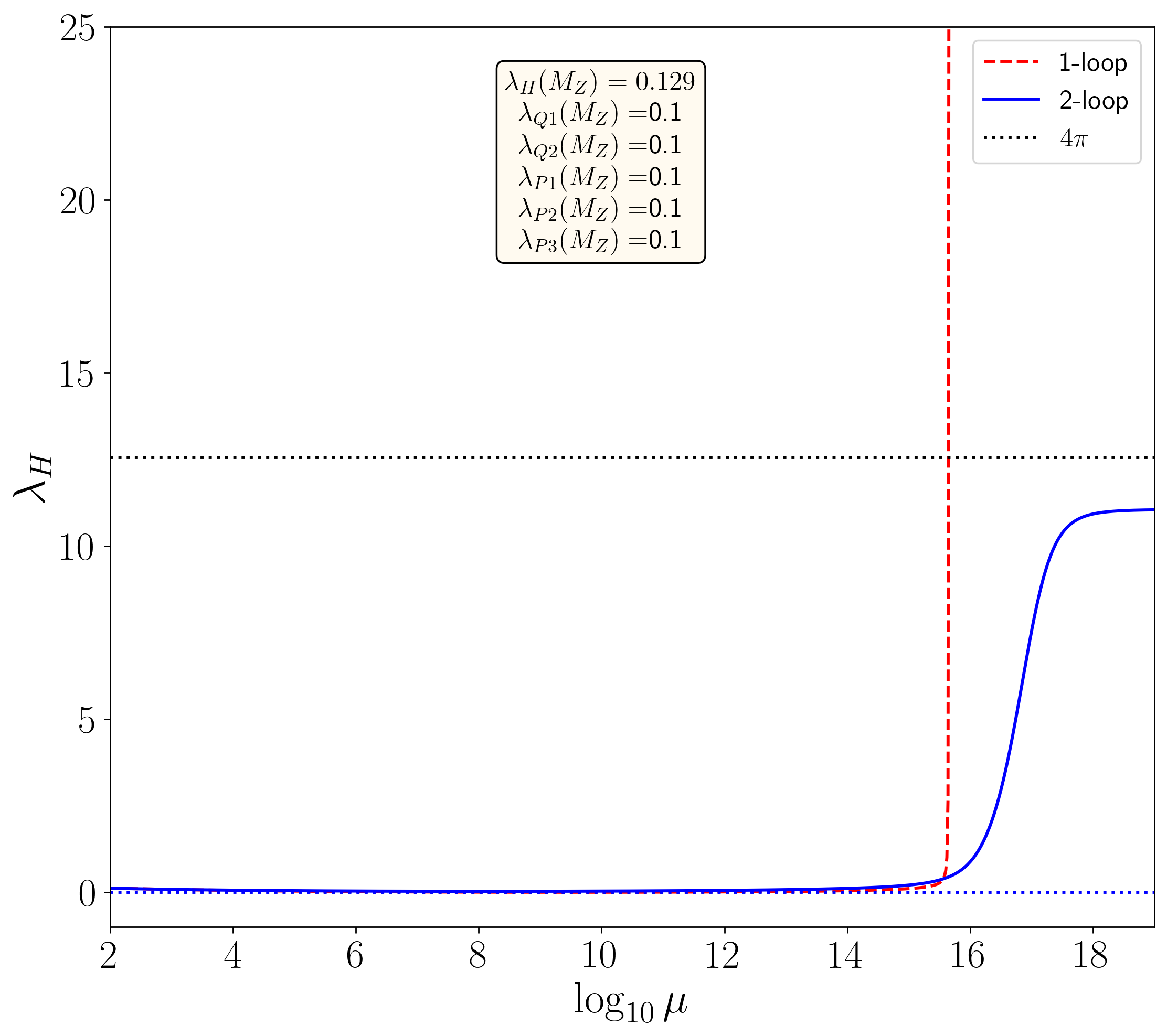}}}
		\mbox{\subfigure[]{\includegraphics[height=0.23\textheight,width=0.4\linewidth,angle=0]{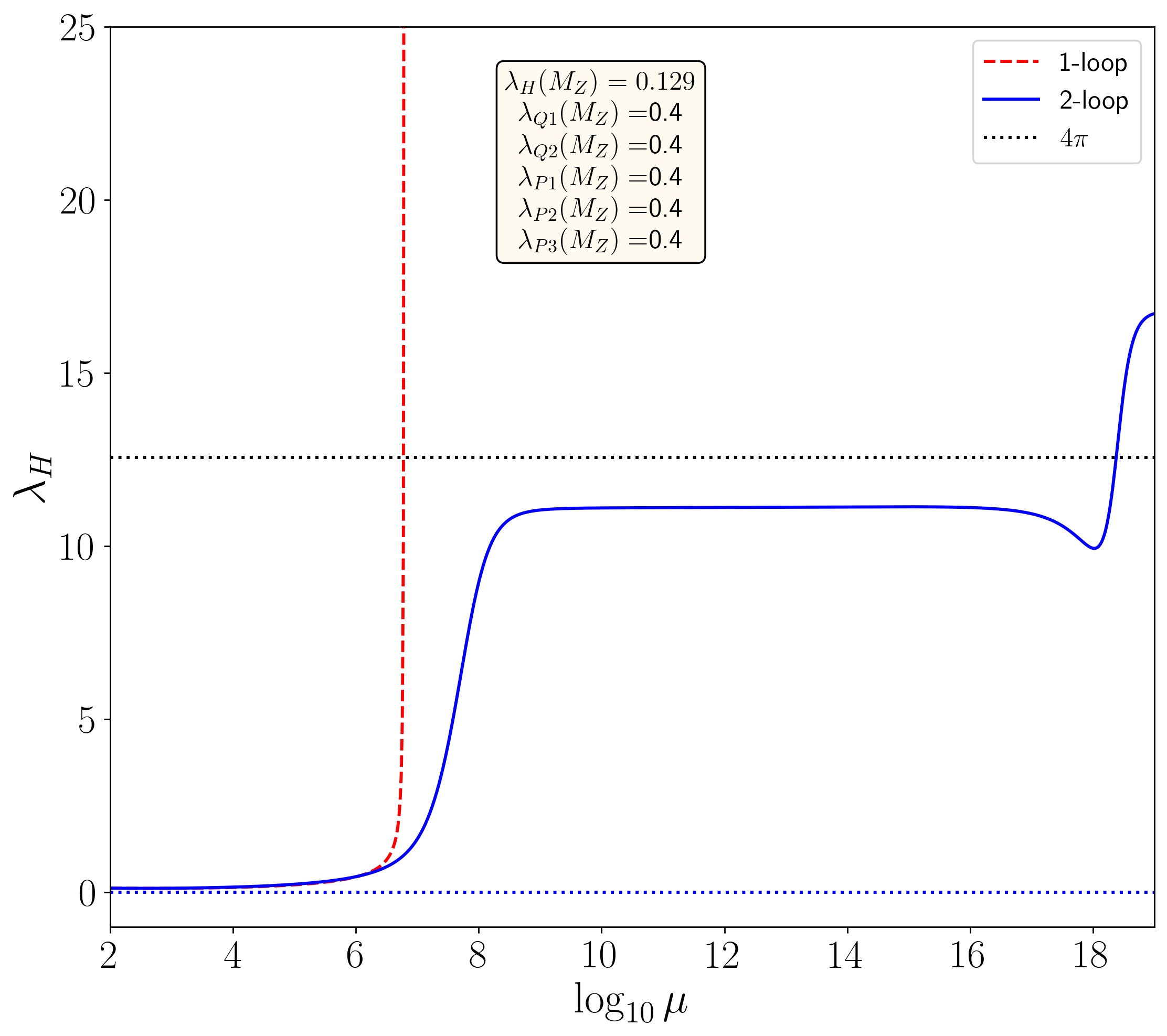}}	
			\hspace*{0.25cm}
			\subfigure[]{\includegraphics[height=0.23\textheight,width=0.4\linewidth,angle=0]{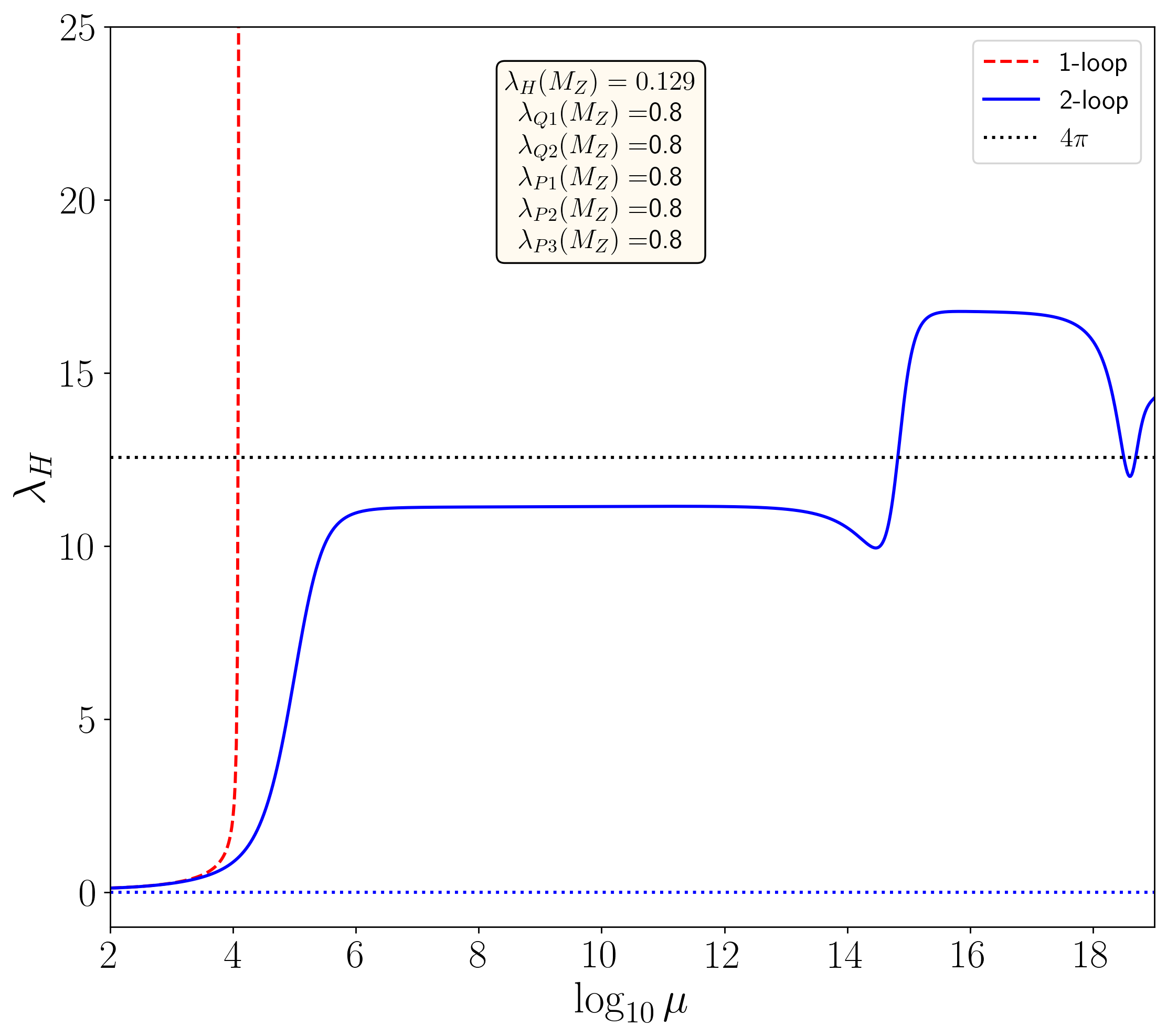}}}
		\caption{$\lambda_H - \log{\mu}$ for I4M  at one-loop and two-loop  where $\lambda_{Q1}(M_Z)=\lambda_{Q2}(M_Z)= \lambda_{P1}(M_Z)=\lambda_{P2}(M_Z)=\lambda_{P3}(M_Z)$ for (a), (b), (c), (d) are $0.01$ (very low), $0.1$ (low), $0.4$ (moderate), $0.8$ (high) respectively where $\lambda_H(M_Z) = 0.129$ for all cases. }\label{fig:I4M_evol_LH}
	\end{center}
\end{figure}

The evolution of SM-like Higgs quartic coupling $\lambda_H$ is shown in \autoref{fig:I4M_evol_LH}. The red dashed curve denotes the one-loop evolution and blue solid curve represents the two-loop evolution. Each sub-figure corresponds to different initial values of new physics quartic couplings, namely $\lambda_{Q1}$, $\lambda_{Q2}$, $\lambda_{P1}$, $\lambda_{P2}$, and $\lambda_{P3}$, where the initial value of SM-like Higgs coupling is taken to be always $\lambda_H(M_Z) = 0.129$ at the electroweak scale. 
\autoref{fig:I4M_evol_LH}(a) shows the evolution of $\lambda_{H}$ for very low initial values of $\lambda_{Q1}(M_Z)=\lambda_{Q2}(M_Z)= \lambda_{P1}(M_Z)=\lambda_{P2}(M_Z)=\lambda_{P3}(M_Z) = 0.01$. The one-loop evolution hits a Landau pole at around $\mu \sim 10^{18}$ GeV, which is very different from SM without any LP.  For achieving such LPs at one-loop for the lower $SU(2)$ multiplets viz.  ISM/IDM/ITM,  we require  relatively larger quartic couplings  \cite{Bandyopadhyay:2025ilx}. Similar is the observation for the Fixed Point. This happens due to  higher degrees of freedom in case of  quartet, the appearances of LP and FP happen at much lower couplings and at much lower scales for relatively higher quartic couplings as can be seen from other graphs in  \autoref{fig:I4M_evol_LH}.

 The one- and two-loop evolution of $\lambda_H$ for SM is given in Figure 3(b) of \cite{Bandyopadhyay:2025ilx}, and the evolution for IDM for an initial value of $\lambda_{i\ne1}(M_Z) = 0.01$ in figure 14(a) of \cite{Bandyopadhyay:2025ilx}. In contrast to these two, the two-loop evolution of I4M is positive for all energy values. However, the evolution begins to increase after $\mu = 10^{18}$ GeV and no FP is present till the Planck scale. In \autoref{fig:I4M_evol_LH}(b), we choose a low initial value of $\lambda_{Q1}(M_Z)=\lambda_{Q2}(M_Z)= \lambda_{P1}(M_Z)=\lambda_{P2}(M_Z)=\lambda_{P3}(M_Z) = 0.1$. Here the one-loop evolution hits a LP around $\mu\sim 10^{15.5}$ GeV and the two-loop evolution attains an FP at an energy scale slightly less than $\mu = 10^{18}$ GeV. This is a drastic change in behaviour compared to the models analyzed in \cite{Bandyopadhyay:2025ilx}. Appearance of FP (and LP) for such low values of quartic coupling is not seen in ISM, ITM, or IDM. For moderate value of the couplings, $\lambda_{Q1}(M_Z)=\lambda_{Q2}(M_Z)= \lambda_{P1}(M_Z)=\lambda_{P2}(M_Z)=\lambda_{P3}(M_Z) = 0.4$, in \autoref{fig:I4M_evol_LH}(c), the one-loop LP appears at a lower energy scale $\mu \sim 10^{6.5}$ GeV and the two-loop FP appears around $\mu \sim 10^{8}$ GeV. The two-loop evolution remains constant till $\mu\sim 10^{17}$ and after that it begin to vary and then increase drastically such that it crosses the value $4\pi$. For higher initial value, $\lambda_{Q1}(M_Z)=\lambda_{Q2}(M_Z)= \lambda_{P1}(M_Z)=\lambda_{P2}(M_Z)=\lambda_{P3}(M_Z) = 0.8$, the one-loop evolution is shown in \autoref{fig:I4M_evol_LH}(d) attains the LP around $\mu \sim 10^{4}$ GeV. The two-loop achieves a FP around $\mu\sim 10^{6}$ GeV and remains constant till $10^{14}$ GeV. After that the curve increases and attains a secondary FP with the $\lambda_H(\mu) > 4\pi$. This is similar to the behaviour shown by $\lambda_H$ in ITM for higher initial values  $\lambda_{HT}(M_Z), \lambda_{T}(M_Z) = 0.8, 1.2$ as shown in figure 10(c), (d) in \cite{Bandyopadhyay:2025ilx}.

\begin{figure}[h]
	\begin{center}
		\mbox{\hspace*{-0.3cm}
			\subfigure[]{\includegraphics[height=0.228\textheight, width=0.42\linewidth,angle=0]{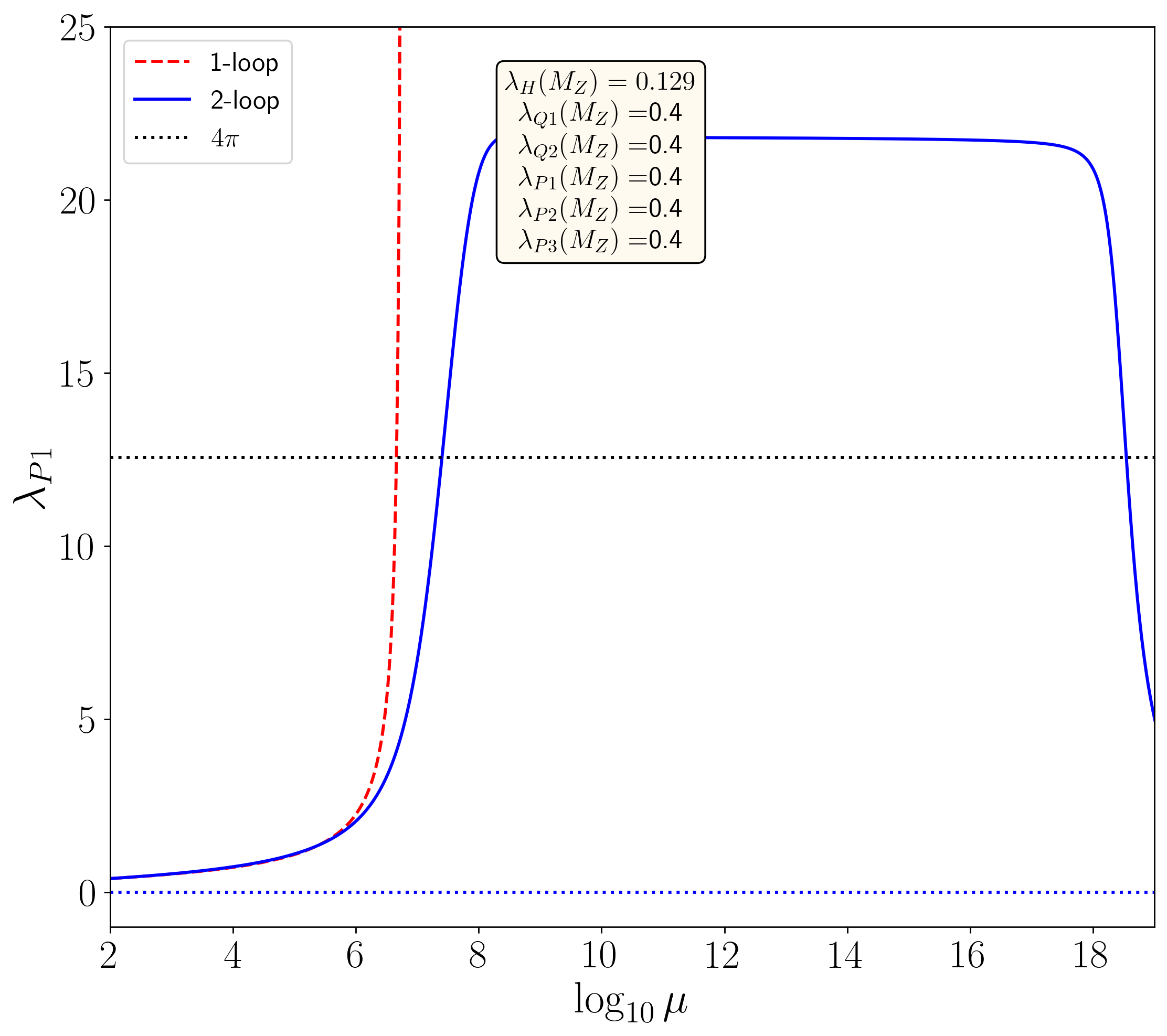}}	
			\hspace*{0.25cm}
			\subfigure[]{\includegraphics[height=0.23\textheight,width=0.4\linewidth,angle=0]{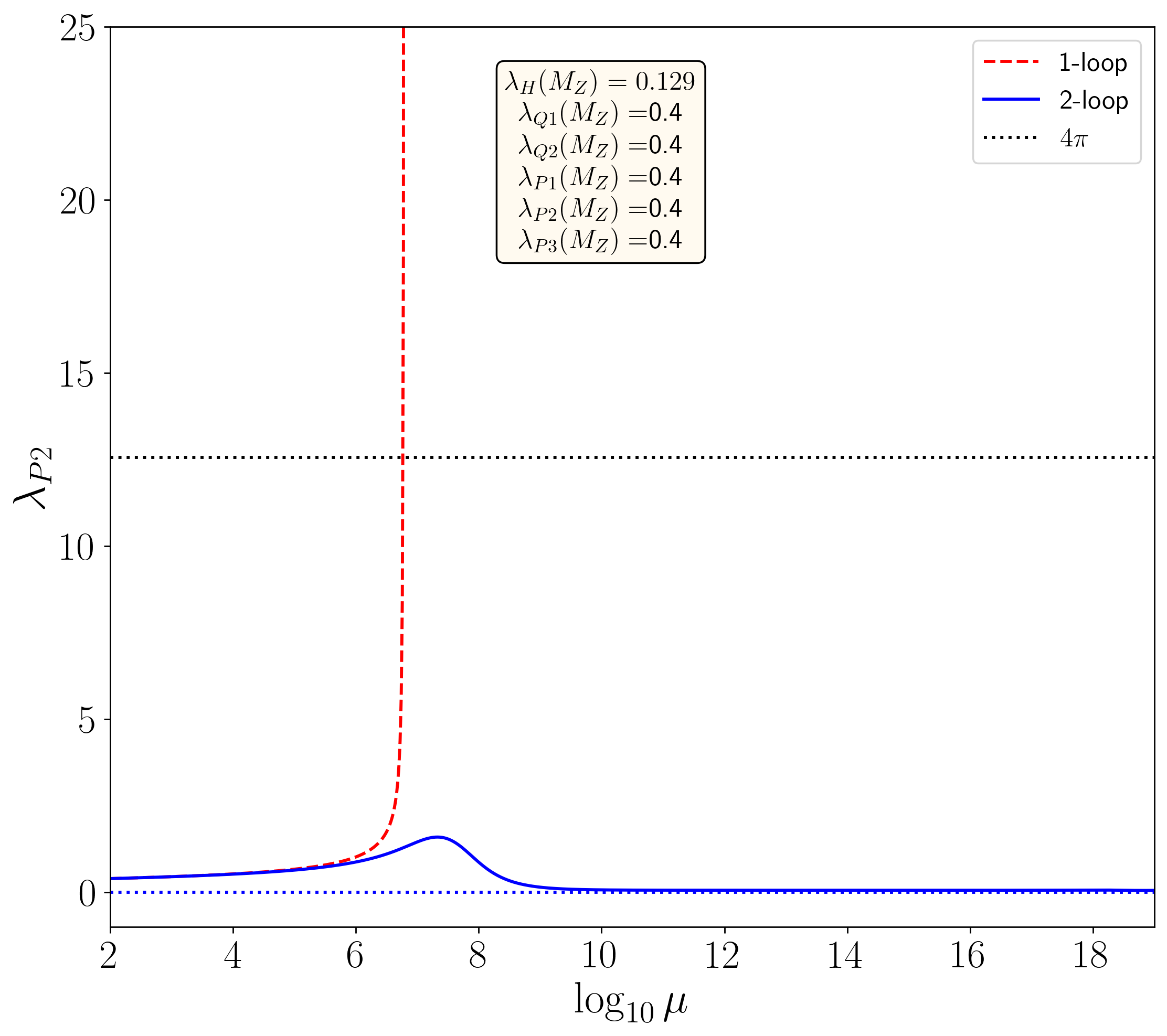}}}
		\mbox{\subfigure[]{\includegraphics[height=0.23\textheight,width=0.4\linewidth,angle=0]{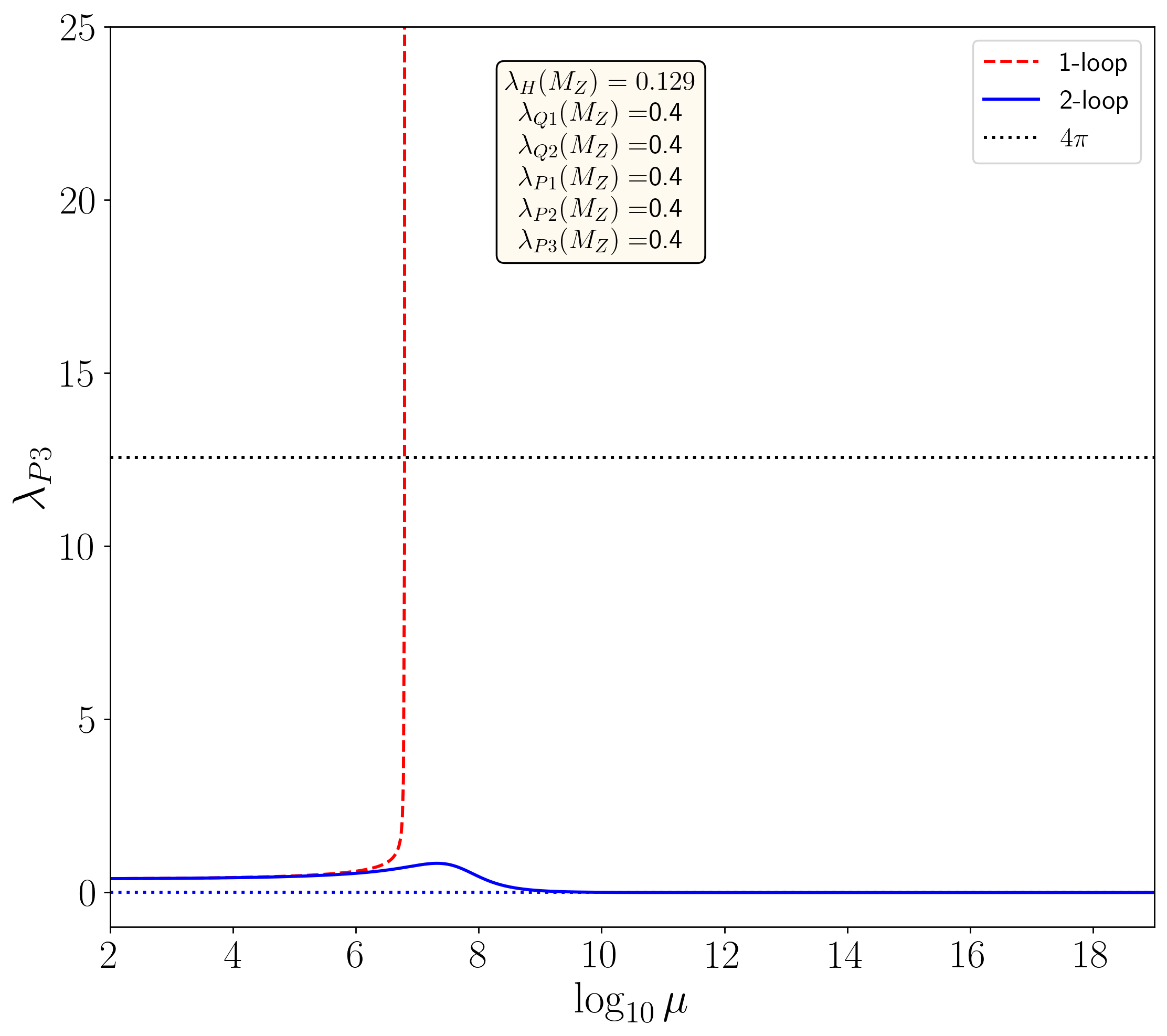}}	
			\hspace*{0.25cm}}
		\caption{$\lambda_{P1} - \log{\mu}$ in (a), $\lambda_{P2} - \log{\mu}$ in (b) and $\lambda_{P3} - \log{\mu}$ in (c) are depicted for I4M  at one-loop and two-loop  for  $\lambda_{Q1}(M_Z)=\lambda_{Q2}(M_Z)= \lambda_{P1}(M_Z)=\lambda_{P2}(M_Z)=\lambda_{P3}(M_Z) = 0.4$ and $\lambda_H(M_Z) = 0.129$. }\label{fig:I4M_evol_LP1_LP2_LP3}
	\end{center}
\end{figure}

In \autoref{fig:I4M_evol_LP1_LP2_LP3}, the one- and two-loop variations of the portal couplings $\lambda_{P1}$, $\lambda_{P2}$, and $\lambda_{P3}$ for the initial values $\lambda_{Q1}(M_Z)=\lambda_{Q2}(M_Z)= \lambda_{P1}(M_Z)=\lambda_{P2}(M_Z)=\lambda_{P3}(M_Z) = 0.4$ are given. In \autoref{fig:I4M_evol_LP1_LP2_LP3}(a), $\lambda_{P1}$ attains a FP behaviour around $\mu \sim 10^{8}$ and continue to have a constant value till $\mu\sim 10^{17}$. However, \autoref{fig:I4M_evol_LP1_LP2_LP3}(b), and (c) shows that $\lambda_{P2}$ and $\lambda_{P3}$ attain a constant value around $\mu \sim 10^{8}$ but it happens to be very small but non-vanishing. We should notice that the FP value of $\lambda_{P1}$ is greater than $4\pi$ which was similar to the discussions in \cite{Bandyopadhyay:2025ilx}, namely $\lambda_{HS}$ in ISM, $\lambda_{HT}$ in ITM, and $\lambda_3$ in IDM.

\begin{figure}[h]
	\begin{center}
		\mbox{\hspace*{-0.3cm}
			\subfigure[]{\includegraphics[height=0.228\textheight, width=0.42\linewidth,angle=0]{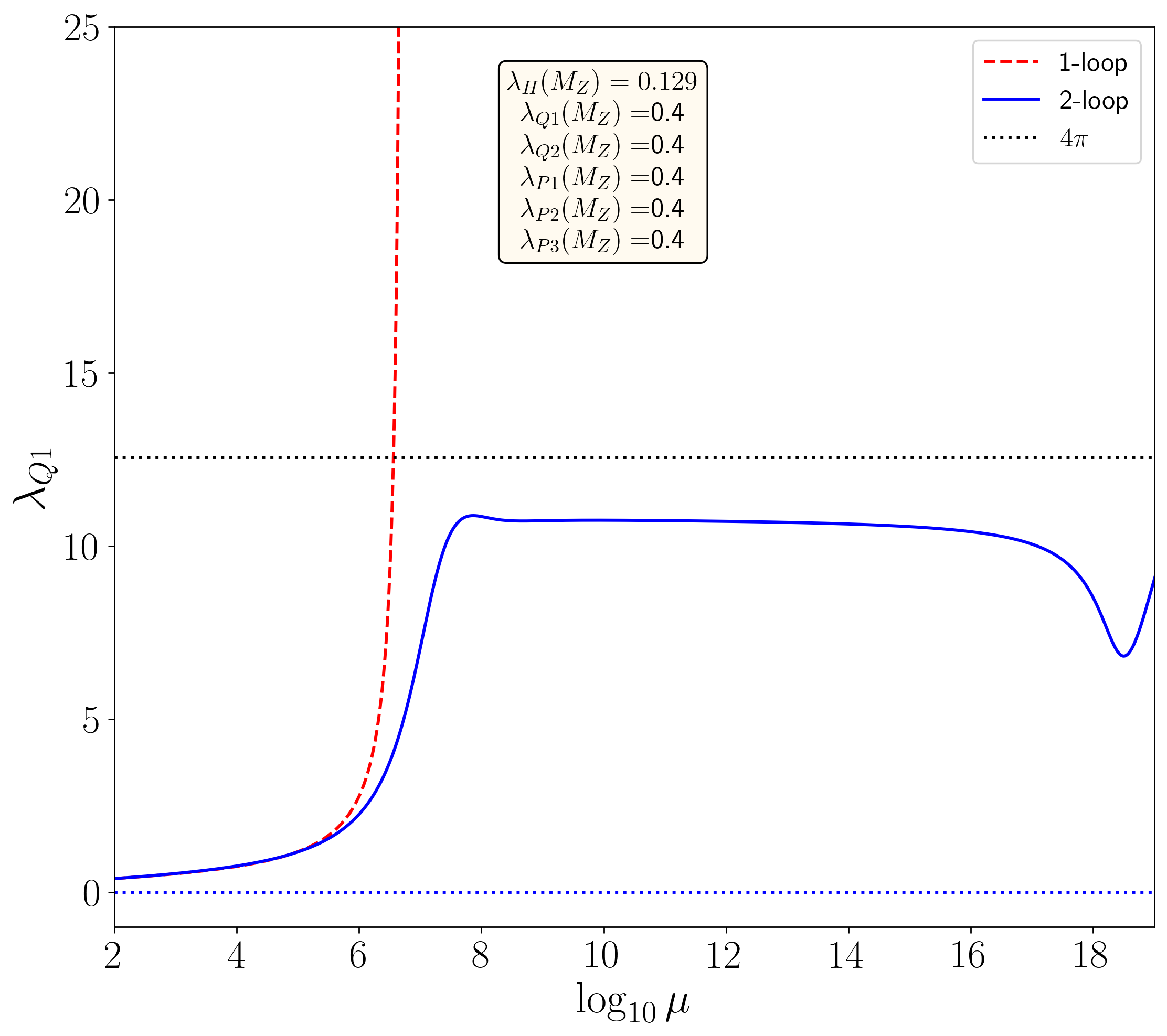}}	
			\hspace*{0.25cm}
			\subfigure[]{\includegraphics[height=0.23\textheight,width=0.4\linewidth,angle=0]{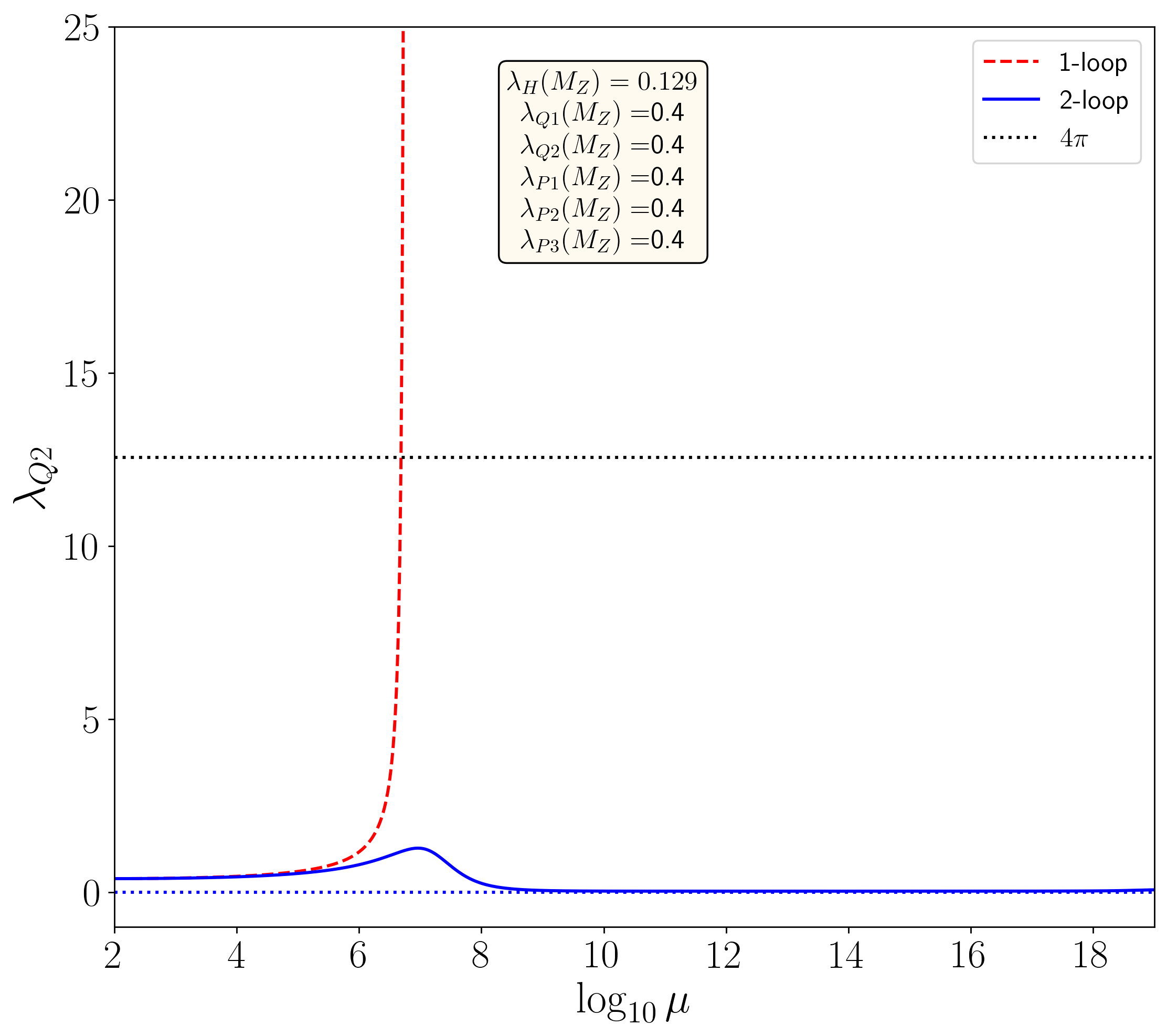}}}
		\caption{Evolution of $\lambda_{Q1}$ and $\lambda_{Q2}$ are depicted in (a) and (b) for I4M  at one-loop and two-loop where $\lambda_{Q1}(M_Z)=\lambda_{Q2}(M_Z)= \lambda_{P1}(M_Z)=\lambda_{P2}(M_Z)=\lambda_{P3}(M_Z) = 0.4$ and $\lambda_H(M_Z) = 0.129$. }\label{fig:I4M_evol_LQ1_LQ2}
	\end{center}
\end{figure}
Similarly, the evolution of self-couplings of the quartet, $\lambda_{Q1}$ and $\lambda_{Q2}$ are given in \autoref{fig:I4M_evol_LQ1_LQ2} for the initial values $\lambda_{Q1}(M_Z)=\lambda_{Q2}(M_Z)= \lambda_{P1}(M_Z)=\lambda_{P2}(M_Z)=\lambda_{P3}(M_Z) = 0.4$. In \autoref{fig:I4M_evol_LQ1_LQ2}(a), $\lambda_{Q1}$ attains a FP around 
$\mu \sim 10^{8}$ and continue to have a constant value till $\mu\sim 10^{17}$ where the value $\lambda_{Q1} < 4\pi$. Meanwhile, the $\lambda_{Q2}$ becomes non-zero but small constant value around the FP energy scale and then continue to stay like that similar to $\lambda_{P2}$ and $\lambda_{P3}$, as shown in  \autoref{fig:I4M_evol_LQ1_LQ2}(b)

In \cite{Bandyopadhyay:2025ilx}, we showed that the terms in the scalar potential that do not have any relative phases among the components of the fields, promotes the occurrence of FP as in the cases of ISM and ITM, whereas the terms with residual phases tend to spoil such occurrences  as in the case of IDM. In order to have a similar analysis we redefine the fields using their complex-polar form as,
\begin{equation}
\begin{split}
	\phi^+ &= |\phi^+|\, \exp(i\theta^+),\ \ \ \phi^0 = |\phi^0|\, \exp(i\theta^0),\ \ \ X^{++} = |X^{++}|\, \exp(i\tilde{\theta}^{++}),\\ 
	X^+ &= |X^+|\, \exp(i\tilde{\theta}^+),\ \ \ X^0 = |X^0|\, \exp(i\tilde{\theta}^0),\ \ \ X^- = |X^-|\, \exp(i\tilde{\theta}^-).
\end{split}
\end{equation}

The quartic terms in the Lagrangian that do not have any residual phases can be written as:
\begin{equation}\label{nophaseterms}
\begin{split}
\lambda_H (\Phi^\dagger \Phi )^2 &= (|{\phi^0}|^2 + |{\phi^+}|^2 )^2,\\
\lambda_{Q1} \sum X^*_{i j k}\ X_{i j k}  \ X^*_{l m n}\  X_{l m n} &= (|{X^0}|^2 + |{X^-}|^2 + |{X^+}|^2 + |{X^{++}}|^2)^2,\\
\lambda_{P1} \sum \Phi^*_{l} \Phi_{l}\ X^*_{ijk}\  X_{ijk} &= |{\phi^0}|^2 |{X^0}|^2 + |{\phi^0}|^2 |{X^-}|^2 + |{\phi^0}|^2 |{X^+}|^2 + |{\phi^0}|^2 |{X^{++}}|^2 \\ &\ \ \ \ + |{\phi^+}|^2 |{X^0}|^2 + |{\phi^+}|^2 |{X^-}|^2 + |{\phi^+}|^2 |{X^+}|^2 + |{\phi^+}|^2 |{X^{++}}|^2.
\end{split}
\end{equation}
Here the $\lambda_H$, $\lambda_{Q1}$, $\lambda_{P1}$-terms solely depend on the absolute values of the fields and do not have any residual phases. However, if we evaluate the $\lambda_{Q2}$, $\lambda_{P2}$, $\lambda_{P3}$-terms, we get terms with residual phases similar to IDM \cite{Bandyopadhyay:2025ilx}. For example let us consider $\lambda_{Q2}$ terms as given below
\begin{equation}\label{Q2phase}
	\begin{split}
		\lambda_{Q2}
		\sum_{} &X^*_{i j k}\  X_{k l m}\ X^*_{l m n}\ X_{n i j} \\ &=\lambda_{Q2}\, \Big( |{X^{++}}|^4 \, +  \frac{5}{9} |{X^{+}}|^4\,+\,\frac{5}{9} |{X^{0}}|^4\, +\,| {X^{-}}|^4\, + \, \frac{16}{9} | {X^{0}}|^2 | {X^{+}}|^2\, \\
		&\ \ \ + 2 | {X^{0}}|^2 | {X^{-}}|^2\,  +\, \frac{2}{3} | {X^{0}}|^2 | {X^{++}}|^2\,+\, \frac{2}{3} | {X^{-}}|^2 | {X^{+}}|^2 \,+\,  2 | {X^{+}}|^2 | {X^{++}}|^2 \\   
		&\ \ \ + \frac{4}{3} | {X^{0}}|  | {X^{-}}|  | {X^{+}}|  | {X^{++}}|  \cos  ({\tilde{\theta}^{0}}-{\tilde{\theta}^{-}}+{\tilde{\theta}^{+}}-{\tilde{\theta}^{++}})\\  
		&\ \ \ + \frac{8}{3 \sqrt{3}}  | {X^{0}}|  | {X^{+}}|^2 | {X^{++}}|\,  \cos({\tilde{\theta}^{0}}-2 {\tilde{\theta}^{+}}+{\tilde{\theta}^{++}}) \\  
		&\ \ \  +\frac{8}{3\sqrt{3}}  | {X^{0}}|^2 | {X^{-}}|  | {X^{+}}|\,  \cos(2 {\tilde{\theta}^{0}}-{\tilde{\theta}^{-}}-{\tilde{\theta}^{+}}) \Big).
	\end{split}
\end{equation}
In this expansions of $\lambda_{Q2}$, there are three terms that contains residual phases out of total twelve terms. These corresponds to three independent residual  phases $(\tilde{\theta}^{0}-{\tilde{\theta}^{-}}+{\tilde{\theta}^{+}}-\tilde{\theta}^{++}),\, (\tilde{\theta}^{0}-2 \tilde{\theta}^{+}+\tilde{\theta}^{++}),\, (\tilde{\theta}^{-}-\tilde{\theta}^{+})$  which are also showed in \autoref{figfeynman}.  $\lambda_{P2}$ expansion has three terms without residual phases out of nine terms and $\lambda_{P3}$ terms have six residual phases as detailed in \autoref{ResPhase}. As we have seen in \cite{Bandyopadhyay:2025ilx} for IDM such terms with residual phases tend to spoil the occurrence of the FP. In contrast to IDM, here there are several individual terms originating from a single term in the potential that has residual phases with even more numbers. In \autoref{rediphase} we present the different relative phases for different quartic couplings coming from I4M part as given by

\begin{equation}
\begin{split}
V_{\mathrm{4-plet}} &+ V_{\mathrm{portal}} = \lambda_{\mathrm{Q1}} \mathrm{( 10-without\, phases,\, 0-phases )}  \\ 
&\, + \lambda_{\mathrm{P1}} \mathrm{( 8-without\, phases,\, 0-phases )}+ \lambda_{\mathrm{Q2}} \mathrm{( 9-without\, phases,\, 3-phases )} \\ 
&\, + \lambda_{\mathrm{P2}} \mathrm{( 6-without\, phases,\, 3-phases )} + \lambda_{\mathrm{P3}} \mathrm{( 0-without\, phases,\, 6-phases )}.
\end{split}\label{rediphase}
\end{equation}

Later in the paragraph we will see that how existence of terms without phases, i.e. modulus only terms enhances the possibility of appearance of FPs, where as the existence of the residual phases tend to spoil such possibilities. Here the situation is little more involved compared to IDM, where certain terms do not have  any residual phases viz., $\lambda_{1,2,3}$ act as an enhancer and $\lambda_{4}$ has two terms without phase and one term with phase acts as moderate spoiler, while $\lambda_{5}$ have three terms with relative phases acts as heavy spoiler\cite{Bandyopadhyay:2025ilx}. Given that $\lambda_{Q2}, \lambda_{P1}$ have both terms with and without phases their behaviour depends on how they appear in the $\beta$-function and gets multiply with other viz., $\lambda_{P3}$.

\begin{figure}[!h]
	\begin{center}
		\begin{tikzpicture}[scalar/.style={dashed},	fermion/.style={dashed, postaction={decorate}, decoration={markings, mark=at position 0.5 with {\arrow{>}}}} ]
			\def\side{1.0}
			\def\pone{5*\side}
			\def\ptwo{10.0*\side}
			\def\pthree{15*\side}						
			
			\draw[fermion] (0,0) -- (-\side,\side);
			\draw[fermion] (-\side,-\side) -- (0,0);
			\draw[fermion] (0,0) -- (\side,\side) ;
			\draw[fermion] (\side,-\side) -- (0,0);
			
			\node at (-\side-0.5,\side) {\Large $X$};
			\node at (-\side-0.5,-\side) {\Large $X^\dagger$};
			\node at (\side+0.5,\side) {\Large $X$};
			\node at (\side+0.5,-\side) {\Large $X^\dagger$}; 
			
			\node[right] at (2.5*\side-0.4+0.2,0) {\Large $=$};
			\node[right] at (7.5*\side-0.4+0.2,0) {\Large $+$};
			\node[right] at (2.9*\side,-3*\side-1.0) {\Large $+$};
			\node[right] at (7.5*\side,-3*\side-1.0) {\Large $+$};

			\draw[scalar] (\pone,0) -- (-\side+\pone,\side);
			\draw[scalar] (-\side +\pone,-\side) -- (\pone,0);
			\draw[scalar]   (\pone,0) -- (\side+\pone,\side) ;
			\draw[scalar] (\side+\pone,-\side) -- (\pone,0);
			
			\node at (-\side+\pone-0.6,\side) {\Large $X^0$};
			\node at (-\side+\pone-0.6,-\side) {\Large $X^-$};
			\node at (\side+\pone+0.6,\side) {\Large $X^+$};
			\node at (\side+\pone+0.6,-\side) {\Large $X^{++}$}; 
			
			\node at (\pone-0.2,-\side-1.0) {\large $\sim \lambda_{Q2} \text{Re}[e^{i ({\tilde{\theta}^{0}}-{\tilde{\theta}^{-}}+{\tilde{\theta}^{+}}-{\tilde{\theta}^{++}})}] $};

			\draw[fermion] (\ptwo,0) -- (-\side+\ptwo,\side);
			\draw[fermion] (-\side +\ptwo,-\side) -- (\ptwo,0);
			\draw[fermion]   (\ptwo,0) -- (\side+\ptwo,\side) ;
			\draw[fermion] (\side+\ptwo,-\side) -- (\ptwo,0);
			
			\node at (-\side+\ptwo-0.6,\side) {\Large $|X^0|$};
			\node at (-\side+\ptwo-0.6,-\side) {\Large $|X^+|$};
			\node at (\side+\ptwo+0.7,\side) {\Large $|X^+|$};
			\node at (\side+\ptwo+0.7,-\side) {\Large ${|X^{++}|}$}; 
			\node at (\ptwo-0.2,-\side-1.0) { $\sim \lambda_5 \text{Re}[e^{i({\tilde{\theta}^{0}}-2 {\tilde{\theta}^{+}}+{\tilde{\theta}^{++}})}] $};
			
			
			\draw[scalar] (\pone+0.6,-4*\side) -- (-\side+\pone+0.6,\side-4*\side);
			\draw[scalar] (-\side +\pone+0.6,-\side-4*\side) -- (\pone+0.6,-4*\side);
			\draw[scalar]   (\pone+0.6,-4*\side) -- (\side+\pone+0.6,\side-4*\side) ;
			\draw[scalar] (\side+\pone+0.6,-\side-4*\side) -- (\pone+0.6,-4*\side);
			
			\node at (-\side+\pone,\side -3*\side-1.0) {\Large $|X^0|$};
			\node at (-\side+\pone,-\side-3*\side-1.0) {\Large $|X^0|$};
			\node at (\side+\pone+1.2,\side-3*\side-1.0) {\Large $|X^+|$};
			\node at (\side+\pone+1.2,-\side-3*\side-1.0) {\Large $|X^+|$}; 
			
			\node at (\pone ,-\side-1.0-3*\side-1.0) { $\sim \lambda_5 \text{Re}[e^{i(2 {\tilde{\theta}^{0}}-{\tilde{\theta}^{-}}-{\tilde{\theta}^{+}})}] $};

			\node[right] at (7.5*\side,-3*\side-1.0) {\Large $+$};
			\node[right] at (7.5*\side+0.5,-3*\side-1.0) { (\text{terms independent of phases})};
		\end{tikzpicture}
		\caption{Diagrams showing relative residual phases associated with $\lambda_{5}$ term.}\label{figfeynman}
	\end{center}
\end{figure}

 To establish further the relationship between the terms with residual phases and the appearance of FPs, we could define the following set of ratios, 
\begin{equation}
\begin{split}
		\alpha_{P1} &= \frac{\lambda_{P1}(M_Z)}{\lambda_{Q1}(M_Z)}, \ \ \ \alpha_{P2} = \frac{\lambda_{P2}(M_Z)}{\lambda_{Q1}(M_Z)},\ \ \  \alpha_{P3} = \frac{\lambda_{P3}(M_Z)}{\lambda_{Q1}(M_Z)},\ \ \ \alpha_{Q2} = \frac{\lambda_{Q2}(M_Z)}{\lambda_{Q1}(M_Z)}.
\end{split}
\end{equation}


\begin{figure}[h]
\begin{center}
		\mbox{
		\subfigure[]{\includegraphics[height=0.17\textheight,width=0.3\linewidth]{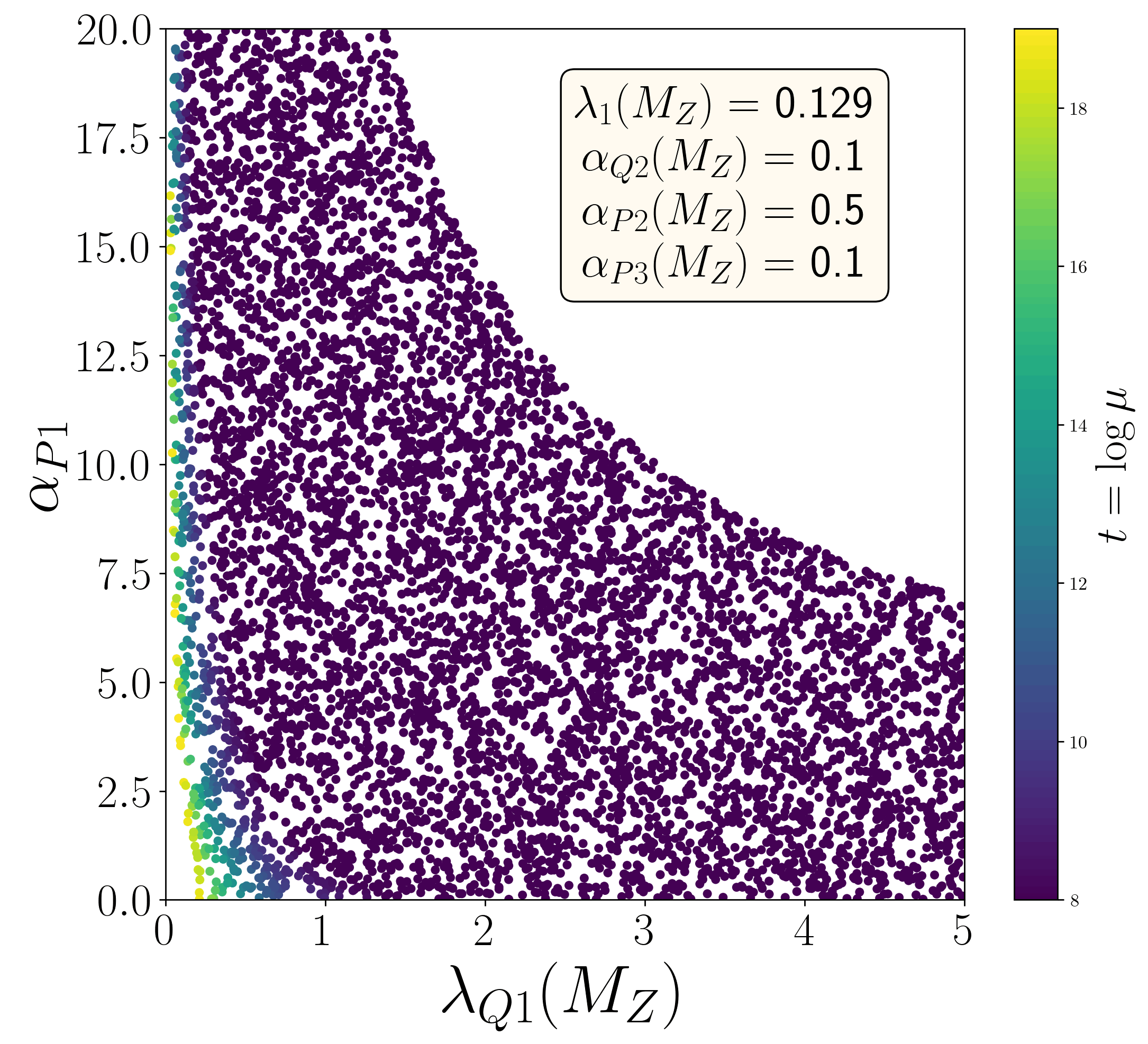}}
		\hspace*{0.25cm}
		\subfigure[]{\includegraphics[height=0.17\textheight,width=0.3\linewidth]{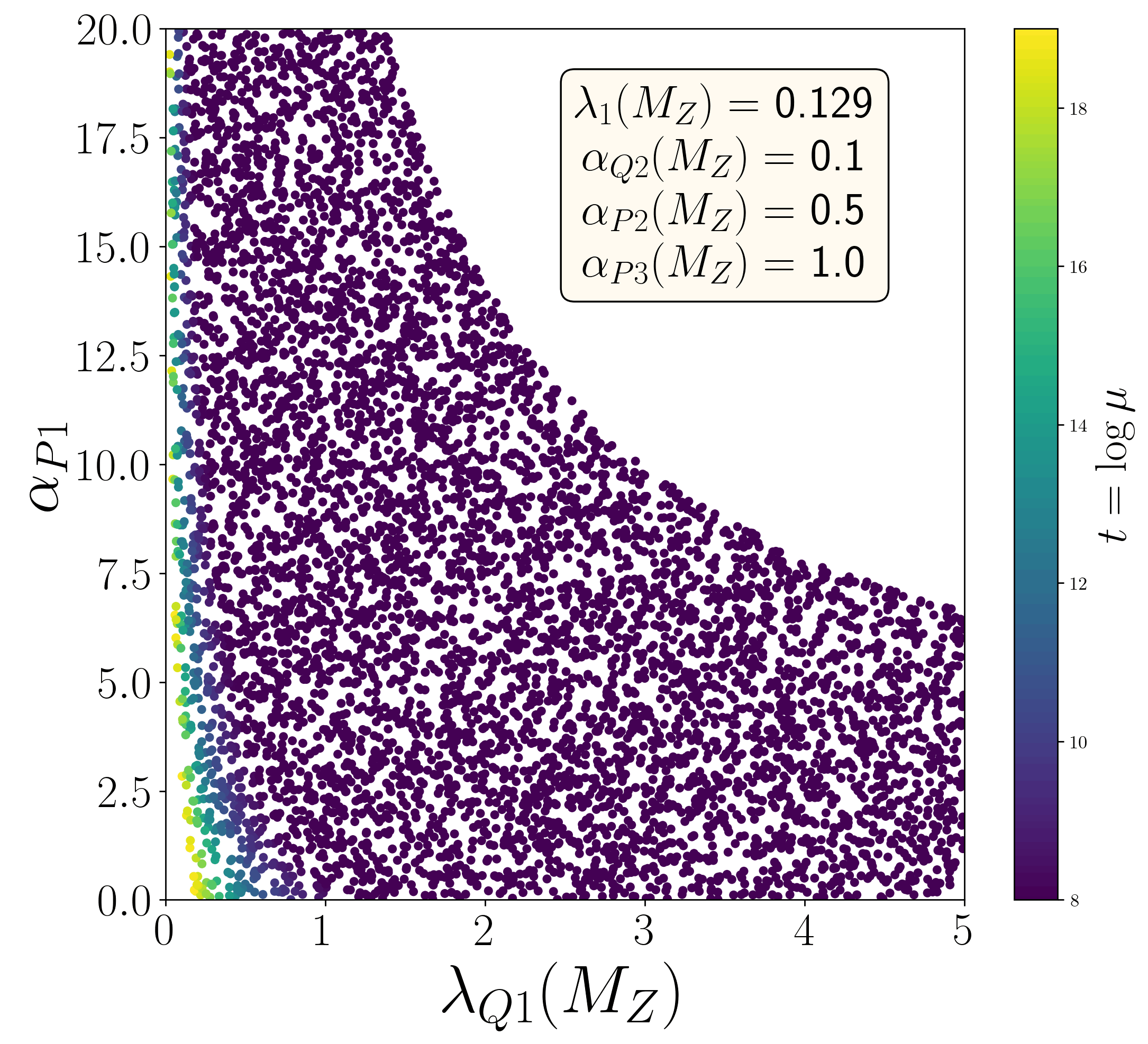}}
		\hspace*{0.25cm}
		\subfigure[]{\includegraphics[height=0.17\textheight,width=0.3\linewidth]{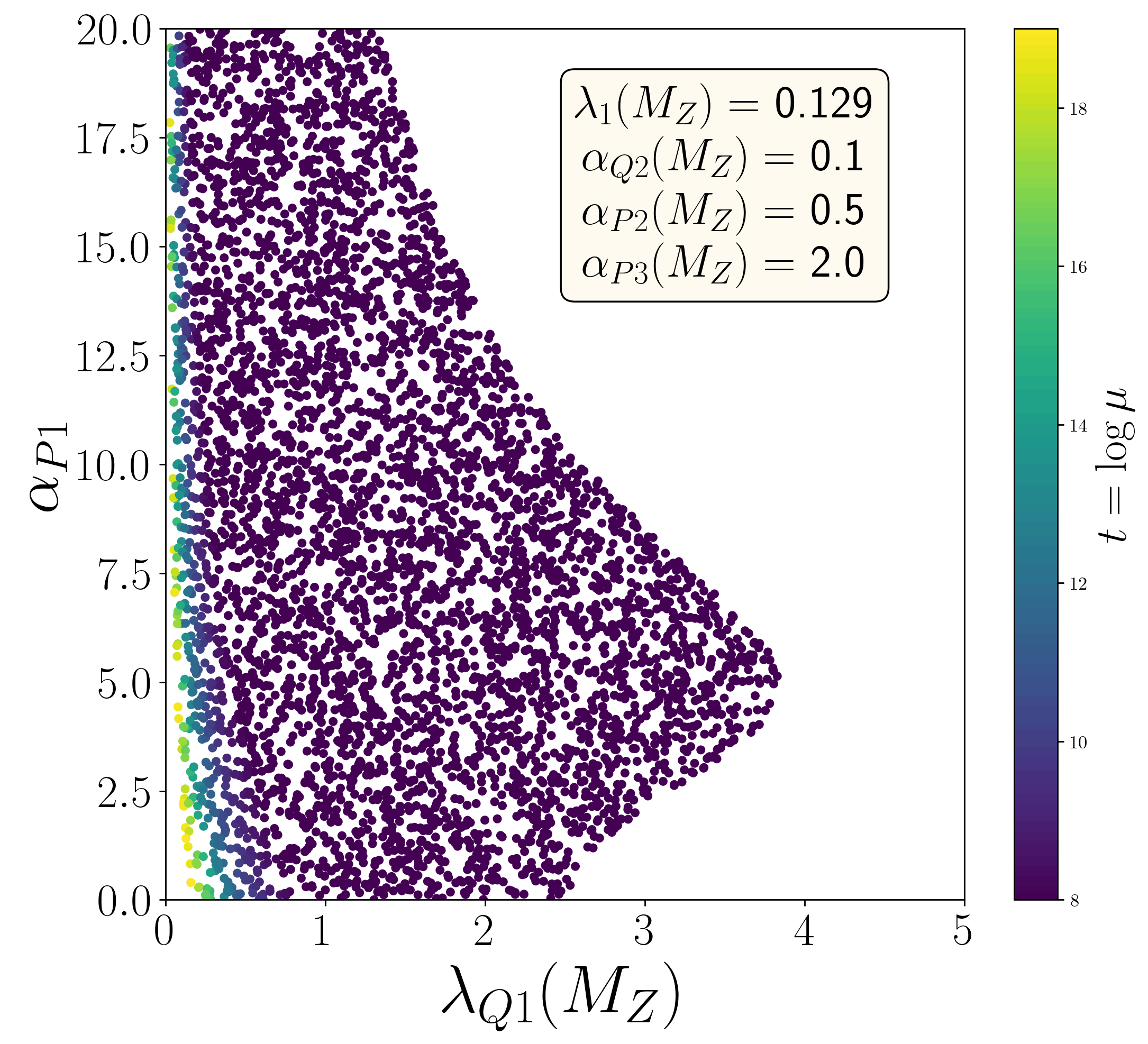}}
	}
	\mbox{
		\subfigure[]{\includegraphics[height=0.17\textheight,width=0.3\linewidth]{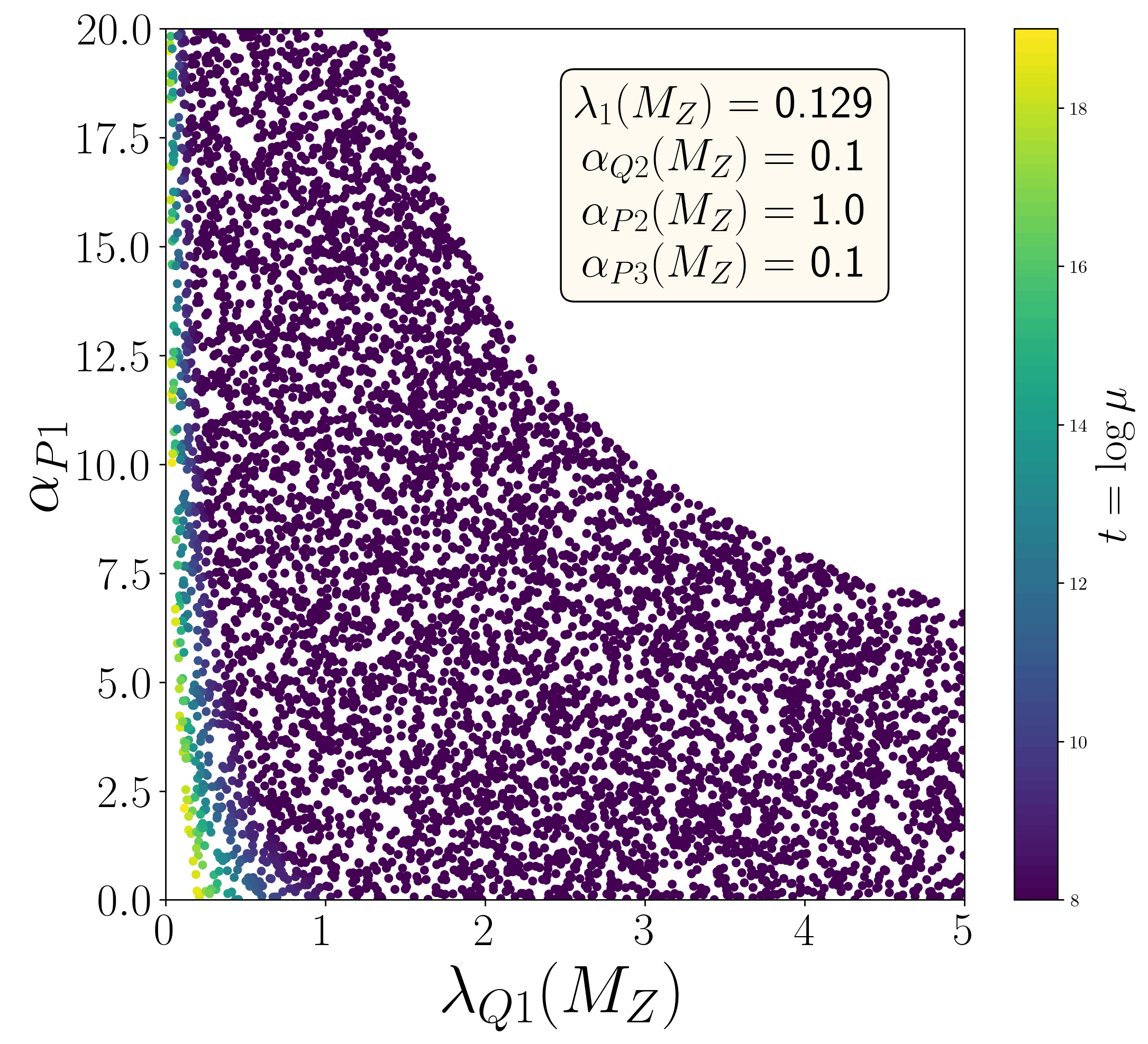}}
		\hspace*{0.25cm}
		\subfigure[]{\includegraphics[height=0.17\textheight,width=0.3\linewidth]{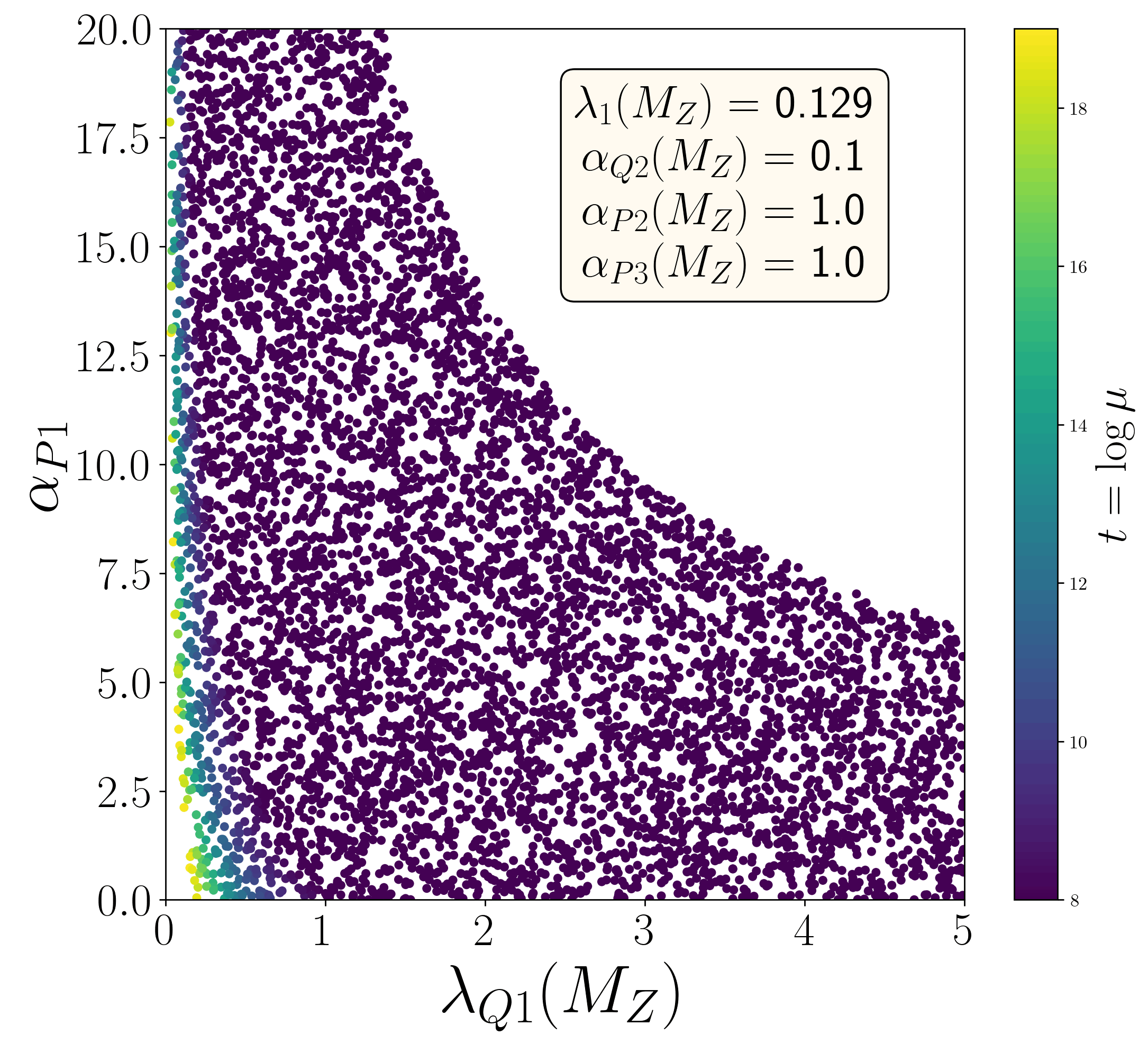}}
		\hspace*{0.25cm}
		\subfigure[]{\includegraphics[height=0.17\textheight,width=0.3\linewidth]{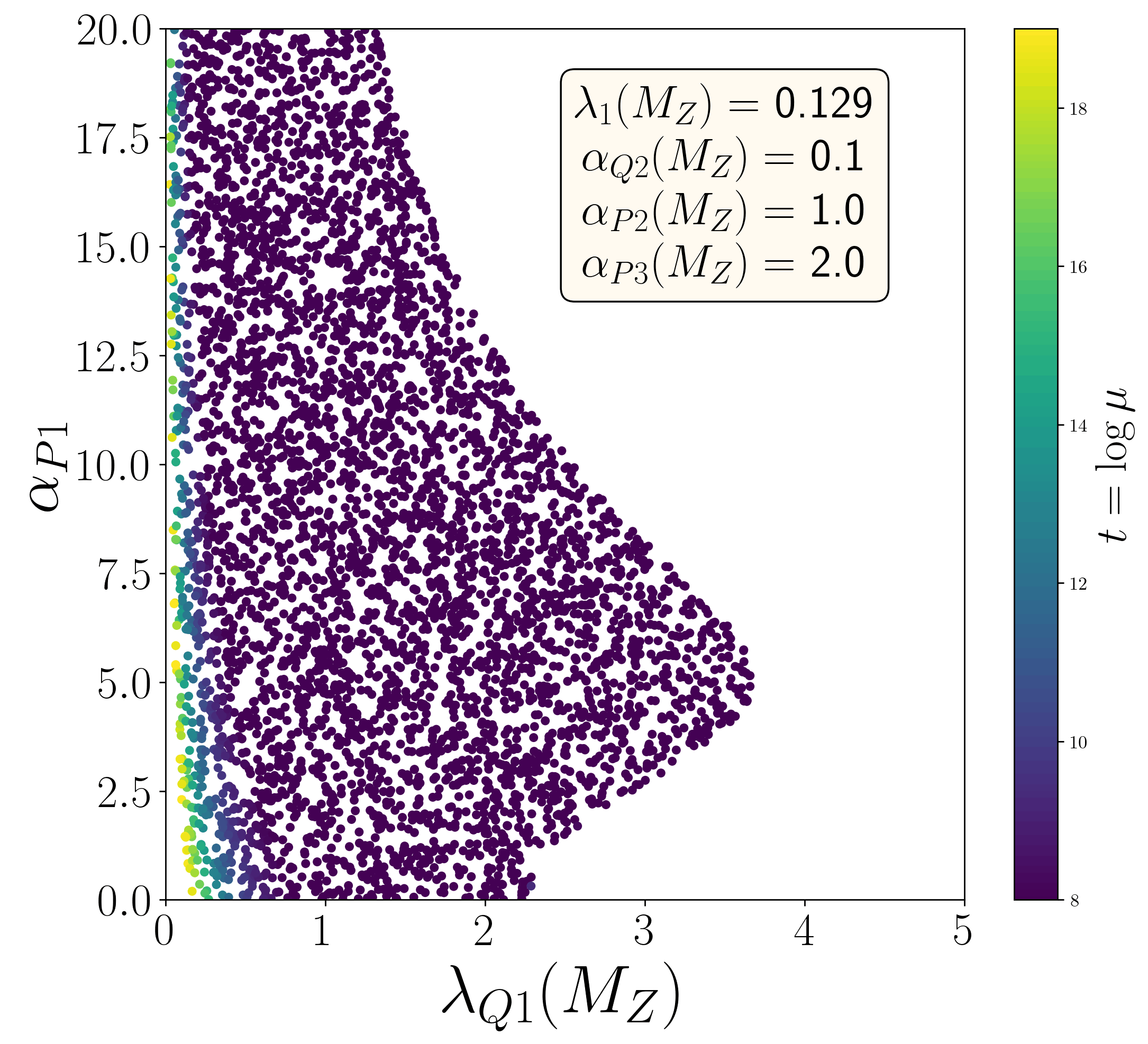}}
	}
	\mbox{
		\subfigure[]{\includegraphics[height=0.17\textheight,width=0.3\linewidth]{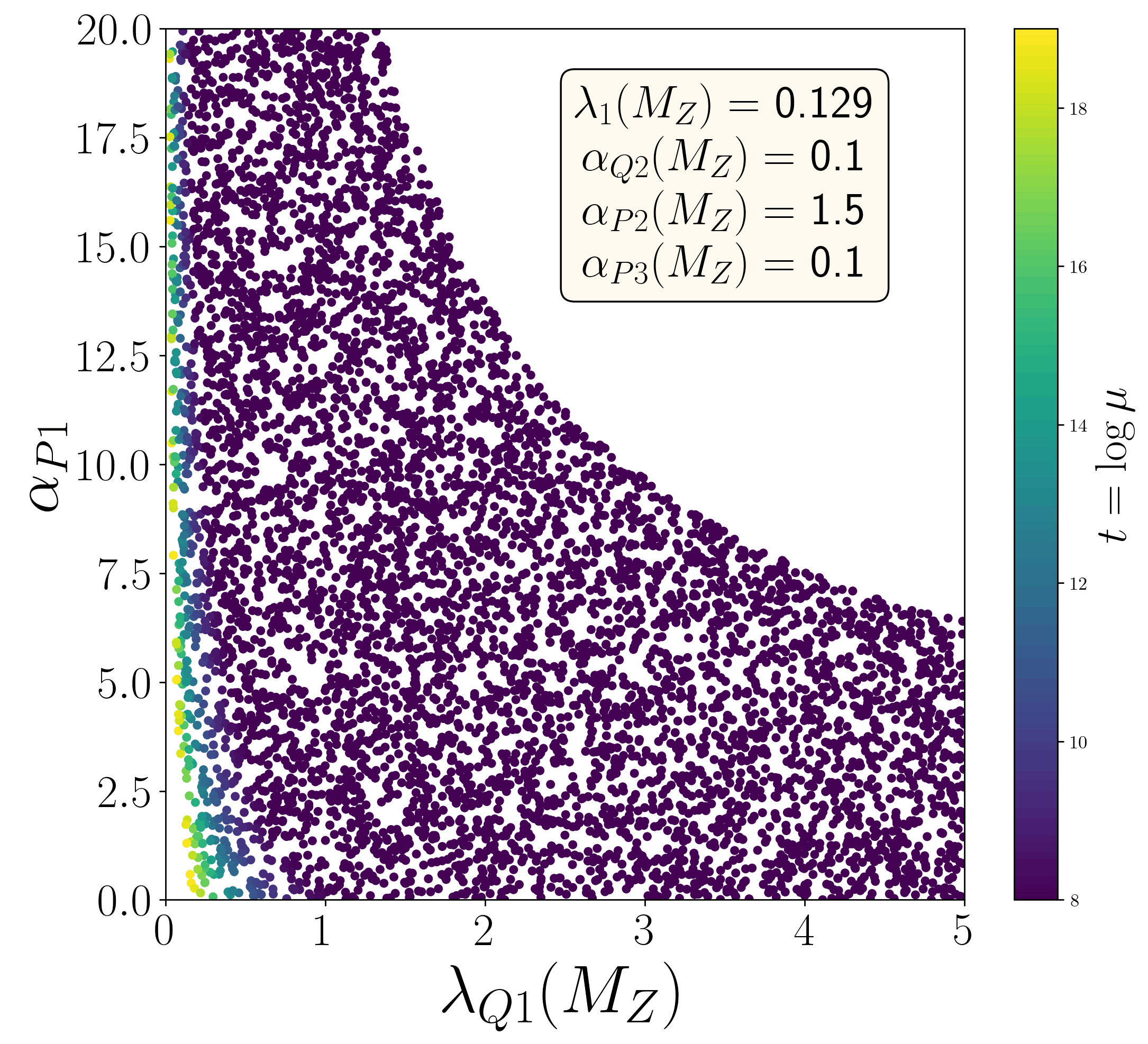}}
		\hspace*{0.25cm}
		\subfigure[]{\includegraphics[height=0.17\textheight,width=0.3\linewidth]{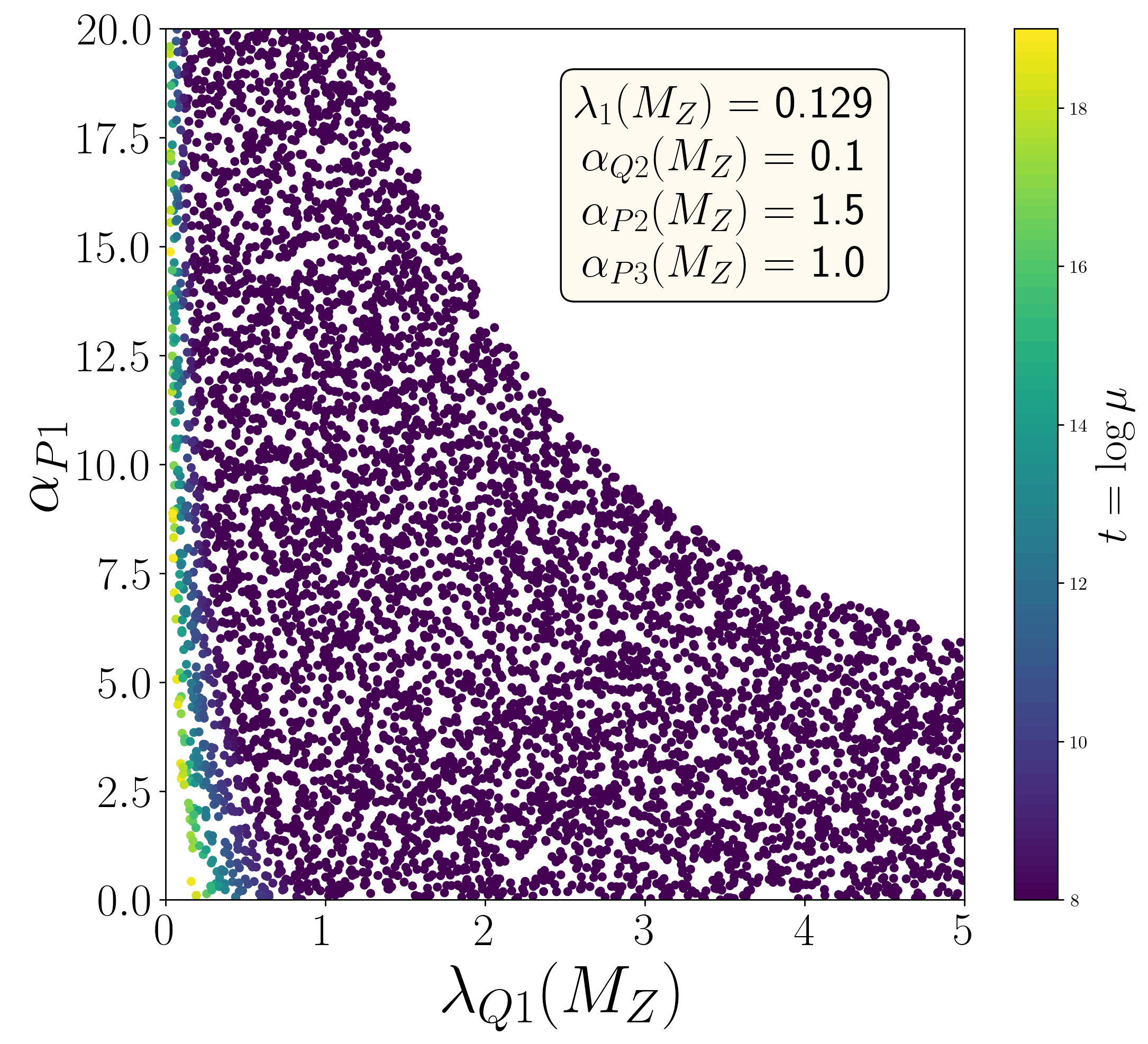}}
		\hspace*{0.25cm}
		\subfigure[]{\includegraphics[height=0.17\textheight,width=0.3\linewidth]{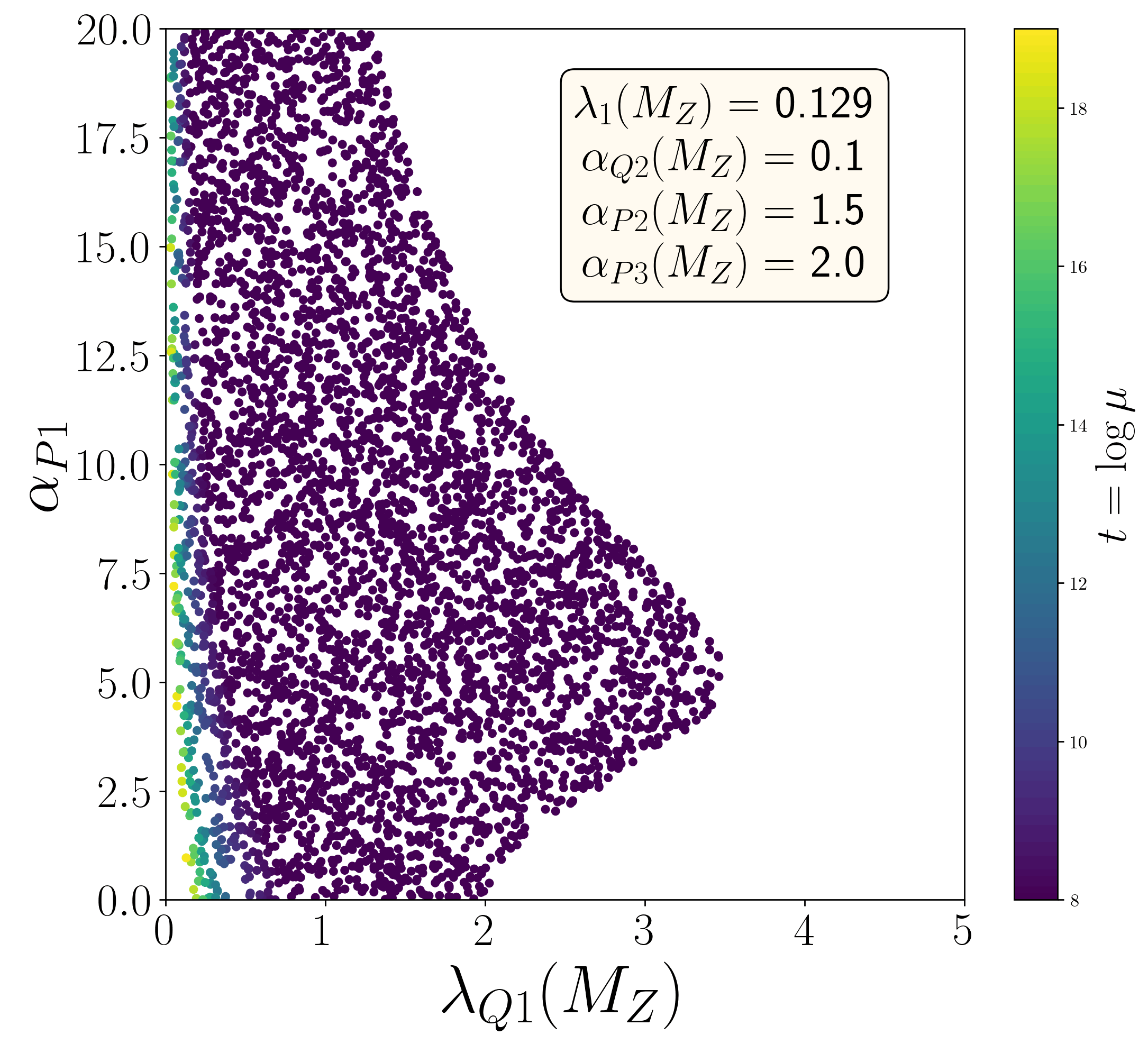}}
	}
\end{center}
\caption{Occurrence of FP at two-loop in $\lambda_{Q1}(M_Z) - \alpha_{P1}$ plane for different values of $\alpha_{P2}$ and $\alpha_{P3}$.  Initial value of Higgs-self quartic coupling $\lambda_{H}(M_Z)  = 0.129$ and $\alpha_{Q2} = 0.1$ for all cases. The values of ($\alpha_{P2}, \alpha_{P3})$ for the cases in (a)–(c), the $\alpha_{P2}$ value is 0.5 and $\alpha_{P3}$ values are 0.1, 1.0, 2.0 respectively. For
		(d)–(f), the $\alpha_{P2}$ value is 1.0 and $\alpha_{P3}$ values are 0.1, 1.0, 2.0 respectively. For (g)–(i), the $\alpha_{P2}$ value is 1.5 and $\alpha_{P3}$ values are 0.1, 1.0, 2.0
		respectively.}
	\label{fig:FP_scan_alphaP1_RUN_aQ2_0p1}
\end{figure}

\autoref{fig:FP_scan_alphaP1_RUN_aQ2_0p1} shows  the values of $\lambda_{Q1}(M_Z)$ and $\alpha_{P1}$, with fixed values of $\alpha_{P2}$ and $\alpha_{P3}$, for which there are  FPs at two-loop in the evolution of $\lambda_{H}(\mu)$. The colour code gives the energy scale at which the FP begins. In all the cases, $\lambda_{H}(M_Z)  = 0.129$ and $\alpha_{Q2} = 0.1$. 

In the \autoref{fig:FP_scan_alphaP1_RUN_aQ2_0p1}(a)-(c), we show the $\lambda_{Q1}(M_Z) - \alpha_{P1}$ plane for some fixed $\alpha_{P3} = 0.1, 1.0, 2.0$ for $\alpha_{P2} = 0.5$. It is clear that as $\alpha_{P3}$ increases, number of FP decreases which is in accordance with a term with residual phases only. Thus $\alpha_{P3}$ behaves like a spoiler. The coupling $\lambda_{Q1}(M_Z)$, which is also associated with terms that has no phases, thus an enhancer, enables the appearance of FP  at even lower energy scales. For  lower $\lambda_{Q1}(M_Z)$ such FPs appear at relatively higher values.  However, as we vary  $\alpha_{P2} $ which has 9-terms without phases and 3-terms with phases, from $0.5$ to $1.5$ row-wise the effect is only visible for larger $\lambda_{P3}=2.0$  for \autoref{fig:FP_scan_alphaP1_RUN_aQ2_0p1}(c),(f),(i). For lower values of $\lambda_{P3}$ the spoiler effect of $\lambda_{P2}$ is not so effective. This is attributed to the competition of terms with and without phases, and we need to look for the cross-terms in the $\beta$-functions involving different $\lambda$s. Finally, by comparing the \autoref{fig:FP_scan_alphaP1_RUN_aQ2_0p1}(g)-(i), where $\alpha_{P3} = 0.1, 1.0, 2.0$, and $\alpha_{P2} = 1.5$ for all three cases, we can conclude that the increase in $\alpha_{P3}$ causes the absence of FP behaviour. Therefore, $\alpha_{P3}$ is clearly a spoiler of FP behaviour which is in agreement with our observation that $\lambda_{P3}$-term has more number of phases. This is very similar to the behaviour of $\gamma$ and $\lambda_5$-term in IDM as discussed in \cite{Bandyopadhyay:2025ilx}.

Another feature is that all these cases in \autoref{fig:FP_scan_alphaP1_RUN_aQ2_0p1}, when the initial values of $\lambda_{Q1}(M_Z)$ is very low, the FP occurs when the $\alpha_{P1}$ value is higher. As the initial values $\lambda_{Q1}(M_Z)$ increases slightly, the FP begin to appear for lower values of $\alpha_1$. The energy scale at which FP appear changes from $\sim 10^{17-18}$ (yellow colour) to $\sim 10^{2-3}$ (blue colour) as the value of $\lambda_{Q1}(M_Z)$ varies from $<0.5$ to $>0.5$.  For $\lambda_{Q1}(M_Z) > 0.5$, the FP appears at lower energy scales $\mu\sim 10^{2-3}$.  But for very high values of $\lambda_{Q1}$ and $\alpha_{P1}$, FP behaviour is absent.

\begin{figure}[h]
	\begin{center}
		\mbox{
	\subfigure[]{\includegraphics[height=0.17\textheight,width=0.3\linewidth]{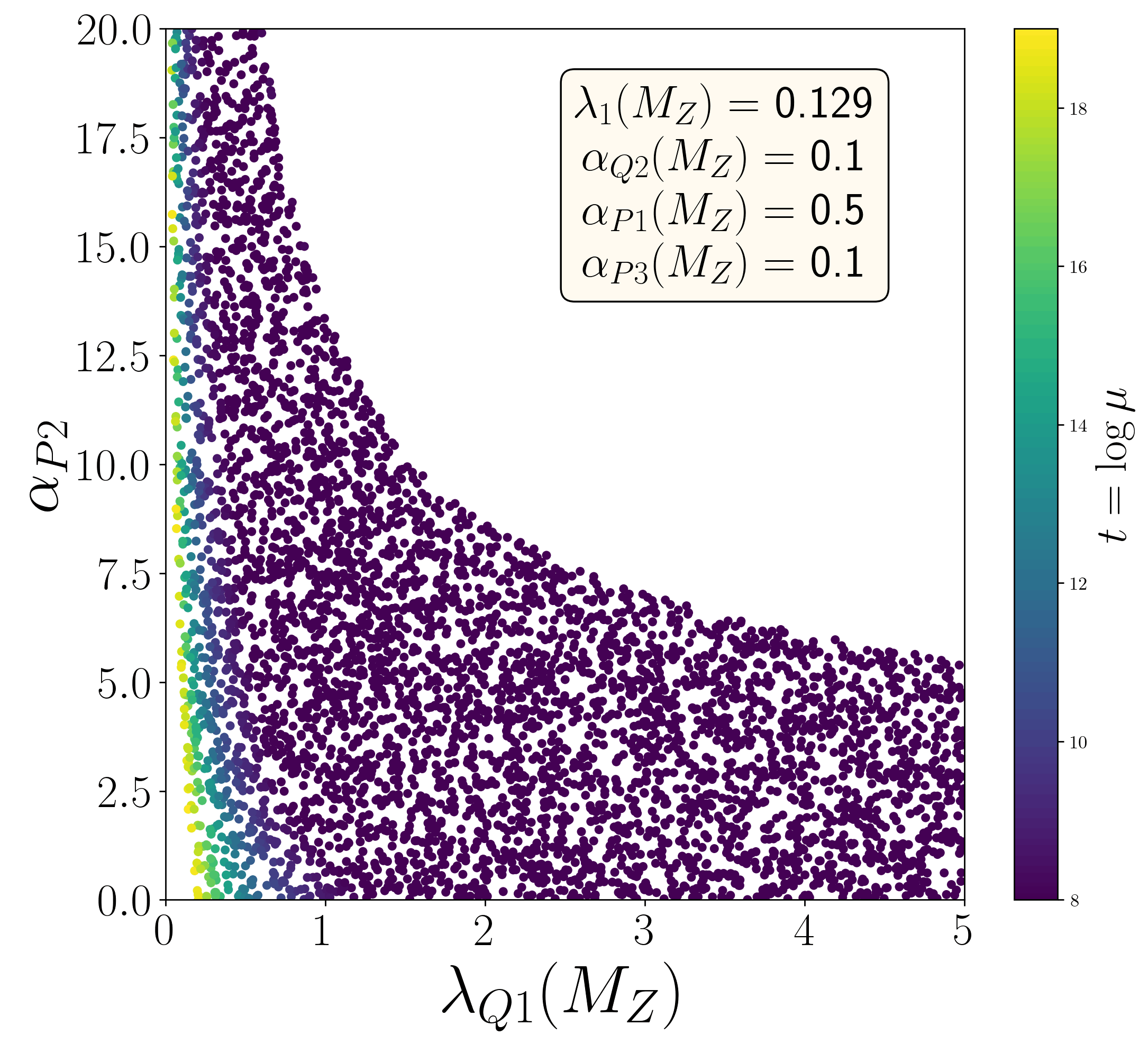}}
	\hspace*{0.25cm}
	\subfigure[]{\includegraphics[height=0.17\textheight,width=0.3\linewidth]{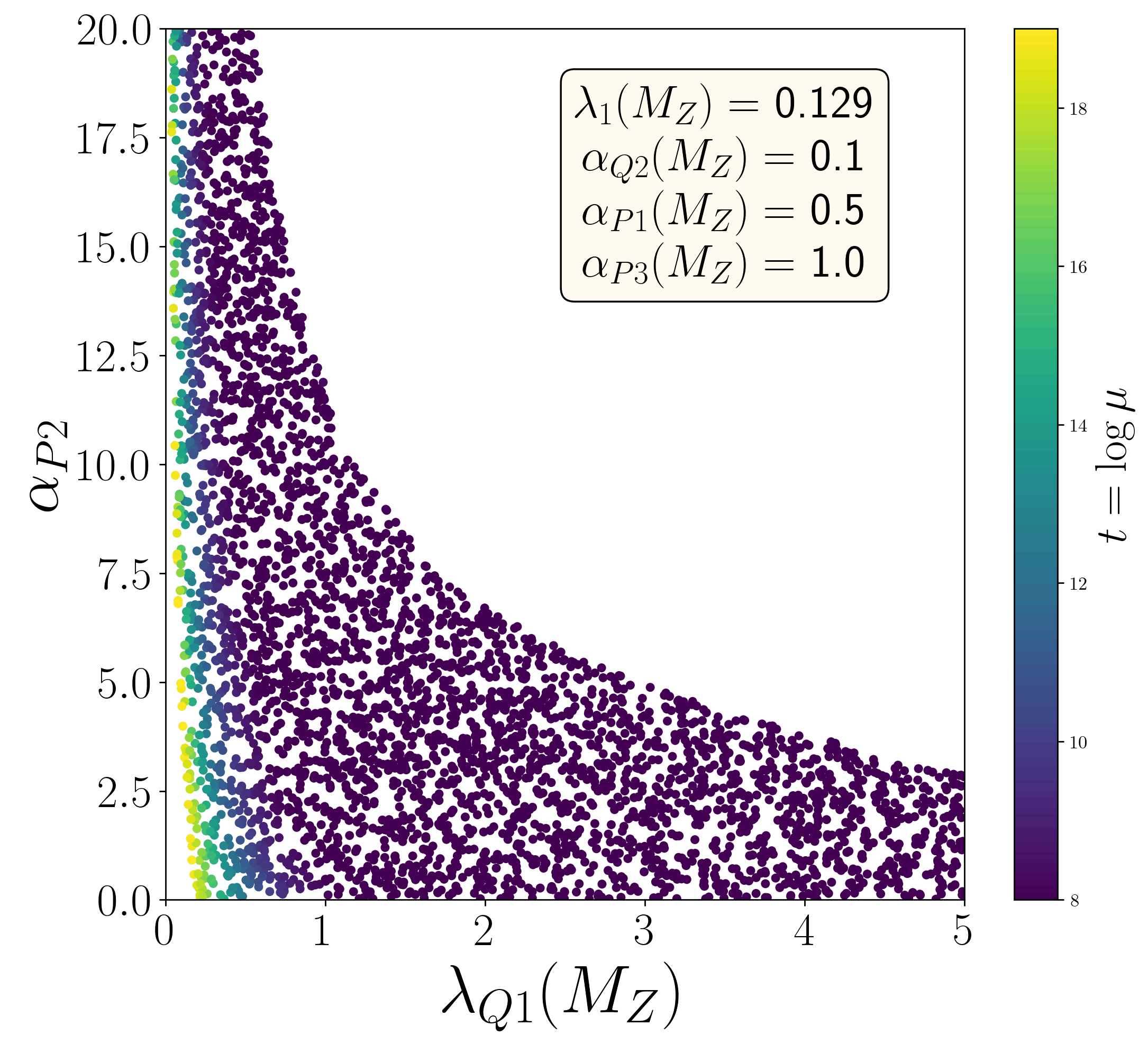}}
	\hspace*{0.25cm}
	\subfigure[]{\includegraphics[height=0.17\textheight,width=0.3\linewidth]{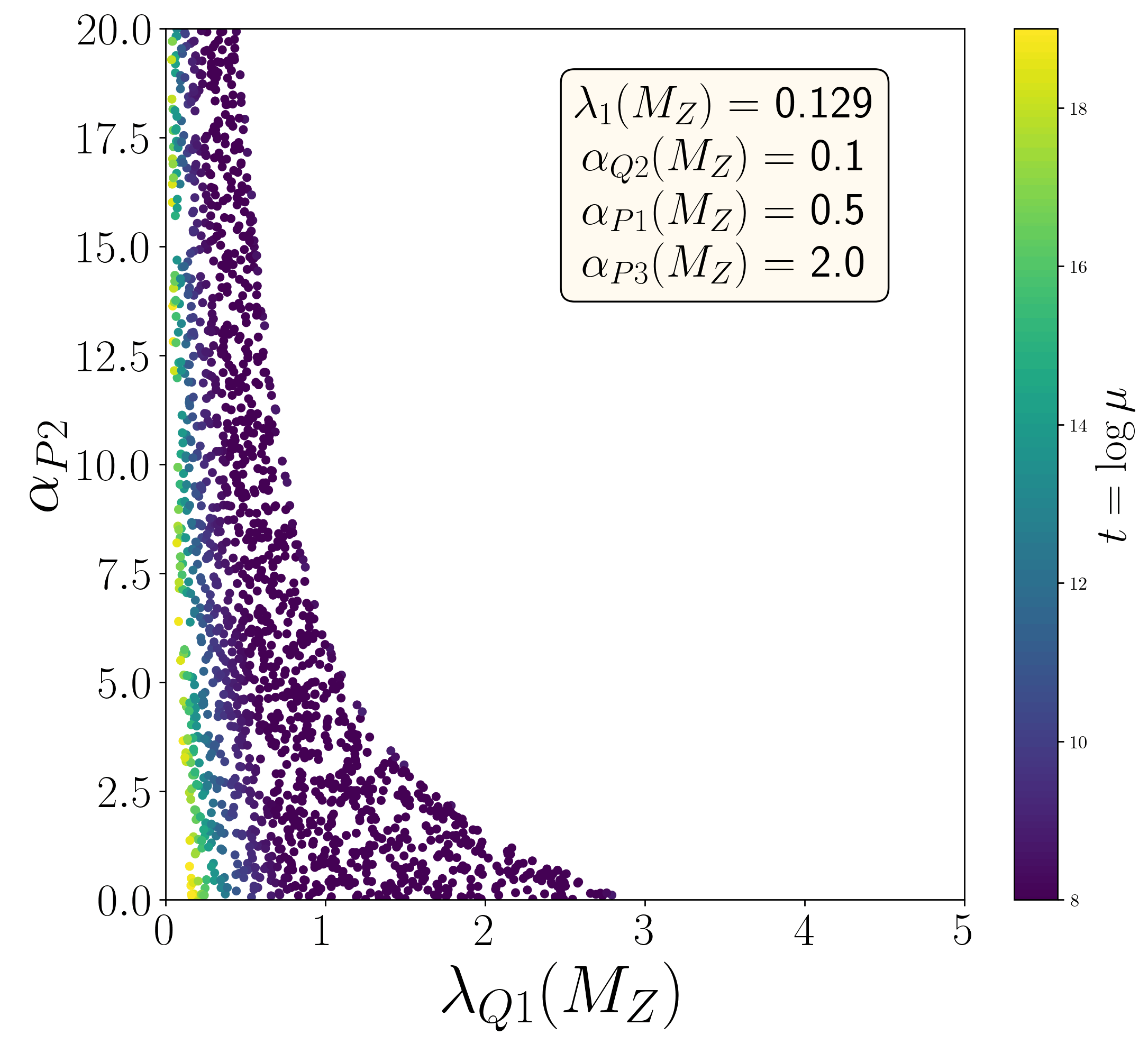}}
}
	\end{center}
	\caption{Occurrence of FP at two-loop in $\lambda_{Q1}(M_Z) - \alpha_{P2}$ plane for different values of $\alpha_{P1}$ and $\alpha_{P3}$.  Initial value of Higgs-self quartic coupling $\lambda_{H}(M_Z)  = 0.129$,  $\alpha_{P2} = 0.5$ and $\alpha_{Q2} = 0.1$ for all cases. The value of $\alpha_{P3}$ in (a), (b), and (c) are 0.1, 1.0, and 2.0 respectively. }
	\label{fig:FP_scan_alphaP2_RUN_aQ2_0p1}
\end{figure}

In \autoref{fig:FP_scan_alphaP2_RUN_aQ2_0p1}, we show the values of $\lambda_{Q1}(M_Z)$ and $\alpha_{P2}$, with fixed values of $\alpha_{P1}$ and $\alpha_{P3}$, for which there are two-loop FPs in the evolution of $\lambda_{H}(\mu)$. The colour code gives the energy scale at which the FP begins. In all the cases, $\lambda_{H}(M_Z)  = 0.129$, $\alpha_{P1} = 0.5$ and $\alpha_{Q2} = 0.1$. The values of $\alpha_{P3}$ for \autoref{fig:FP_scan_alphaP2_RUN_aQ2_0p1}(a), (b) and (c)  are $0.1$, $1.0$  and $2.0$ respectively. We can clearly see that as $\alpha_{P3}$ increases, the Fixed Points are absent for higher values of $\lambda_{Q1}$. This not only strengthens our conclusion that $\alpha_{P3}$ is a spoiler but also points that $\alpha_{Q1}$ with  6 terms without phases (see \autoref{nophaseterms})  is relatively smaller FP enhancer compared to $\alpha_{P2}$  which has 9 terms without any phases \autoref{rediphase}. This behaviour is also present when the $\lambda_{Q1}$ is relatively small, Fixed Points are present for larger $\alpha_{P2}$ also.  However, if both $\lambda_{Q1}$ and $\alpha_{P2}$ are higher, the Fixed Points disappear as the interference terms in the $\beta$-function enhance  the phase effect coming from $\alpha_{P2}$.

\begin{figure}[h]
	\begin{center}
	\mbox{
		\subfigure[]{\includegraphics[height=0.17\textheight,width=0.3\linewidth]{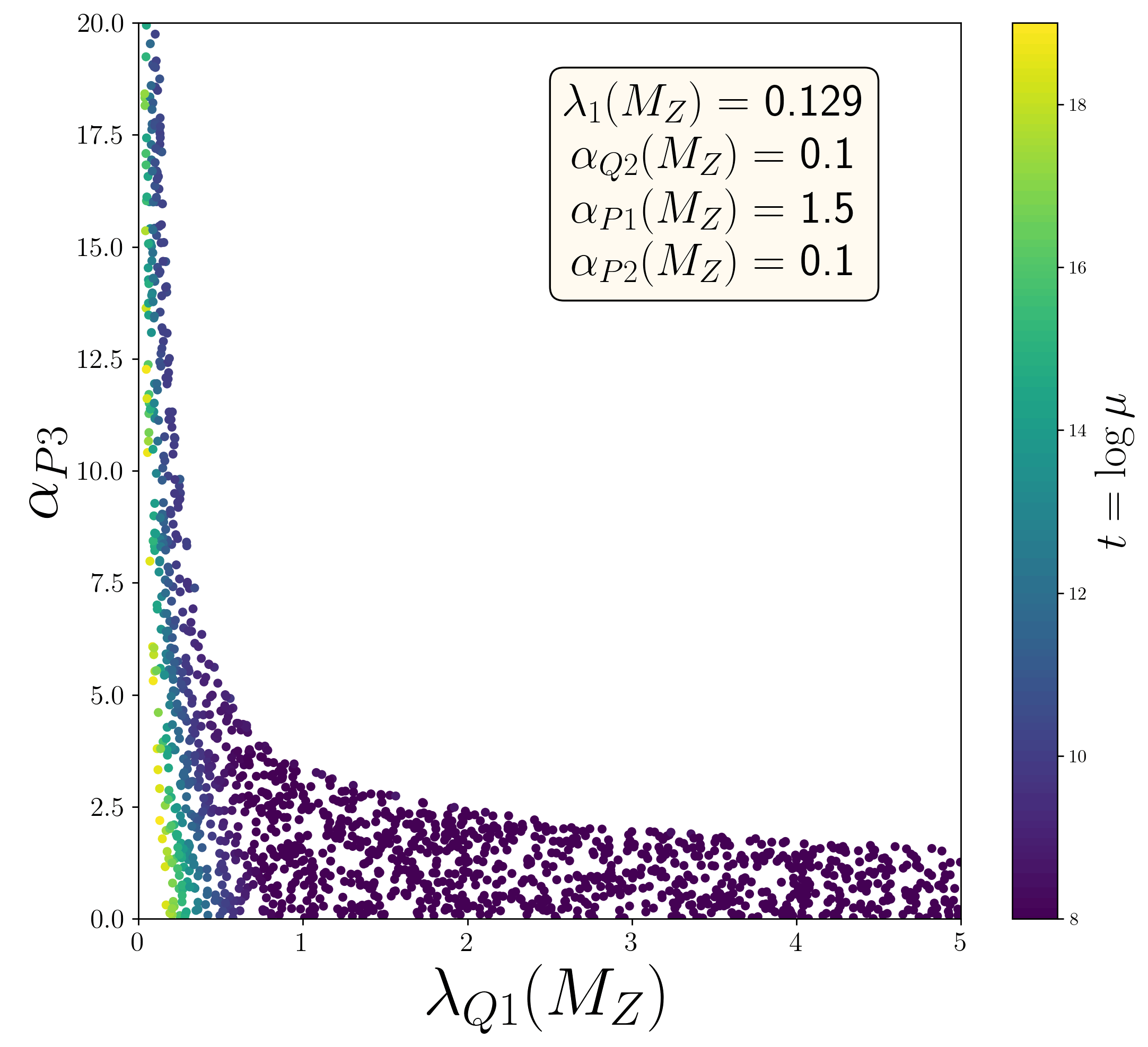}}
		\hspace*{0.25cm}
		\subfigure[]{\includegraphics[height=0.17\textheight,width=0.3\linewidth]{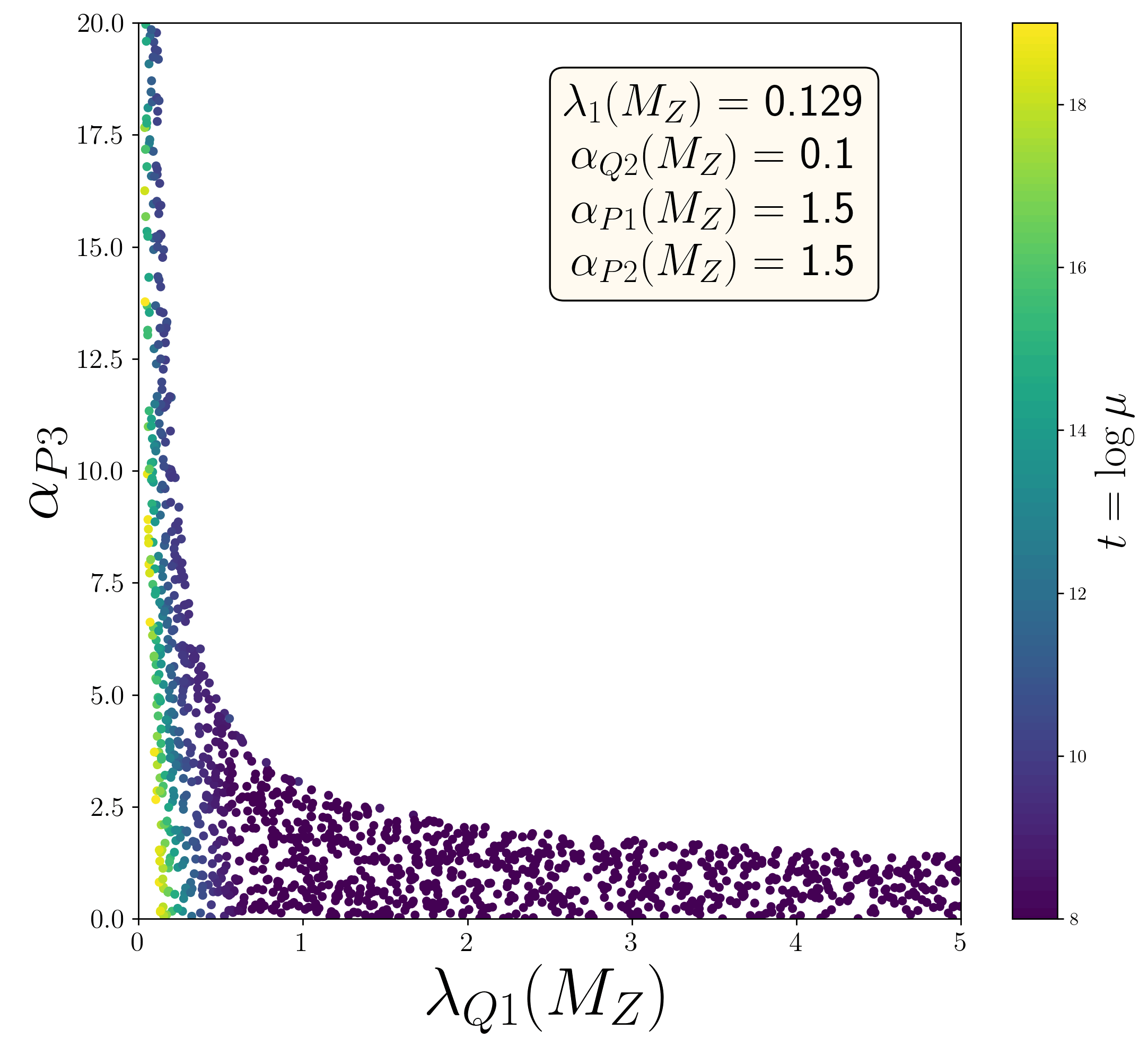}}
		\hspace*{0.25cm}
		\subfigure[]{\includegraphics[height=0.17\textheight,width=0.3\linewidth]{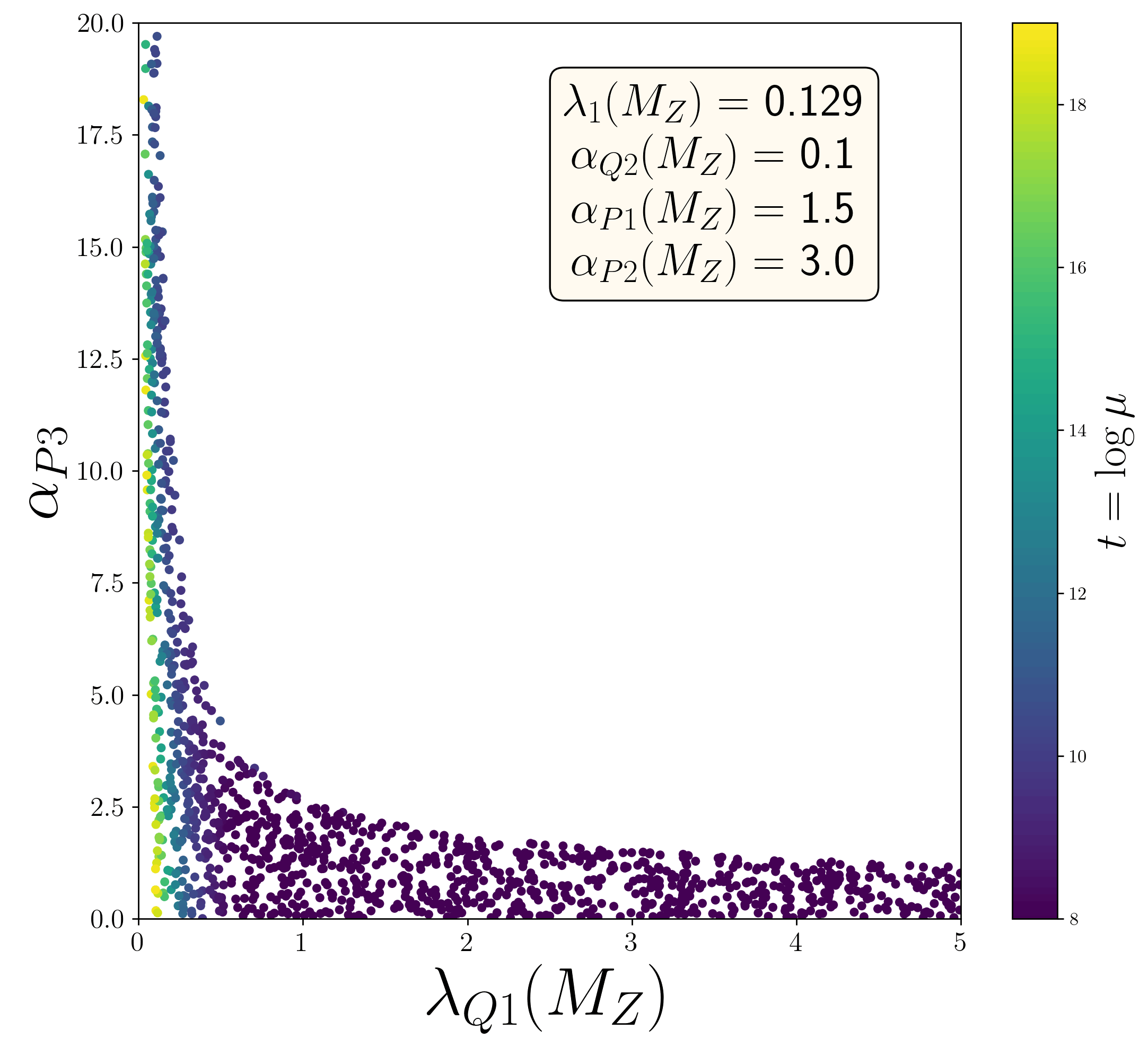}}
	}
	\end{center}
	\caption{Occurrence of FP at two-loop in $\lambda_{Q1}(M_Z) - \alpha_{P3}$ plane for different values of $\alpha_{P1}$ and $\alpha_{P2}$.  Initial value of Higgs-self quartic coupling $\lambda_{H}(M_Z)  = 0.129$,  $\alpha_{P1} = 1.5$ and $\alpha_{Q2} = 0.1$ for all cases. The value of $\alpha_{P2}$ in (a), (b), and (c) are 0.1, 1.5, and 3.0 respectively. }
	\label{fig:FP_scan_alphaP3_RUN_aQ2_0p1}
\end{figure}

In \autoref{fig:FP_scan_alphaP3_RUN_aQ2_0p1}, we show the values of $\lambda_{Q1}(M_Z)$ and $\alpha_{P3}$ for which Fixed Points appear in the evolution of $\lambda_{H}(\mu)$ for fixed values of $\alpha_{P1}$ and $\alpha_{P2}$. \autoref{fig:FP_scan_alphaP3_RUN_aQ2_0p1}(a), (b), and (c) has $\alpha_{P2}$ values of $0.1$, $1.5$, and $3.0$ respectively. In all three cases, $\lambda_{H}(M_Z)  = 0.129$, $\alpha_{P1} = 1.5$ and $\alpha_{Q2} = 0.1$. Here the behaviour is dominated by $\alpha_{P3}$ after some critical values of $\lambda_{Q1, Q2}$. Thus , except when $\lambda_{Q1}$ is very small, the appearance of Fixed Point is spoiled by the increase in $\alpha_{P3}$. We can see that most of the $\lambda_{Q1}-\alpha_{P3}$ plane is devoid of Fixed Points which is in agreement with our finding that $\alpha_{P3}$ is a heavy spoiler.

\begin{figure}[h]
	\begin{center}
		\mbox{
			\subfigure[]{\includegraphics[height=0.17\textheight,width=0.3\linewidth]{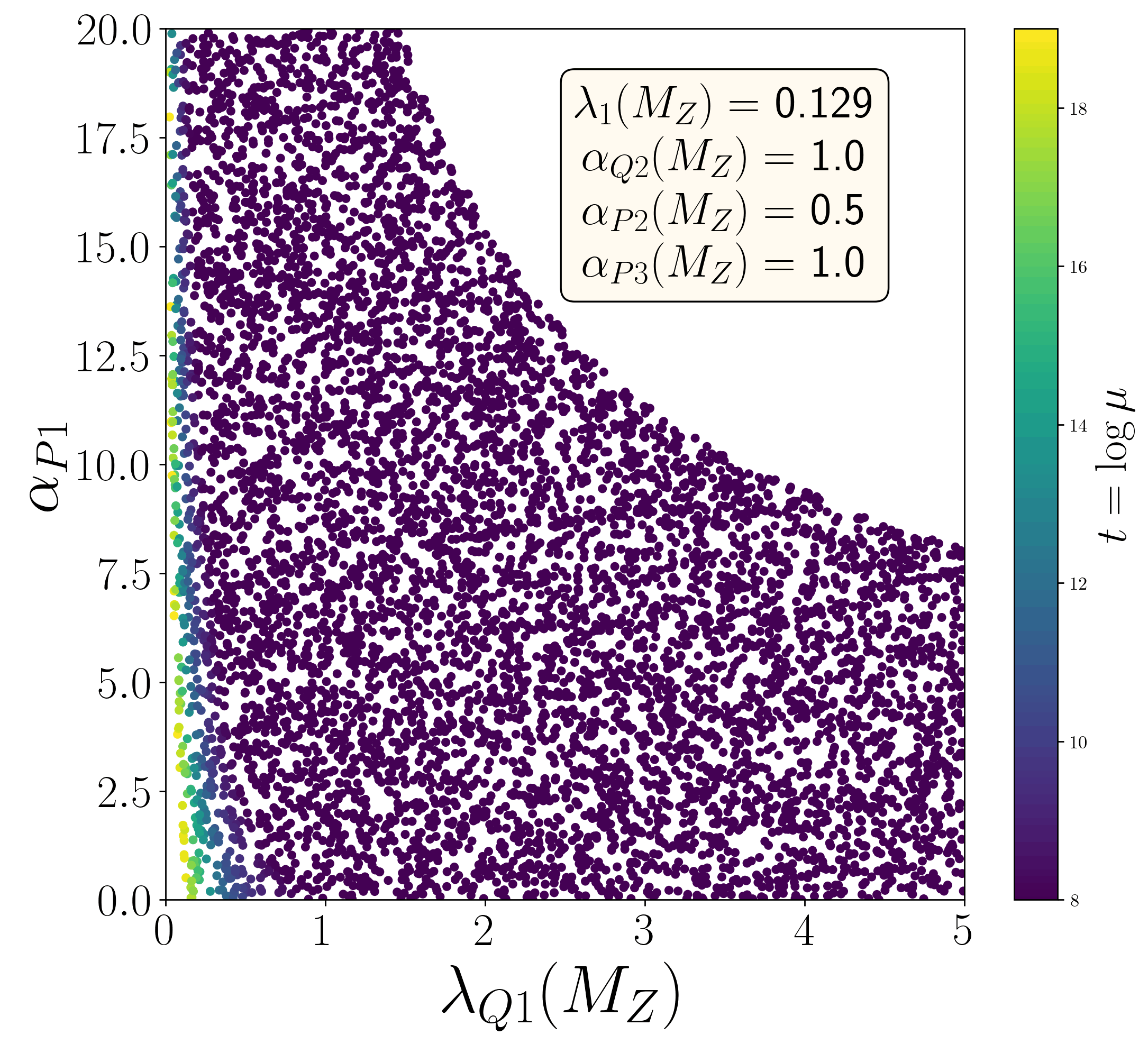}}
			\hspace*{0.25cm}
			\subfigure[]{\includegraphics[height=0.17\textheight,width=0.3\linewidth]{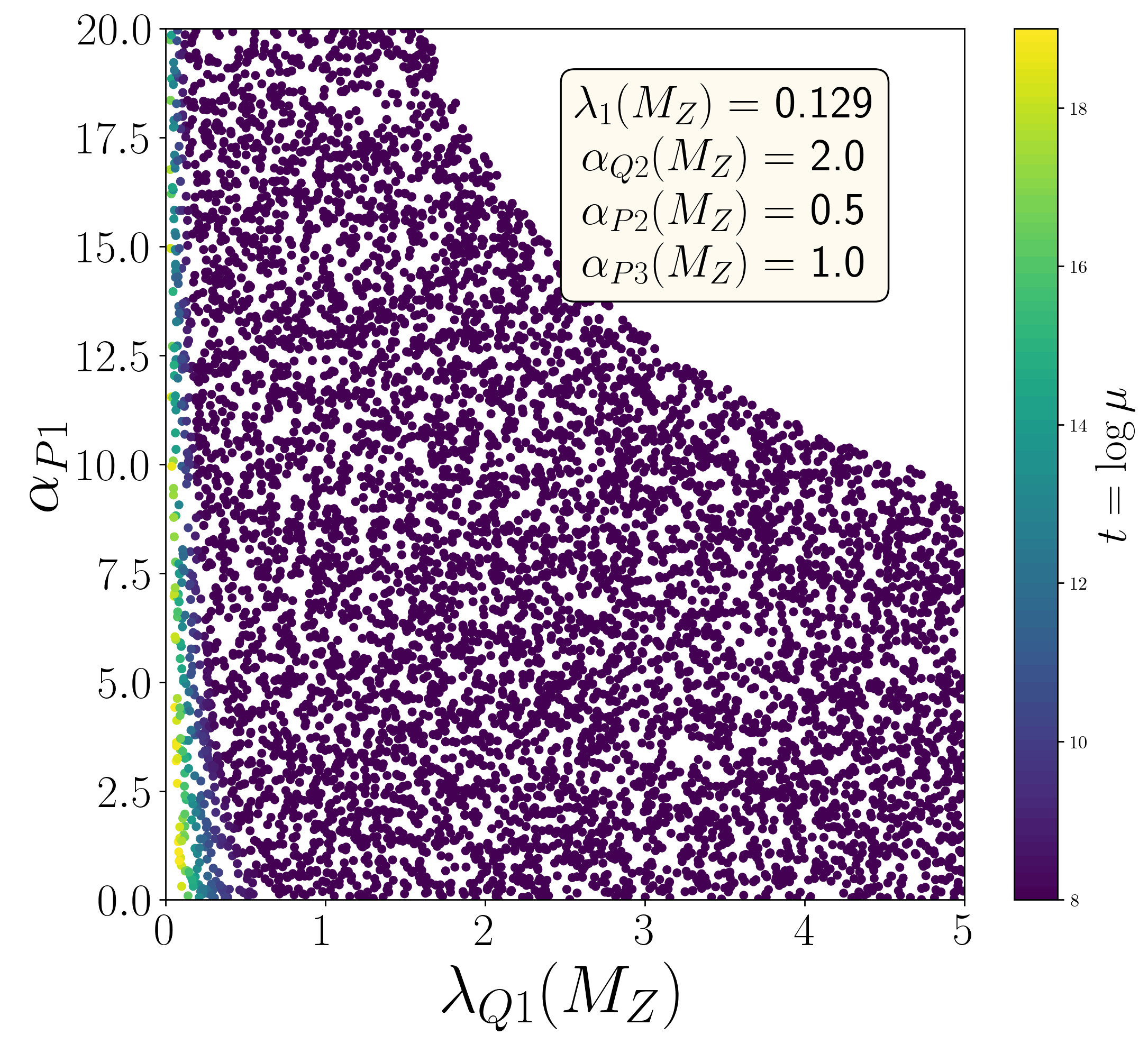}}
		}
			\mbox{
	\subfigure[]{\includegraphics[height=0.17\textheight,width=0.3\linewidth]{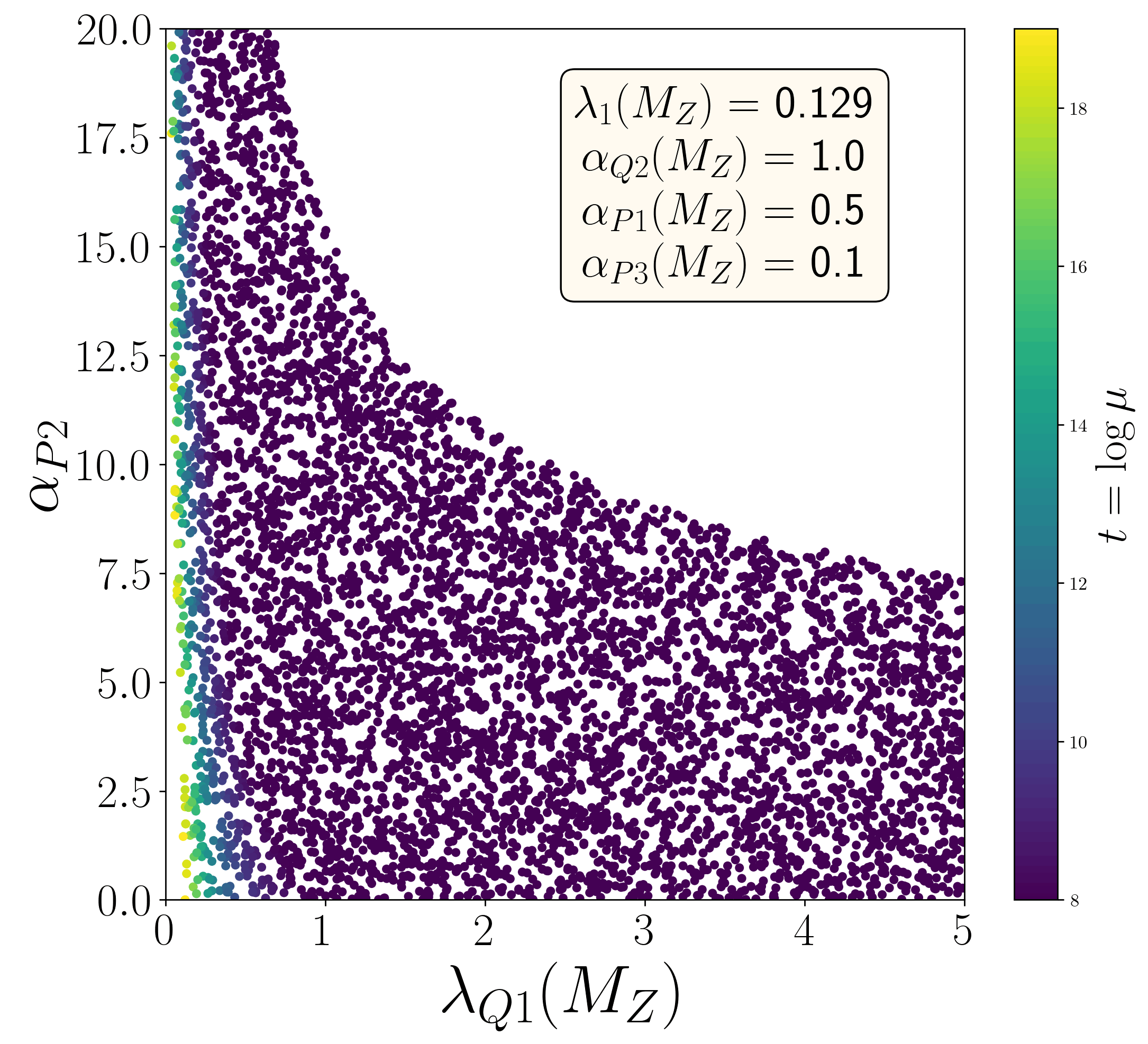}}
			\hspace*{0.25cm}
	\subfigure[]{\includegraphics[height=0.17\textheight,width=0.3\linewidth]{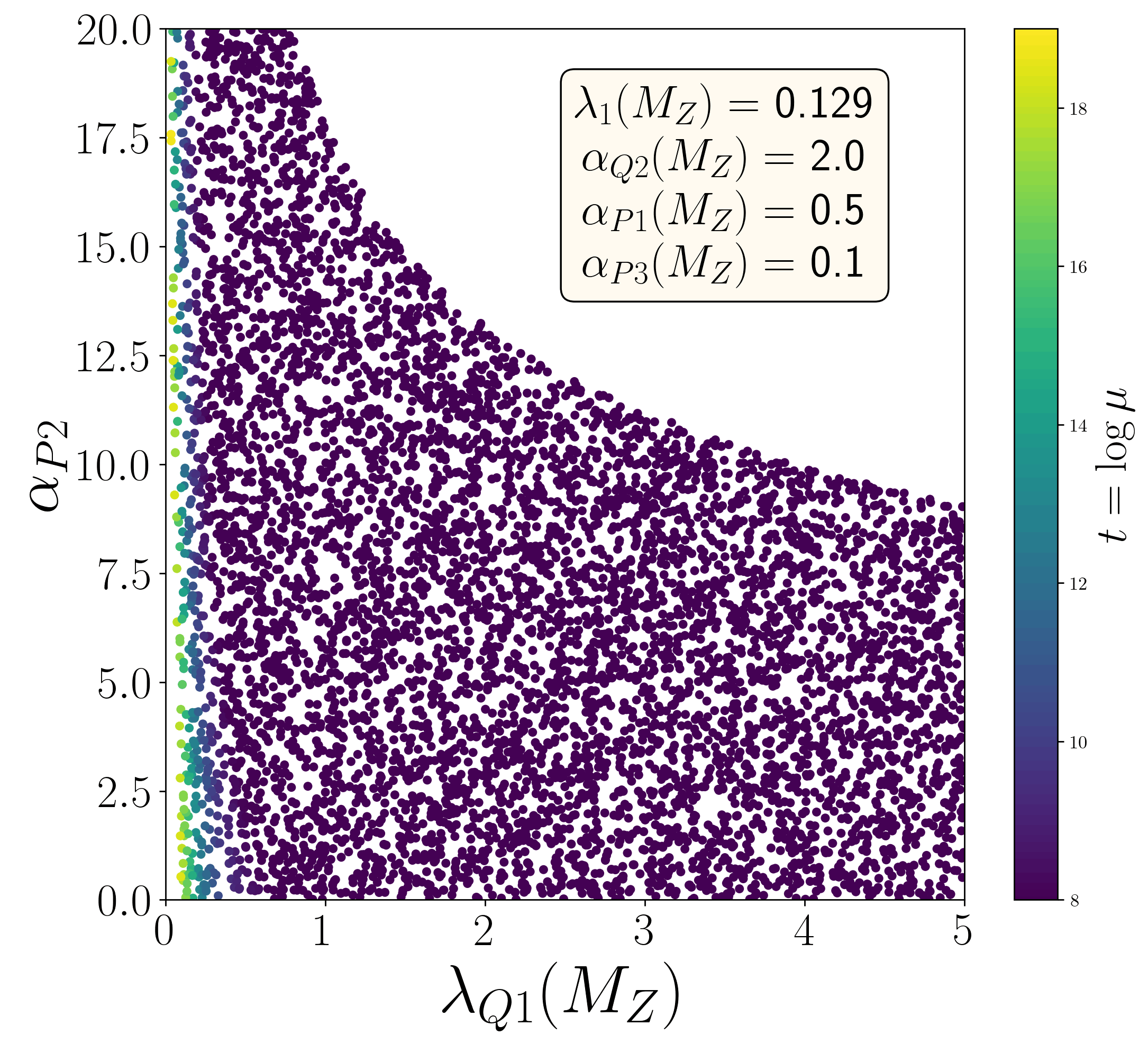}
		}}
	\end{center}
	\caption{Occurrence of FP at two-loop in $\lambda_{Q1}(M_Z) - \alpha_{P1}$ plane in (a), (b) with $\alpha_{P2} = 0.5,\,\alpha_{P3} = 1.0$  and in $\lambda_{Q1}(M_Z) - \alpha_{P2}$ plane in (c), (d) with $\alpha_{P1} = 0.5, \,\alpha_{P3} = 0.1$. For left and right panel we have  $\alpha_{Q2} = 1.0, \, 2.0$, respectively and Higgs-self quartic coupling $\lambda_{H}(M_Z)  = 0.129$ for all cases. }\label{fig:FP_scan_alphaP1_RUN_aQ2_different}
\end{figure}

In \autoref{fig:FP_scan_alphaP1_RUN_aQ2_different} (a), (b) we present the variation in $\lambda_{P1}-\lambda_{Q1}$ plane for fixed values of $\alpha_{Q2}, \alpha_{P2}, \alpha_{P3}$. 
In comparison with \autoref{fig:FP_scan_alphaP1_RUN_aQ2_0p1}(b) with $\alpha_{Q2} = 0.1$, we can see that number of allowed points in the $\lambda_{Q1}-\alpha_{P1}$ plane increases significantly as we change the $\alpha_{Q2}$ to $1.0$ and $2.0$ in \autoref{fig:FP_scan_alphaP1_RUN_aQ2_different}(a) and (b) respectively, while keeping $\alpha_{P2} = 0.5$ and $\alpha_{P3} = 1.0$ fixed. Similarly, when we compare \autoref{fig:FP_scan_alphaP2_RUN_aQ2_0p1}(b), \autoref{fig:FP_scan_alphaP1_RUN_aQ2_different}(a) and \autoref{fig:FP_scan_alphaP1_RUN_aQ2_different}(b), $\alpha_{Q2}$ = $0.1$, $1.0$, and $2.0$, while keeping with fixed values of $\alpha_{P1} = 0.5$ and $\alpha_{P3} = 0.1$, we can see that the $\lambda_{Q1}-\alpha_{P1}$ plane will contain more number of allowed points as $\alpha_{Q2}$ increases.  This is also evident from $\lambda_{Q1}-\alpha_{P2}$ plane in   \autoref{fig:FP_scan_alphaP1_RUN_aQ2_different}(c) and \autoref{fig:FP_scan_alphaP1_RUN_aQ2_different}(d)  for $\alpha_{Q2}=1.0,\, 2.0$, respectively, validating  the fact that $\alpha_{Q2}$ is also an enhancer of FPs due to large number of (nine) terms without any phases. 

\section{Mass spectrum at tree and one-loop level}\label{massspectrum}

\subsection{Tree-level mass spectrum:}
After EWSB, the tree-level masses of the BSM particles $X_R$, $X_I$, $X^{++}$, and $X^+_{1,2}$ in terms of the Higgs vev and the Lagrangian parameters $\mu_X^2$, $\lambda_{P1, P2, P3}$ are given in \autoref{eq:4plet-neutral-masses}, \autoref{eq:4plet-doubly-charged-masses} and \autoref{eq:4plet-singly-charged-masses}. The expressions of the masses represented in terms of the neutral scalar masses after eliminating the bare mass parameter $\mu_X^2$ is given in the \autoref{tab:masses-in-terms-of-cand-DM}.
From this table, we can see that in general, at tree-level, the masses of the particles are non-degenerate. However, if the portal coupling $\lambda_{P3}$ is zero, then the neutral scalars $X_R$ and $X_I$ become degenerate. More importantly, one of the singly charged scalar $X^\pm_1$ and the doubly charged scalar $X^{\pm\pm}$ will have masses less than the neutral scalars by an amount   $\frac{1}{6}v_h^2\lambda_{P2}$ and $\frac{2}{6}v_h^2\lambda_{P2}$ respectively, thereby making the doubly charged scalar $X^{\pm\pm}$ as the lightest stable particle (LSP) as it is $\mathbb{Z}_2$-odd. This will give rise to an electro-magnetically charged dark matter, which is not viable. Thus, this set of parameter at the tree-level is excluded phenomenologically. In order to have a tree-level viable neutral dark matter we definitely need  $\lambda_{P3}\neq 0$ . For $\lambda_{P3}> 0$, $X_R$ is the LSP and thus dark matter and for $\lambda_{P3}< 0$, $X_I$ plays the role of the dark matter. The demand of  $X_R$ being the LSP, we get the necessary conditions $\lambda_{P3} > 0$ and $2 \lambda_{P3} > |\lambda_{P2}|$, where
the latter condition ensure the doubly charged particle $X^{++}$ is heavier than the $X_R$, in addition to the singly charged particles. For $2 \lambda_{P3} = \lambda_{P2}$, then the mass of the doubly charged scalar $M_{X^{++}}$ will be equal to $M_{X_R}$.

\begin{table}[htbp]
	\centering
	\renewcommand{\arraystretch}{2.0}
	\setlength{\tabcolsep}{10pt}
	
	\begin{tabular}{|c|p{8.2cm}|}
		\hline
		\textbf{Particle} 
		& \textbf{Mass squared in terms of $M_{X_R}$}  \\
		\hline
		
		$X_R$
		& $\displaystyle M_{X_R}^2 =  M_{X_R}^2$
		 \\
		\hline
		
		$X_I$
		& $\displaystyle M_{X_I}^2 = \displaystyle M_{X_R}^2 + \frac{4}{3}\lambda_{P3} v_h^2$
		\\
		\hline
		
		$X^{++}$
		& $\displaystyle M_{X^{++}}^2 =  M_{X_R}^2
		+ \frac{v_h^2}{6}\big( 4\lambda_{P3} - 2\lambda_{P2}  \big)$
		\\
		\hline
		
		$X_1^{+}$
		& $\displaystyle M_{X_1^+}^2 =  M_{X_R}^2
		+ \frac{v_h^2}{6}\Big(4\lambda_{P3} 
		- \sqrt{\lambda_{P2}^2 + 12\lambda_{P3}^2}
		\Big)$
		\\
		\hline
		
		$X_2^{+}$
		& $\displaystyle M_{X_2^+}^2 =  M_{X_R}^2
		+ \frac{v_h^2}{6}\Big(4\lambda_{P3}
		+ \sqrt{\lambda_{P2}^2 + 12\lambda_{P3}^2}\Big)$
		\\
		\hline
	\end{tabular}
	\caption{Expressions for tree-level masses of I4M mass eigenstates in terms of $M_{X_R}$ and the portal couplings. }
	\label{tab:masses-in-terms-of-cand-DM}
\end{table}

\subsection{Mass splitting at one-loop:}
Since the 4-plet has gauge interactions, the masses of the particles can acquire quantum corrections from different orders of perturbation theory. Here we shall consider the leading quantum correction to the masses coming from one-loop.  Here we follow \cite{Cirelli:2005uq} for the approximate corrections, where the masses of gauge bosons are much smaller than the masses of $\mathbb{Z}_2$-odd multiplets in estimating the one-loop mass splitting. In this approximation the difference between the masses of the charged component with electric charge $Q$ and the neutral component with charge zero, due to one-loop corrections, is given by \cite{Cirelli:2005uq},
\begin{equation}
	M_Q - M_0 \simeq Q \big( Q + \frac{1}{\cos \theta_W}\big) \Delta M,
\end{equation}
where $\Delta M = (166 \pm 1)$ MeV and the Weinberg angle $\theta_W \approx 29^\circ $. This results into the  relative mass corrections for  $X^{\pm\pm}$, $X_{1,2}^\pm$, $X_I$  with $X_R$  as $1043.59, \,  355.80, \, 0.00$ MeV, respectively. This gives rise to positive corrections to the charged states compared to the neutral ones, leaving a scope for changing the respective mass hierarchies. Therefore, if the tree-level LSP is neutral, then it will remain lightest even at one-loop level. But if the tree level LSP is charged, it may or may not remain lightest after the one-loop correction. Additionally, the next-to-lightest $\mathbb{Z}_2$-odd particle (NLSP) at tree level may or may not remain the NLSP at one-loop.

\begin{figure}[h]
	\begin{center}
		\mbox{
			\subfigure[]{\includegraphics[height=0.22\textheight,width=0.45\linewidth]{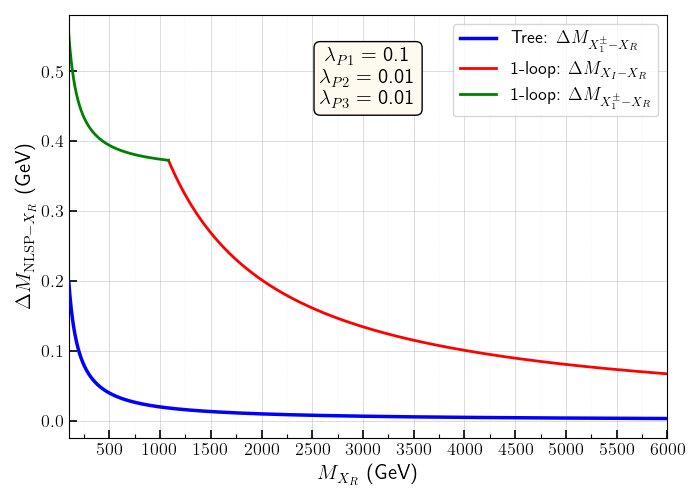}}
			\hspace*{0.25cm}	\subfigure[]{\includegraphics[height=0.22\textheight,width=0.45\linewidth]{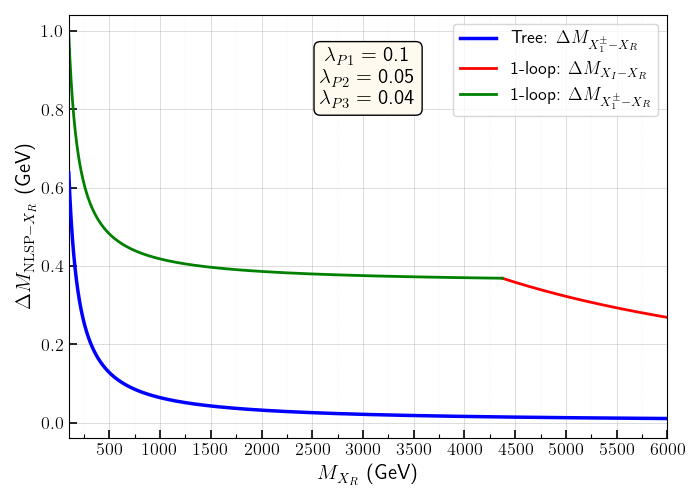}}
		}
		\mbox{
			\subfigure[]{\includegraphics[height=0.22\textheight,width=0.45\linewidth]{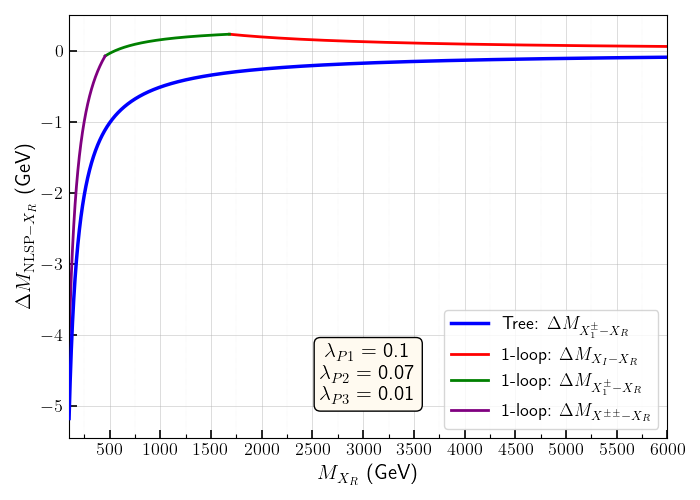}}
			\hspace*{0.25cm}
			\subfigure[]{\includegraphics[height=0.22\textheight,width=0.45\linewidth]{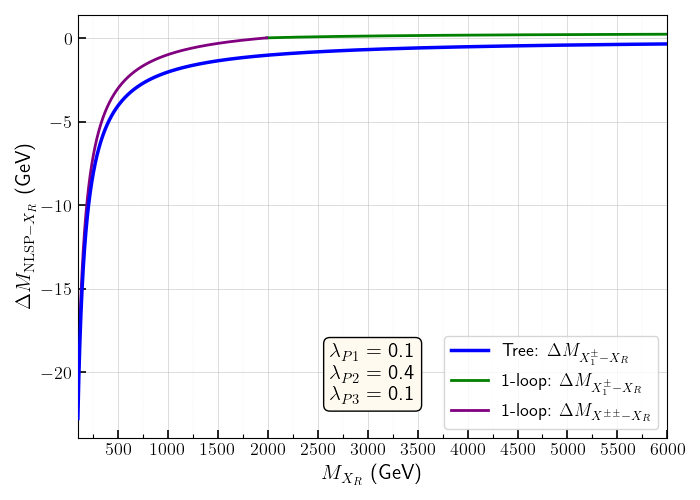}}
		}
	\end{center}
	\caption{The tree-level and one-loop level mass splitting between the NLSP and LSP for different choices of  the Higgs portal couplings $\lambda_{P1, P2, P3}$. The Y-axis shows the mass splitting $\Delta M _{\rm NLSP-LSP} \equiv \Delta M = M_{X} - M_{X_R}$, where $M_{X} = \min(M_{X^{++}}, M_{X_{1,2}^+}, M_{X_{I}})$ and the x-axis shows the $M_{X_R}$.}
	\label{fig:mass-differences}
\end{figure}
In \autoref{fig:mass-differences}, we show the difference between the masses of NLSP and LSP i.e. $\Delta M _{\rm NLSP-LSP}$(y-axis) plotted against the mass of neutral LSP, i.e. $M_{X_R}$ (x-axis),  for different values of $\lambda_{P1, P2, P3}$, at tree- and one-loop levels. In all the cases though the chosen neutral LSP is $X_R$, the nature of  NLSP may change, especially at one-loop level as we detailed them in \autoref{fig:mass-differences}. Thus possible mass splitting can be seen as $\Delta M _{\rm NLSP-LSP} \equiv \Delta M = M_{X} - M_{X_R}$, where $M_{X} = \min(M_{X^{++}}, M_{X_{1,2}^+}, M_{X_{I}})$.

In \autoref{fig:mass-differences}(a), we choose the portal couplings to have fixed values  $\lambda_{P1} = 0.1, \, \lambda_{P2} = 0.01$, and $\lambda_{P3} = 0.01$. In this case, the tree-level NLSP is $X^\pm_1$ and the corresponding mass difference $\Delta M_{X^\pm_1 - X_R} = M_{X^\pm_1} - M_{X_R}$ is shown in the blue curve. The $\Delta M_{X^\pm_1 - X_R}$ stays around 100 MeV at the tree-level for $M_{X_R} \lsim 200$ GeV. However, once it crosses 200 GeV, $\Delta M_{X^\pm_1 - X_R}$ decreases rapidly and becomes the order of a few MeV. The corresponding $\Delta M_{\rm NLSP-LSP}$  at one-loop is  given in the green curve. A significant change NLSP nature as the NLSP changes from $X_1^+$, to  $X_I$ around $M_{X_R}\gsim 1090$ GeV as shown by the red curve. For the region where $X_1^\pm$ is NLSP, the $\Delta M_{X^\pm_1 - X_R}$ decreases rapidly from $\sim$ 0.55 GeV to 0.38 GeV as $M_{X_R}$ increases. However, after the flip of NLSP to $X_I$ the $\Delta M_{X_I - X_R}$  starts from $\sim 0.38$ GeV and reduces gradually as $M_{X_R}$ increases to  $70$ MeV. 
	
Similarly we  consider the case for $\lambda_{P1, P2, P3} =\, $ 0.1, 0.05, and 0.04, respectively  in \autoref{fig:mass-differences}(b). The tree-level behaviour is similar to the previous case as  $X_1^\pm, \, X_R $ are the NLSP and LSP, and  the corresponding mass splitting  starts from $\sim 500$ MeV and gradually reduces  to a few MeV.  As we have seen the NLSP change in \autoref{fig:mass-differences}(a), \autoref{fig:mass-differences}(b) also shows similar change at one-loop level but relatively higher masses $M_{X_R} \gsim 4370$ GeV, where the NLSP changes from $X^\pm_1$ to $X_I$. In this case at one-loop the mass splitting  mostly stays around $400-300$ MeV with a gradual  reduction with the increase of  $M_{X_R}$.

In \autoref{fig:mass-differences}(c), we consider the parameter choice $\lambda_{P1,P2,P3} = 0.1,\,0.07,$ and $0.01$, respectively. In this case, the tree-level mass ordering is qualitatively different from the previous cases since the neutral scalars are heavier than the charged scalar $X_1^\pm$ and hence the mass difference $\Delta M_{X^\pm_1 - X_R} = M_{X_1^\pm} - M_{X_R}$ is negative, as shown by the blue curve. This is not surprising since $2\lambda_{P3} = 0.02 < |\lambda_{P2}| = 0.07$ which causes the tree-level LSP to be charged rather than neutral ones. For lower values of  $M_{X_R}$, the magnitude of $\Delta M_{X^\pm_1 - X_R}$ is around a few GeV, and this decreases rapidly as $M_{X_R}$ increases. For $M_{X_R} \sim$1000 GeV, the tree-level  $\Delta M_{X_1^\pm - X_R}$ becomes less negative and slowly approaches zero but never really touches it. It is worth mentioning here, since the electro-magnetically charged state is the LSP here, this region of parameter space cannot provide the desired dark matter candidate. Interestingly, at one-loop the mass hierarchy changes as the charged states get more quantum corrections as described by the violet curve in \autoref{fig:mass-differences}(c). It shows that for lower values of  $M_{X_R}$ the doubly charged scalar $X^{\pm\pm}$ is lighter than the neutral scalar but when $M_{X_R}\gsim$ 455 GeV, the singly charged scalar $X^\pm_1$ becomes LSP, shown as green curve, and stays like that till $M_{X_R} \lesssim 540$ GeV. After this $X_R$ becomes the LSP as the mass hierarchy flips between $X^\pm_1$ and $X_R$. Around $M_{X_R}\sim 1680$ GeV,  $X_I$ becomes the NLSP keeping $X_R$ as the LSP  as shown via the red curve with typical mass gap around a few 100 MeV. 

Finally, in \autoref{fig:mass-differences}(d), the portal couplings are $\lambda_{P1} = 0.1$, $\lambda_{P2} = 0.4$, and $\lambda_{P3} = 0.1$.  In this case also, the tree-level mass hierarchy is such that $X^\pm_1$ is the LSP and $X_R$ is the NLSP. Like the previous case, here also the mass difference $\Delta M_{X^\pm_1 - X_R}$  at tree-level is always negative as shown in the blue curve, and  it does not provide any tree-level dark matter candidate. At one-loop, although the hierarchy changes, the scenario remains the same  as $X^{\pm \pm}$  becomes the LSP and $X_R$ the NLSP as shown by the violet curve. Till $M_{X_R}=1940$ GeV, $ X^{\pm \pm}$  stays as LSP  making the region unphysical.  From $M_{X_R}=1940$ to $1990$ GeV as the mass gap becomes positive,  $ X^{\pm \pm}$ becomes NLSP, having $X_R$ the neutral dark matter, while keeping the mass gap as low as $27$ MeV. However,  for $M_{X_R} > 1990$ TeV, $X^\pm_1 $ becomes NLSP while a neutral $X_R$ becomes LSP,  making the scenario possible for a dark matter, even though the mass difference between them is very low i.e. around $\lesssim 245$ MeV.

The one-loop corrections play a crucial role in changing the tree-level mass hierarchy at one-loop level and opening regions in the parameter space that are excluded at tree level due to a charged LSP. Nevertheless, even at one-loop the mass spectrum stays very compressed one from a few MeV to a few hundreds of MeV. When the mass gap between the charged NLSP and LSP is above the $\pi^\pm$ mass, it can give rise to displaced pions \cite{Chun:2009mh,Bandyopadhyay:2010wp, Jangid:2020qgo,SabanciKeceli:2018fsd}. However, due to very low energy this pions are very difficult to be tracked at colliders giving rise to disappearing charged tracks \cite{Bandyopadhyay:2024plc, Bandyopadhyay:2023joz}. However, we have regions where the mass gaps are less than pion  and/or muon masses giving rise to have displaced very soft charged muon or only electrons  final states, the detectability of such soft charged particle needs a detailed study.

\begin{figure}[h]
	\begin{center}
		\mbox{
			\subfigure[]{\includegraphics[height=0.17\textheight,width=0.3\linewidth]{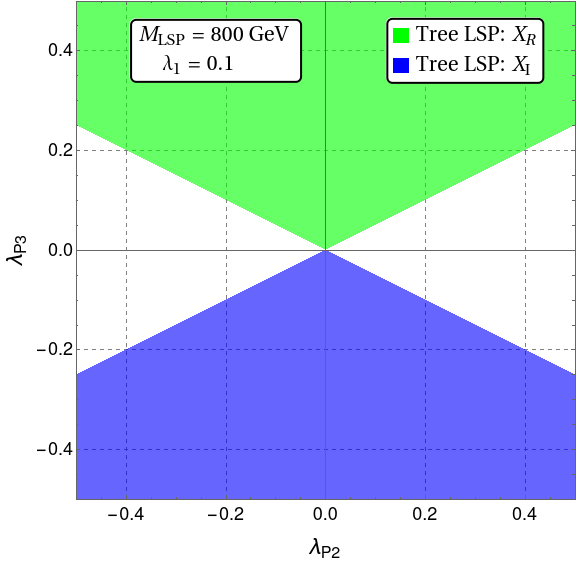}}
			\hspace*{0.25cm}
			\subfigure[]{\includegraphics[height=0.17\textheight,width=0.3\linewidth]{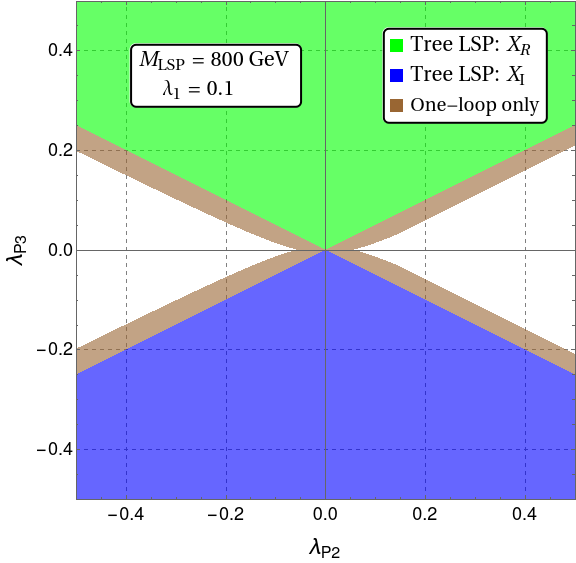}}
			\hspace*{0.25cm}
			\subfigure[]{\includegraphics[height=0.17\textheight,width=0.3\linewidth]{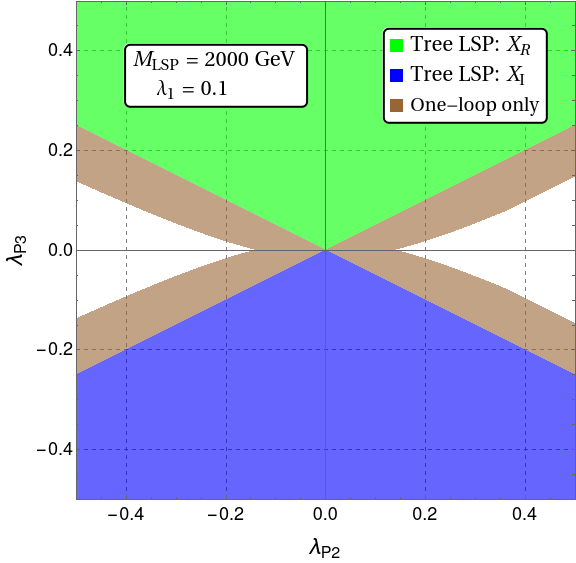}}
		}
		\mbox{
			\subfigure[]{\includegraphics[height=0.17\textheight,width=0.3\linewidth]{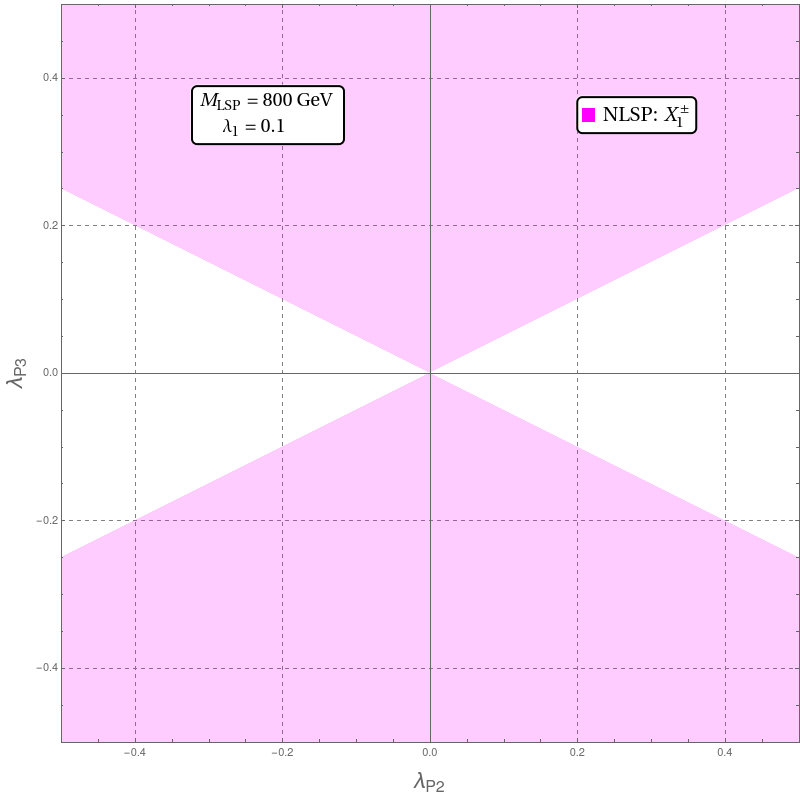}}
			\hspace*{0.25cm}
			\subfigure[]{\includegraphics[height=0.17\textheight,width=0.3\linewidth]{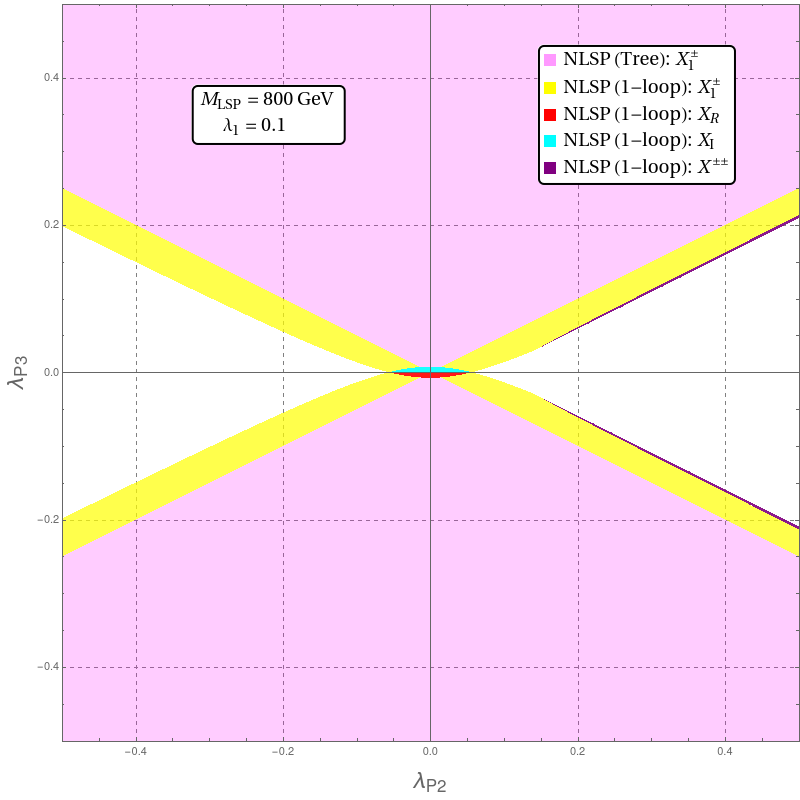}}
			\hspace*{0.25cm}
			\subfigure[]{\includegraphics[height=0.17\textheight,width=0.3\linewidth]{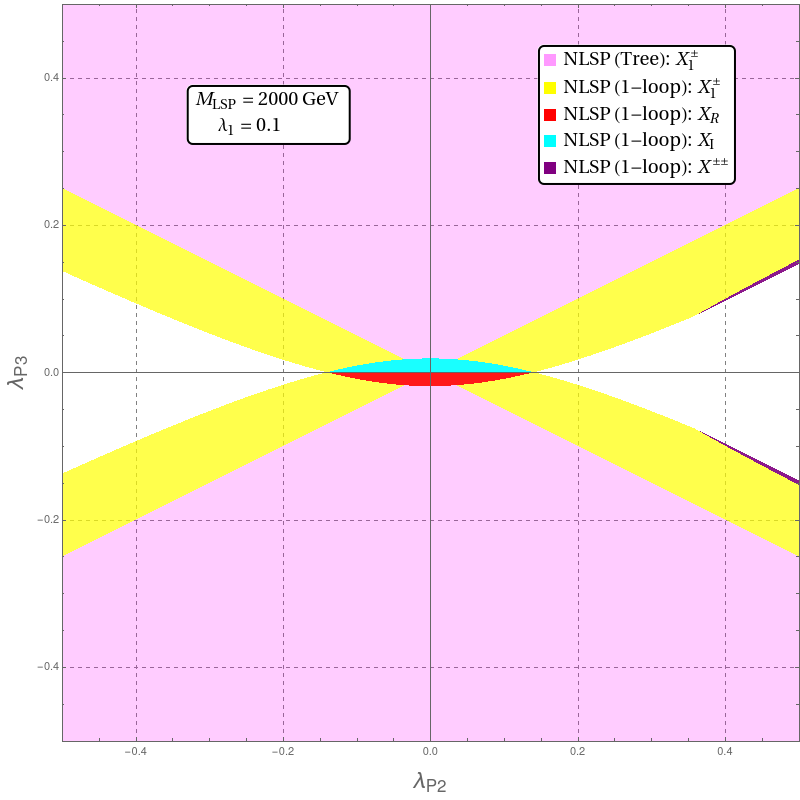}}
		}
	\end{center}
	\caption{Panels (a), (b), and (c) show the regions in the $\lambda_{P2}$--$\lambda_{P3}$ plane where a neutral scalar DM candidate exists for $M_{X_R} = 800$ GeV, $800$ GeV, and $2000$ GeV, respectively. The green and blue regions correspond to $X_R$ and $X_I$ being the DM candidate, respectively. The brown regions in panels (b) and (c) denote parameter regions where $X_R$ or $X_I$ is the DM candidate at one-loop level, but not at tree level. Panels (d), (e), and (f) show the corresponding NLSP regions in the $\lambda_{P2}$--$\lambda_{P3}$ plane for the same benchmark masses. The magenta regions correspond to the NLSP being $X_1^\pm$ at tree level. Yellow, red, cyan, and violet regions respectively corresponds to $X_1^\pm$, $X_R$, $X_I$, and $X^{\pm\pm}$ at one-loop level but not at tree-level.}
	\label{fig:phases-lightest-particles}
\end{figure}

After the discussion on the mass gaps between NLSP and LSP at both tree- and one-loop level, let us focus on the feasible parameter space for LSP and NLSP in the Higgs portal coupling planes. \autoref{fig:phases-lightest-particles}(a),  (b) and (c) show the regions in the $\lambda_{P2}$--$\lambda_{P3}$ plane where the lightest $\mathbb{Z}_2$-odd particle (LSP) is a neutral scalar for  $M_{\rm LSP}=800$ GeV for (a), (b) and $2000$ GeV for (c), respectively. In these plots, the green region ($\lambda_{P3} > 0$) corresponds to the parameter space where the CP-even scalar $X_R$ is the LSP, while the blue region ($\lambda_{P3} < 0$) corresponds to the CP-odd scalar $X_I$ being the LSP. \autoref{fig:phases-lightest-particles}(a) represent entirely a  the tree-level LSP region, satisfying the condition $2\,\lambda_{P3} > |\lambda_{P2}|$.  The origin $\lambda_{P2}=\lambda_{P3}=0$ marks the point where all scalar components are degenerate at tree level. The brown regions in  \autoref{fig:phases-lightest-particles}(b),  (c) correspond to one-loop corrected parameter space where we have neutral LSP due to the change of mass hierarchy at one-loop as explained in the previous paragraphs. The regained regions of neutral LSP as dark matter regions at one-loop get enhanced as we move from  $M_{\rm LSP}=800$ GeV in \autoref{fig:phases-lightest-particles}(a) to  $M_{\rm LSP}=2000$ GeV at  \autoref{fig:phases-lightest-particles}(b), respectively.

Similar to LSP case, now we show the regions of NLSP and their nature in 
\autoref{fig:phases-lightest-particles}(d), (e), and (f) in the $\lambda_{P2}$--$\lambda_{P3}$ plane. Here  \autoref{fig:phases-lightest-particles}(d) shows a tree-level NLSP  region for $M_{\rm LSP}=800$ GeV given in magenta colour, where $X^\pm_1$ remains a NLSP for the whole regions.  \autoref{fig:phases-lightest-particles}(e), (f) represent the NLSP regions for $M_{\rm LSP}=800,\, 2000$ GeV, respectively considering one-loop mass corrections. It can be seen at one-loop regions that the $X^\pm_1$ as NLSP region extends in the yellow regions, which were earlier not physically viable. However, there are parameter space where the nature of NLSP changes as explained in the previous paragraphs. It can be seen near the origin of the planes  as the red and cyan  regions, where $X_R$ and $X_I$ become the neutral NLSPs,  respectively and these region enhance as we move to higher LSP mass from 800 GeV in (e) to  2000 GeV in(f). However, the violet regions, which appear as a narrow band at the right edge of the yellow region, shows the doubly charged $X^{\pm\pm}$ as NLSP, and these region reduces as we move to higher LSP mass from 800 GeV in (e) to  2000 GeV in(f).

\section{Dark matter phenomenology}\label{DMpheno}
The inert models are simple extensions of Standard Model that provide dark matter (DM) candidates. However, most of the parameter space for candidate DMs in the models ISM, ITM and IDM provide dark matter masses lesser than a few TeV \cite{Jangid:2020qgo, Bandyopadhyay:2024plc, Bandyopadhyay:2025ilx,Bandyopadhyay:2023joz, Bandyopadhyay:2025jlg,Bandyopadhyay:2024gyg}. These parameter spaces have been now highly constrained  by the data coming from recent direct detection experiments like LUX-ZEPLIN \cite{LZ:2024zvo}.  Hence, it is worth investigating the possibility of DM in models with higher $SU(2)$ multiplets in detail. In I4M,  we  have two neutral components, the CP-even real scalar $X_R$ and CP-odd pseudo-scalar $X_I$, lightest of which can be the  dark matter candidate. For this study we focus on $X_R$ as the dark matter. This when in thermal equilibrium will go through possible annihilation and co-annihilation processes. The dominant annihilation modes are shown in \autoref{fig:annihilation} where DM annihilates to $ZZ, \, W^\pm W^\mp$. Depending on the nature of NLSP, different  co-annihilations are possible. In \autoref{fig:coannihilations}  we summarize $X_R-X^\pm_1, \, X_R-X^{\pm\pm}, \, X_R-X_I$  co-annihilation modes. Due to the compressed nature of the spectra, for most of the scanned points co-annihilation is dominant over the annihilation. There are dominant modes where the other particles from the $I4M$ annihilate or co-annihilate to SM particles as shown in \autoref{Feyn:DManni}.

\begin{figure}[h]
	\begin{tikzpicture}[
		baseline=(current bounding box.center),
		scale=0.8,
		transform shape,
		fermion/.style={solid, thick},
		boson/.style={decorate, decoration={snake}, thick},
		scalar/.style={dashed, thick}
		]
		\begin{scope}		
			\draw[scalar] (-0.7,1) -- (0,0);
			\draw[scalar] (-0.7,-1) -- (0,0);
			\draw[scalar] (0,0) -- (1.2,0) node[midway, above] {$h$};
			\draw[boson] (1.2,0) -- (1.8,0.9) node[right] {$Z$, $W^+$};
			\draw[boson] (1.2,0) -- (1.8,-0.9) node[right] {$Z$, $W^-$};
			\node[left] at (-1,1) {$X_R$};
			\node[left] at (-1,-1) {$X_R$};
		\end{scope}
		\hspace*{4cm}
		\begin{scope}[xshift=1cm]
			\draw[scalar] (-1,1) -- (0,1);
			\draw[scalar] (-1,-1) -- (0,-1);
			\draw[scalar] (0,1) -- (0,-1) node[midway, right] {$X_I$};
			\draw[boson] (0,1) -- (1,1) node[right] {$Z$, $W^+$};
			\draw[boson] (0,-1) -- (1,-1) node[right] {$Z$, $W^-$};
			\node[left] at (-1,1) {$X_R$};
			\node[left] at (-1,-1) {$X_R$};
		\end{scope}
		\begin{scope}[xshift=6cm]
			\draw[scalar] (0,0) -- (-0.8,1) node[left] {$X_R$};
			\draw[scalar]  (0,0) -- (-0.8,-1) node[left] {$X_R$};
			\draw[boson] (0,0) -- (0.8,1) node[right] {$Z$, $W^+$};
			\draw[boson] (0,0) -- (0.8,-1) node[right] {$Z$, $W^-$};
		\end{scope}
			\begin{scope}[xshift=10cm]
			\draw[scalar] (0,0) -- (-0.8,1) node[left] {$X_R$};
			\draw[scalar]  (0,0) -- (-0.8,-1) node[left] {$X_R$};
			\draw[scalar] (0,0) -- (0.8,1) node[right] {$h$};
			\draw[scalar] (0,0) -- (0.8,-1) node[right] {$h$};
		\end{scope}
	\end{tikzpicture}
	\caption{Dominant annihilation modes $X_R\, X_R  \rightarrow Z Z,\ W^+ W^-,\ h h$}
	\label{fig:annihilation}
\end{figure}
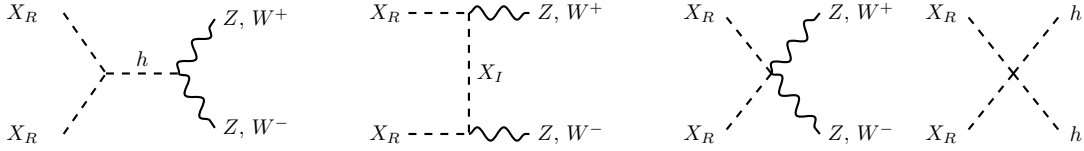

\begin{figure}[H]
	\begin{tikzpicture}[
		baseline=(current bounding box.center),
		scale=0.8,
		transform shape,
		fermion/.style={dashed, thick},
		boson/.style={decorate, decoration={snake}, thick},
		scalar/.style={dashed, thick}
		]
		\hspace*{-1.5cm}
		\begin{scope}
			\draw[scalar] (0,1) -- (-0.8,1)  node[left] {$X_R$};
			\draw[scalar] (0,-1) -- (-0.8,-1) node[left] {$X_1^\pm$};
			\draw[scalar] (0,1) -- (0,-1) node[midway, right] {$X_1^\pm, X_2^\pm$};
			\draw[boson] (0,1) -- (0.8,1) node[right] {$W^\pm$};
			\draw[boson] (0,-1) -- (0.8,-1) node[right] {$Z$};
		\end{scope}
			
		\begin{scope}[xshift=4cm]
			\draw[scalar] (0,1) -- (-0.8,1)  node[left] {$X_R$};
			\draw[scalar] (0,-1) -- (-0.8,-1) node[left] {$X^{++}$};
			\draw[scalar] (0,1) -- (0,-1) node[midway, right] {$X_1^+, X_2^+$};
			
			\draw[boson] (0,1) -- (0.8,1) node[right] {$W^+$};
			\draw[boson] (0,-1) -- (0.8,-1) node[right] {$W^+$};
		\end{scope}
		\begin{scope}[xshift=8cm]
			\draw[scalar] (0,1) -- (-0.8,1)  node[left] {$X_R$};
			\draw[scalar] (0,-1) -- (-0.8,-1) node[left] {$X_I$};
			\draw[scalar] (0,1) -- (0,-1) node[midway, right] {$X_1^+, X_2^+$};
			\draw[boson] (0,1) -- (0.8,1) node[right] {$W^+$};
			\draw[boson] (0,-1) -- (0.8,-1) node[right] {$W^-$};
		\end{scope}
		\begin{scope}[xshift=12cm]
			\draw[scalar] (0,1) -- (-0.8,1)  node[left] {$X_R/X^\pm_{1,2}$};
			\draw[scalar] (0,-1) -- (-0.8,-1) node[left] {$X^\pm_{1,2}/X_R$};
			\draw[scalar] (0,1) -- (0,-1) node[midway, right] {$X_R/X^\pm_{1,2}$};
			\draw[scalar] (0,1) -- (0.8,1) node[right] {$h$};
			\draw[boson] (0,-1) -- (0.8,-1) node[right] {$W^\pm$};
		\end{scope}
		\begin{scope}[xshift=16 cm]
			\draw[scalar] (0,1) -- (-0.8,1)  node[left] {$X_R/X_I$};
			\draw[scalar] (0,-1) -- (-0.8,-1) node[left] {$X_I/X_R$};
			\draw[scalar] (0,1) -- (0,-1) node[midway, right] {$X_R/X_I$};
			\draw[scalar] (0,1) -- (0.8,1) node[right] {$h$};
			\draw[boson] (0,-1) -- (0.8,-1) node[right] {$Z$};
		\end{scope}
	
%
\begin{scope}[xshift=-1.5cm,yshift=-4cm]
	\draw[scalar] (0,0) -- (-0.8,1) node[left] {$X_R$};
	\draw[scalar]  (0,0) -- (-0.8,-1) node[left] {$X^{++}$};
	\draw[boson] (0,0) -- (0.8,1) node[right] {$W^+$};
	\draw[boson] (0,0) -- (0.8,-1) node[right] {$W^+$};
\end{scope}	
\begin{scope}[xshift=2.cm,yshift=-4cm]
	\draw[scalar] (0,0) -- (-0.8,1) node[left] {$X_R$};
	\draw[scalar] (0,0) -- (-0.8,-1) node[left] {$X^\pm_{1,2}$};
	\draw[boson] (0,0) -- (0.8,1) node[right] {$W^\pm$};
	\draw[boson] (0,0) -- (0.8,-1) node[right] {$Z/\gamma$};
\end{scope}	
\begin{scope}[xshift=5.cm,yshift=-4cm]
	\draw[scalar] (0,0) -- (-0.6,1)  node[left] {$X_R$} ;
	\draw[scalar] (0,0) --(-0.6,-1)  node[left] {$X_I$};
	\draw[boson] (0,0) -- (1.5,0) node[midway, above] {$Z$};
	\draw[boson] (1.5,0) -- (2.2,0.9) node[right] {$W^+$};
	\draw[boson] (1.5,0) -- (2.2,-0.9) node[right]  {$W^-$};
\end{scope}
\begin{scope}[xshift=9.5cm,yshift=-4cm]
	\draw[scalar] (0,0) -- (-0.6,1)  node[left] {$X_R$} ;
\draw[scalar] (0,0) --(-0.6,-1)  node[left] {$X^\pm_{1,2}$};
\draw[boson] (0,0) -- (1.5,0) node[midway, above] {$W^\pm$};
\draw[boson] (1.5,0) -- (2.2,0.9) node[right] {$W^\pm$};
\draw[boson] (1.5,0) -- (2.2,-0.9) node[right]  {$Z/\gamma$};
\end{scope}
\begin{scope}[xshift=14cm,yshift=-4cm]
	\draw[scalar] (0,0) -- (-0.6,1)  node[left] {$X_R$} ;
	\draw[scalar] (0,0) --(-0.6,-1)  node[right] {$X^\pm_{1,2}/X_I$};
	\draw[boson] (0,0) -- (1.5,0) node[midway, above] {$W^\pm/Z$};
	\draw[boson] (1.5,0) -- (2.2,0.9) node[right] {$h$};
	\draw[scalar] (1.5,0) -- (2.2,-0.9) node[right]  {$W^\pm/Z$};
\end{scope}
	\end{tikzpicture}
	\caption{Contribution via different co-annihilation modes $X_R-X^{\pm\pm}, \, X_R-X^\pm_{1,2}, \,  X_R-X_I$.}
	\label{fig:coannihilations}
\end{figure}
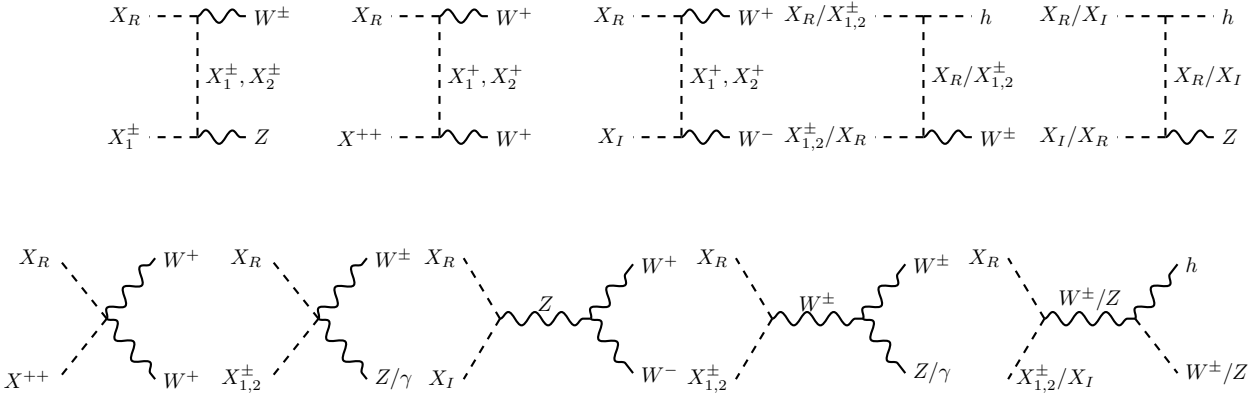
For the calculations of these annihilations and co-annihilations of dark matter to SM particles, we used the 
model files generated via \texttt{SARAH} 4.15.1\cite{Staub:2013tta} and implemented them in {\tt micrOMEGAs} 6.2.3 \cite{Belanger:2013oya,Belyaev:2012qa}. Here instead of the tree-level masses we implemented the one-loop masses for the quartet as explained in the previous section.  The dark matter relic is calculated using  {\tt micrOMEGAs}, where the following Boltzmann equation for the DM number density is followed:
\begin{equation}
	\frac{d n_{\rm DM}}{dt}\,+ 3 H\, n_{\rm DM}=  -\langle \sigma v \rangle ( n^2_{\rm DM} - n^2_{\rm DM, \,eq}),
\end{equation}
where $n_{\rm DM}$ is the number density of the dark matter, $n_{\rm DM, \,eq}$ is the same in the thermal equilibrium, H is the Hubble constant and $\langle \sigma v \rangle$  is the thermal average of the total annihilation cross-section \cite{Belanger:2013oya}.

Here the $\langle \sigma v \rangle$   depends on the annihilation, co-annihilations of the dark matter and other scatterings as shown in  \autoref{fig:annihilation}, \autoref{fig:coannihilations} and \autoref{Feyn:DManni}, respectively, which depend mostly on the Higgs portal couplings as described in \autoref{eq:4plet-portal-potential}. However, a certain combination of these portal couplings makes the effective Higgs-DM-DM coupling, which is  vev times the $\lambda_{X_R}$ as can be seen below
\begin{equation}
	\lambda_{X_R} \equiv \frac{1}{6} \big(3 \lambda_{P1} + 2 \lambda_{P2} - 4 \lambda_{P3}\big),
\end{equation} 
which plays an important role in the annihilation and co-annihilation of the DM as briefed in \autoref{fig:annihilation}, \autoref{fig:coannihilations}. The corresponding effective quartic coupling of Higgs-Higgs-DM-DM is equal to $\frac{\lambda_{X_R}}{2}$, which also plays a crucial role in the DM annihilation at early epochs. As we discuss the variation of  DM relic with the mass of DM, as well as, the constraint coming from the direct dark matter experiment, we consider different ratio of the portal couplings $\lambda_{P1},\,\lambda_{P2},\, \lambda_{P3} $ to demonstrate the portal coupling dependency to these measurements. This ratio can be written as 
\begin{equation}
	\lambda_{P1} : \lambda_{P2} : \lambda_{P3} = r_1 : r_2 : r_3.
\end{equation}
The corresponding values of portal couplings are 
\begin{equation}
	\lambda_{P1} = \frac{r_1}{N} \lambda_{X_R},\ \ \  \lambda_{P2} = \frac{r_2}{N} \lambda_{X_R}, \ \ \  \lambda_{P3} = \frac{r_3}{N} \lambda_{X_R},
\end{equation}
where the normalization constant $N = \frac{1}{6}(3 r_1 + 2 r_2 - 4 r_3)$. 
\subsection{Relic density }
Armed with all the theoretical set up we now plan to scan the parameter space for the DM relic which is allowed by the recent measurement of Planck data  $\Omega h^2_{\text{obs}} = 0.1200 \pm 0.0012$ \cite{Planck:2018vyg}. For this,  $M_{X_R}$ and $\lambda_{X_R}$ are chosen as the input parameters for the analysis with a variation range of $M_{X_R} \in [10, 15000]$ GeV and $\lambda_{X_R} \in [ -5.0, \,5.0]$, respectively. Here we considered points with $\lambda_{P3} > 0$ as we will be focussing on the case where $X_R$ is the candidate DM. Although, from the expressions of the masses and vertices, one can conclude that the analysis will hold good for $\lambda_{P3} < 0$ where $X_I$ will be the candidate DM. 

\begin{figure}[H]
	\begin{center}
		\mbox{
			\subfigure[]{\includegraphics[height=0.25\textheight,width=0.5\linewidth]{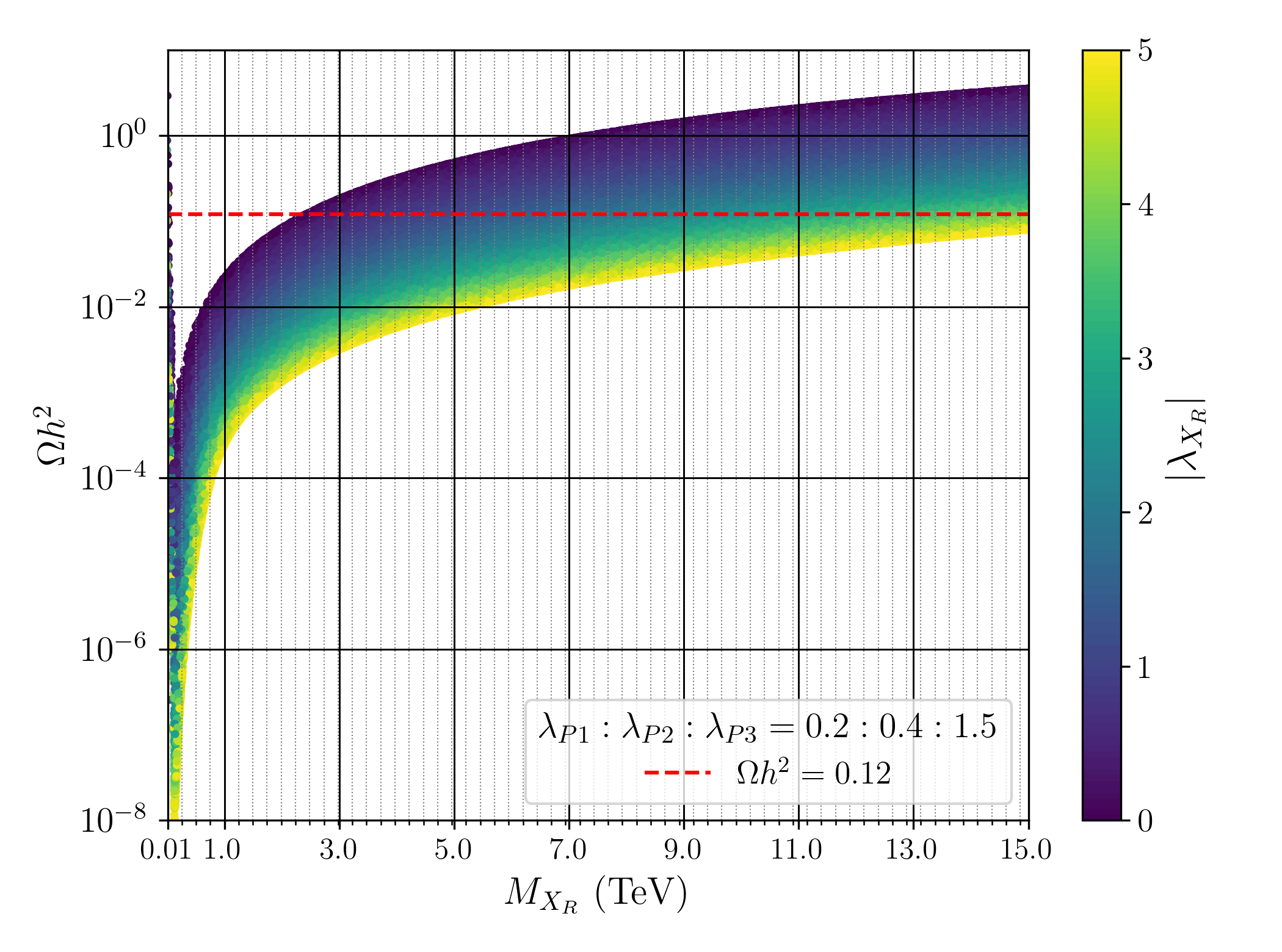}}
			\hspace*{0.25cm}
			\subfigure[]{\includegraphics[height=0.25\textheight,width=0.5\linewidth]{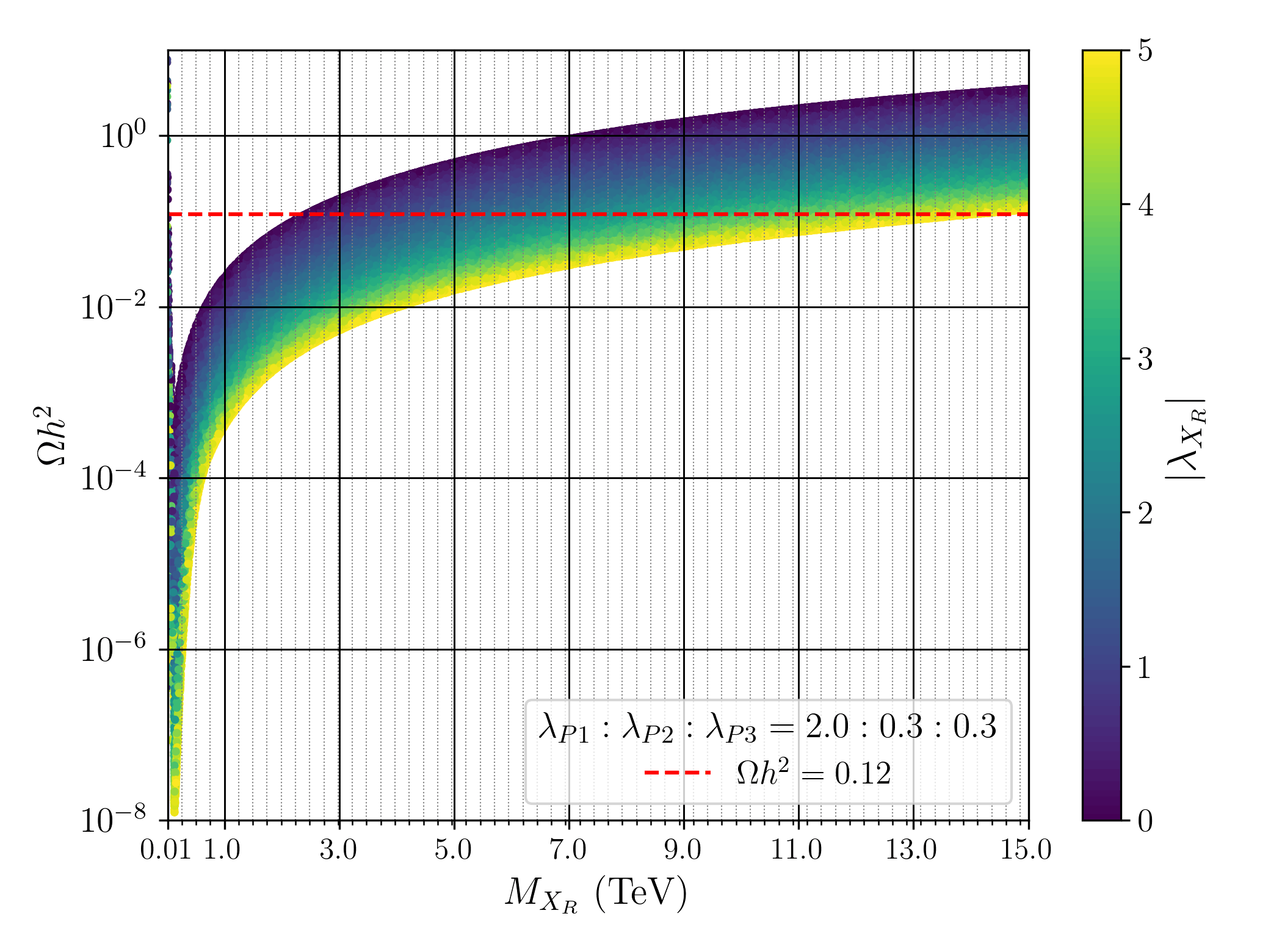}}
		}
		\mbox{
			\subfigure[]{\includegraphics[height=0.25\textheight,width=0.5\linewidth]{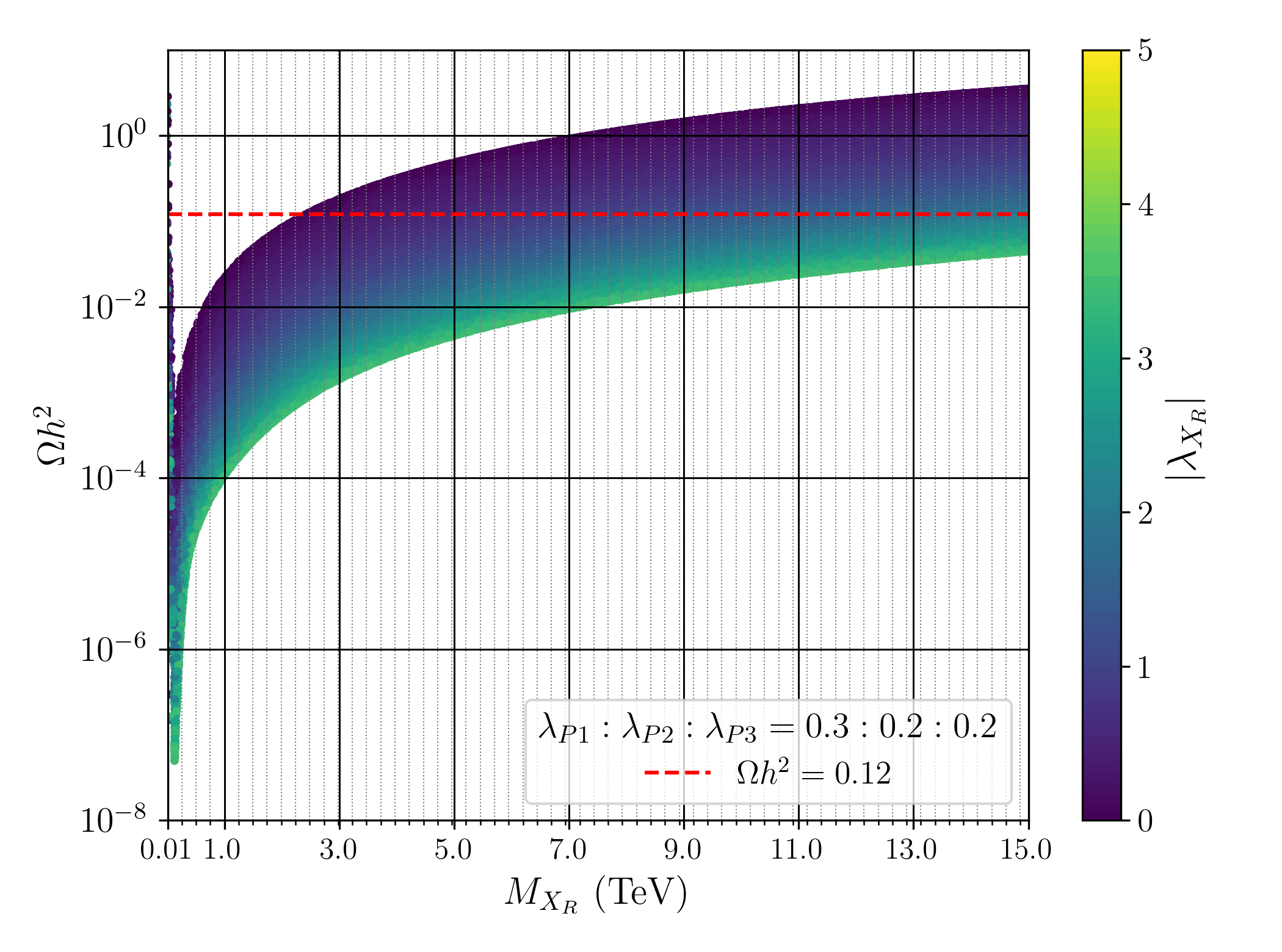}}
			\hspace*{0.25cm}
			\subfigure[]{\includegraphics[height=0.25\textheight,width=0.5\linewidth]{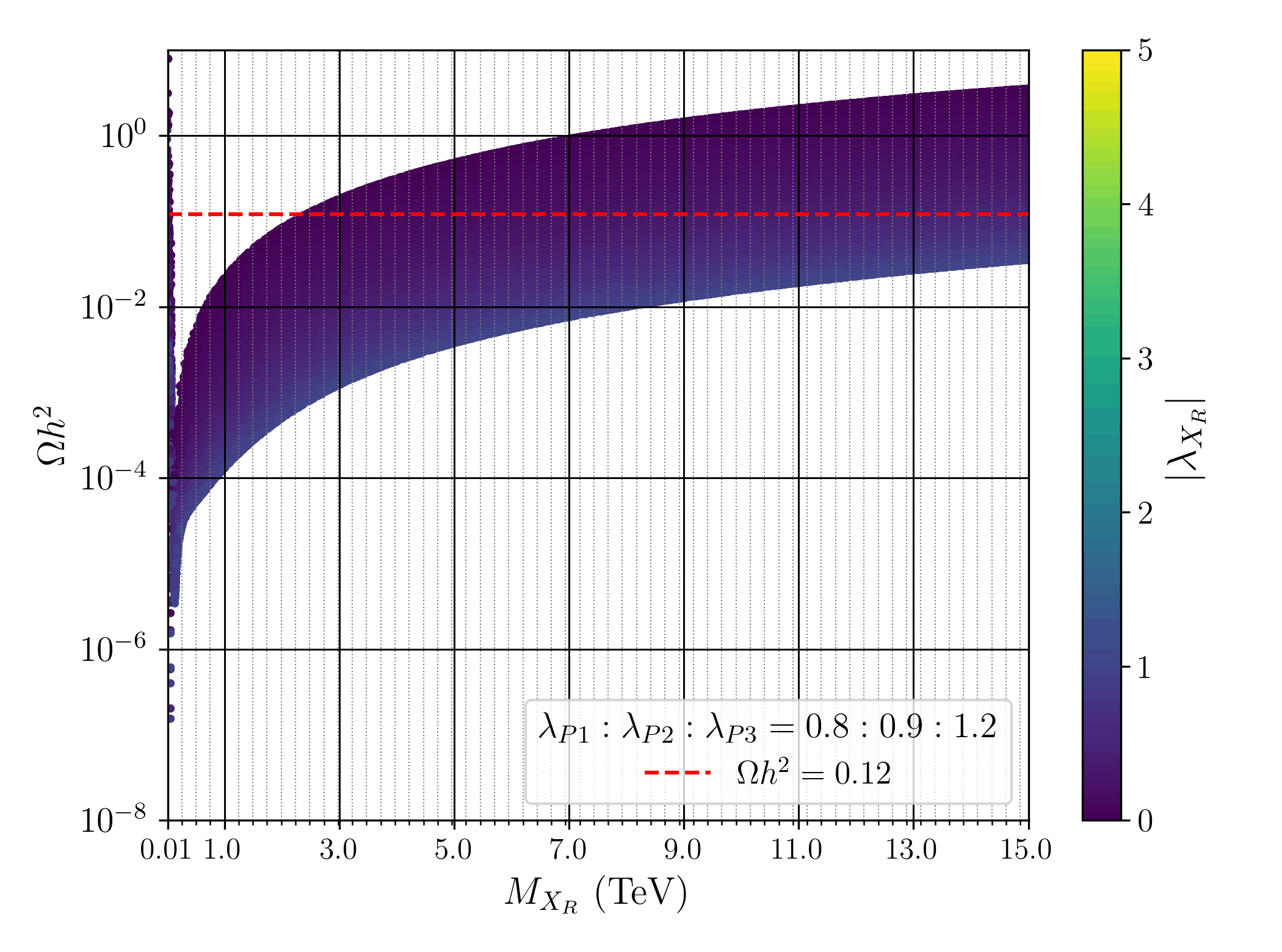}}
		}
	\end{center}
	\caption{DM relic density $\Omega h^2$ verses the DM mass $M_{X_R}$ with different values of $\lambda_{X_R}$ for the four ratios of
		$(r_1 : r_2 : r_3)=(0.2:.0.4:1.5),\, (2.0:0.3:0.3),\, (0.3:0.2:0.2), (0.8:0.9:1.2) $ described in (a), (b), (c) and (d).}\label{fig:DM-relic}
\end{figure}

Since the mass spectrum is very compressed both annihilation and co-annihilations will contribute. We plotted the DM relic for the four ratios of 	$(r_1 : r_2 : r_3)=(0.2:0.4:1.5),\, (2.0:0.3:0.3),\, (0.3:0.2:0.2), (0.8:0.9:1.2) $ in \autoref{fig:DM-relic}(a), (b), (c), (d), respectively. In the x-axis we vary the DM mass from 10 GeV  to 15000 GeV.  The dominant annihilation  modes are $X_R, X_R \to W^+ W^-, \,  Z Z$, where the $W^\pm W^\mp$ mode is totally dependent on the gauge couplings, however the  $ZZ$ modes also  gets contribution from the Higgs portal coupling $\lambda_{X_R}$. The same portal coupling for its large absolute  values can make $hh$ annihilation mode comparable. This is apparent from the values of the of  $\lambda_{X_R}$, where lower values are given by blue points and the higher absolute  values are given by green or lighter yellow. We enhance the absolute  values of $\lambda_{X_R}$  when we change $\lambda_{X_R} = -0.1\, \rm{to}\, -4.0$ in  \autoref{fig:DM-relic}(a). For $M_{X_R}=6000$ GeV,  we see that for lower  $|\lambda_{X_R}|$, the DM annihilates dominantly into  $W^\pm W^\mp$ which is not enough to keep the DM relic within the allowed bounds giving rise to an over abundant point (see the top point of the blue band). Whereas taking a larger  absolute value of  $\lambda_{X_R}=-4.0$, not only enhances $ZZ$ annihilation but also make the annihilation to $hh$ comparable. These enhance the total annihilation cross-section giving rise to an yellow underabundant point as can be seen from \autoref{fig:DM-relic}(a). There are negative interference terms due to the  $\lambda_{X_R}<0$, however,  quadratic  and quartic  terms  coming from  Higgs portal and contact  annihilation dominate over this interference term, enhancing the total annihilation cross-section. Moreover,  the enhancement in the co-annihilation processes are more for  the increment of  $|\lambda_{X_R}|$ in this case.

In  \autoref{fig:DM-relic}(b) we consider the parameter space where $\lambda_{P1}$ is large by choosing $r_1:r_2:r_3= 2.0:0.3:0.3$, In this case the DM relic points are for positive $\lambda_{X_R}$. Here again we see the blue points are with smaller DM annihilation and co-annihilation cross-sections giving rise to mostly larger DM relic and even over abundant points.  Interestingly, due to the choice of  $\lambda_{X_R} > 0$, the interference terms involving $\lambda_{X_R}$ are positive contributions  for  the annihilation. Contrary to the previous case, thus the annihilation increases more as we go for larger $\lambda_{X_R}$. For example , here at $M_{X_R}=6000$ GeV, as we change the $\lambda_{X_R}=0.1\, \rm{to}\, 4.0$, we reach to an underabundant point from an over abundant point in DM relic. However, the enhancement in the co-annihilation  for such increment in $\lambda_{X_R}$ are relatively smaller, making the high DM mass cut off for correct DM relic in almost similar range as in \autoref{fig:DM-relic}(a).

\autoref{fig:DM-relic}(c)  we present the ratio as $(0.3:0.2:0.2)$, which corresponds to the individual portal couplings with positive and relatively close values  to each other.  The observed relic is satisfied from $M_{X_R} \gsim 2000$ GeV for smaller magnitude of $\lambda_{X_R}$. As the $M_{X_R}$ increases, the red line, which corresponds to the observed DM relic,  passes through relatively larger values of $\lambda_{X_R}$. However, unlike the previous cases there are more underabundant points for larger masses, making  the over abundant cut-off for higher DM mass values.
	
Similarly, in  \autoref{fig:DM-relic}(d) the fixed ratio is  $( 0.8 : 0.9 : 1.2)$ with  $\lambda_{X_R}<0$. Notice that, the $\lambda_{X_R}$ is relatively restricted as for $|\lambda_{X_R}| \gsim 1.04$, the individual portal couplings couplings cross perturbativity limit ($\lambda_{P1,P2,P3} \gsim 4 \pi$), which makes the green and yellow points absent from the figure unlike the other scenarios.. Nevertheless, there are relatively more number of under-abundant points for this fixed ratio compared to previous cases. Apart from the $s$-channel resonance point around $M_{X_R}=60$ GeV,  the correct DM relic can be obtained for  $M_{X_R} \sim 2500$ GeV for the smaller magnitude of $\lambda_{X_R}$.  For higher portal couplings DM cut-off for correct relic goes beyond 15 TeV as depicted in 	\autoref{fig:DM-relic}(d).
	
It is interesting to see that around $M_{X_R}\approx60$ GeV, we can see the Higgs s-channel resonance point satisfying the correct relic in all four cases, and it is also observed for other scalar extended DM models. However, the lower multiplets of $SU(2)$  i.e. singlet, doublet and triplet get an upper mass bounds around a TeV to a few TeV \cite{Jangid:2020dqh,Bandyopadhyay:2025jlg, Bandyopadhyay:2024gyg, Bandyopadhyay:2023joz, Bandyopadhyay:2024plc}. However, such upper DM mass bounds is relaxed beyond 15 TeV in the case of $I4M$.  In the next section we will observe how the direct dark matter cross-section put bounds on the DM mass and the effective Higgs portal coupling  $\lambda_{X_R}$.

\subsection{Bounds from direct detection}\label{DDM}
In this section we will focus on restricting the DM mass and the effective Higgs portal coupling via the direct detection experiments. In this case the scalar  DM couples to the nucleus via the Higgs boson making  $\lambda_{X_R}$ the effective coupling for the direct detection.  There are several direct detection experiments like  XENONnT \cite{XENON:2023iku}, LZ (LUX-ZEPLIN) \cite{LZ:2024zvo}, etc., that are currently running. However, since the LUX-ZEPLIN 2024 data is the most stringent one we use that to constrain the parameter space. The spin-independent direct detection cross-section of the DM and nucleon is given by
\begin{equation}
		\sigma_{\text{SI}} = \frac{\lambda_{X_R}^2 f_N^2}{(4\pi) M_h^4} \frac{M_N^2}{(M_N + M_{X_R})^2},
\end{equation}
	which is proportional to the $\lambda_{X_R}^2 $, the DM-Higgs effective coupling,  $f_N \approx 0.30$ represents the nucleon form factor \cite{Belanger:2013oya,Cline:2013gha}, $M_h$ is the mass of the Higgs boson, and $M_N$ is the mass of the nucleon.  We used  \texttt{micrOMEGAs} to estimate this cross-section for the scanned parameter space. However, it it interesting to notice that for the underabundant DM relic points as shown in \autoref{fig:DM-relic}, only the fraction which contribute as DM will be detected via the direct detection experiments. Thus the effective direct detection cross-section becomes 
	\begin{equation}\label{sigmanSIscaled}
		\sigma^{\text{scaled}}_{\text{SI}} = \frac{\relic}{\relicobs}\sigma_{\text{SI}},
	\end{equation}
where the $\sigma_{\text{SI}} = \frac{1}{2}(\sigma^{\text{proton}}_{\text{SI}} + \sigma^{\text{neutron}}_{\text{SI}})$ is the average  spin-independent scattering cross section per nucleon. 
\begin{figure}[h]
	\begin{center}
		\mbox{
			\subfigure[]{\includegraphics[height=0.2\textheight,width=0.32\linewidth]{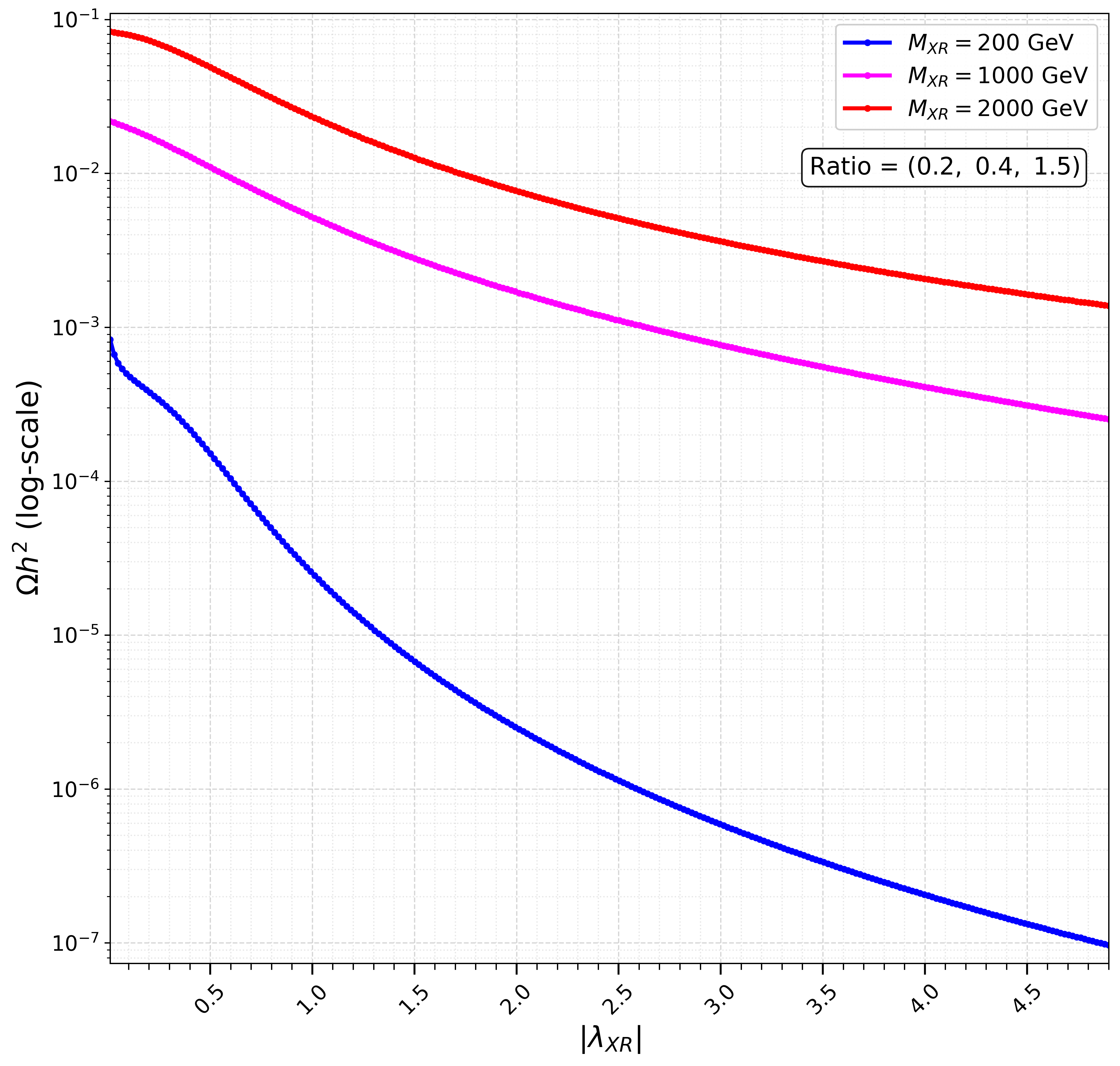}}
			\hspace*{0.25cm}
			\subfigure[]{\includegraphics[height=0.2\textheight,width=0.32\linewidth]{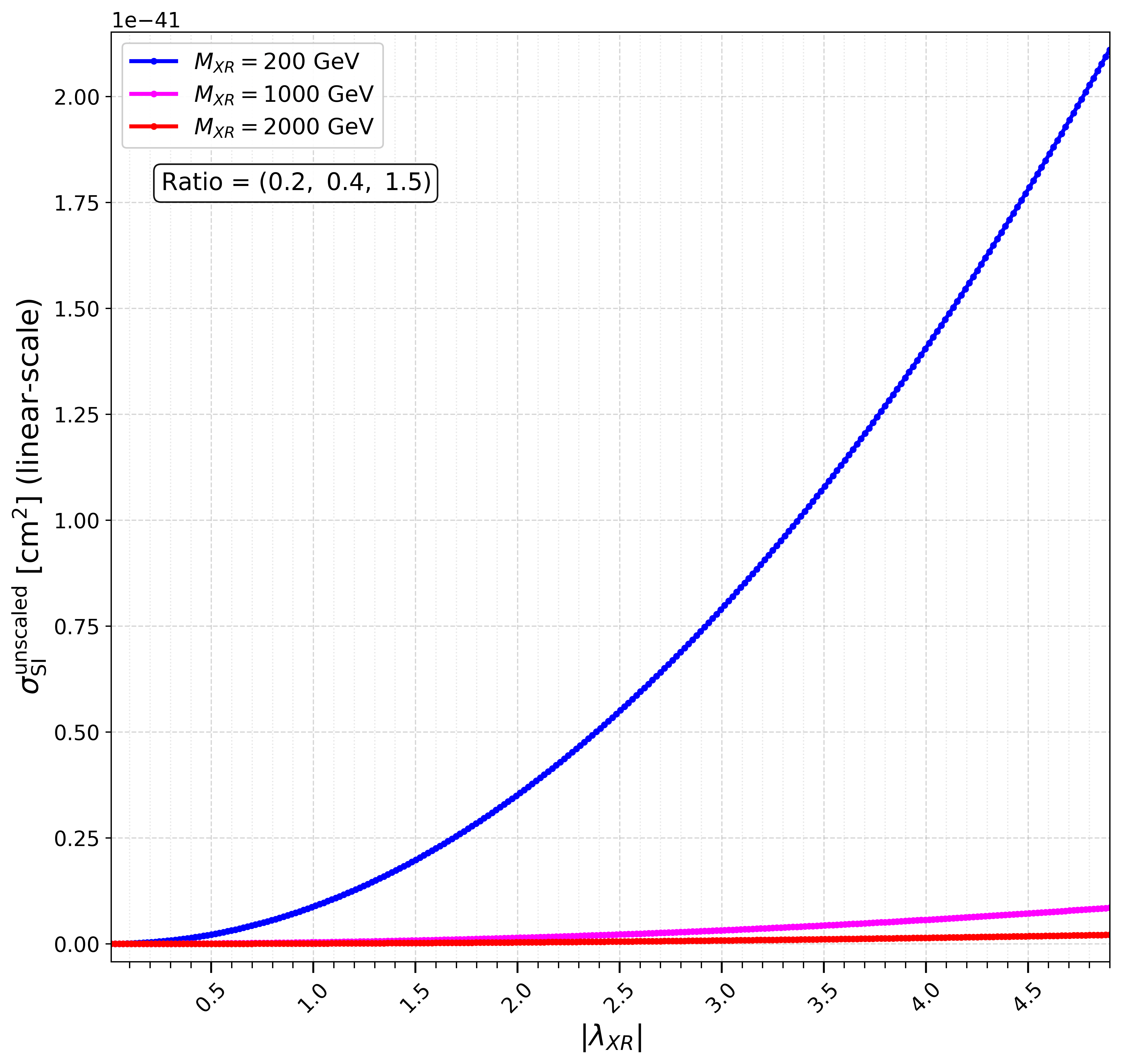}}
			\hspace*{0.25cm}
			\subfigure[]{\includegraphics[height=0.2\textheight,width=0.32\linewidth]{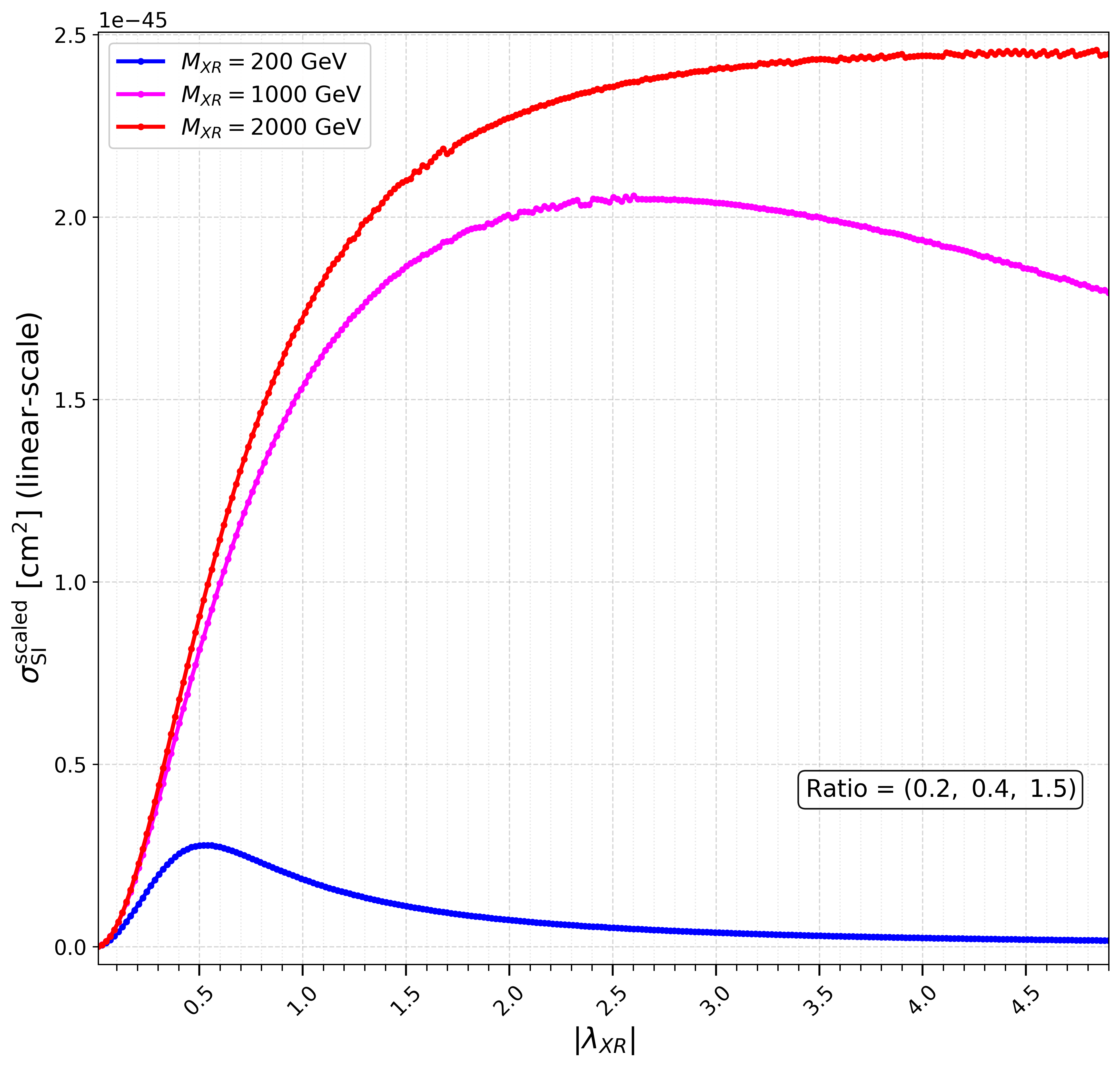}}}
	\end{center}
	\caption{The variation of relic density $\Omega h^2$, unscaled cross-section $\sigma_{\mathrm{SI}}^\mathrm{unscaled}$, scaled cross-section $\sigma_{\mathrm{SI}}^\mathrm{scaled}$ with $|\lambda_{X_R}|$ for $M_{X_R}=200, \, 1000,\, 2000$ GeV in red, magenta and blue, respectively for the ratio (0.2 : 0.4 : 1.5).}
	\label{fig:DM-sigmaSI-fixed-mass}
\end{figure}

\begin{figure}[H]
	\begin{center}
		\mbox{
			\subfigure[]{\includegraphics[height=0.25\textheight,width=0.5\linewidth]{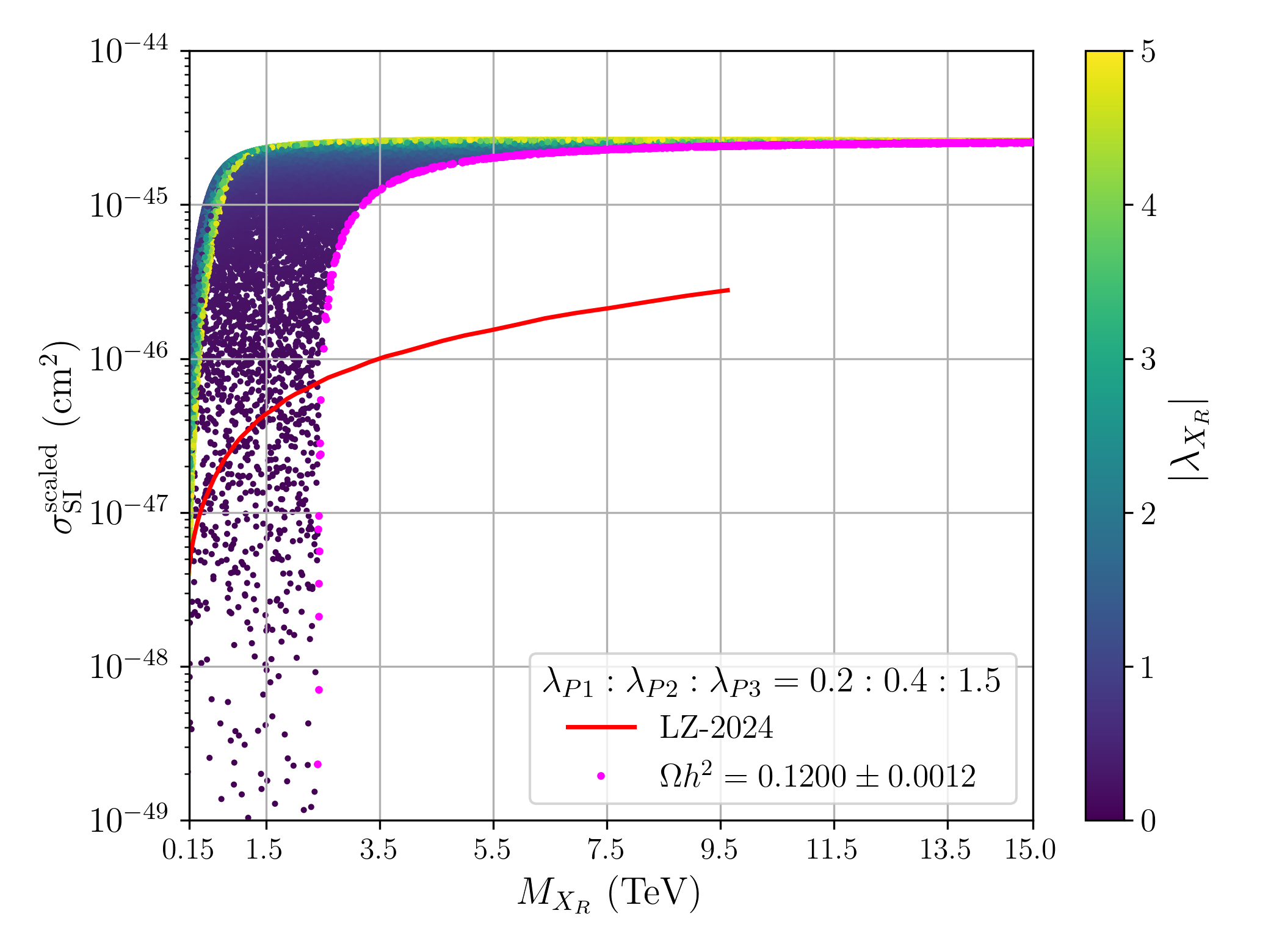}}
			\hspace*{0.25cm}
			\subfigure[]{\includegraphics[height=0.25\textheight,width=0.5\linewidth]{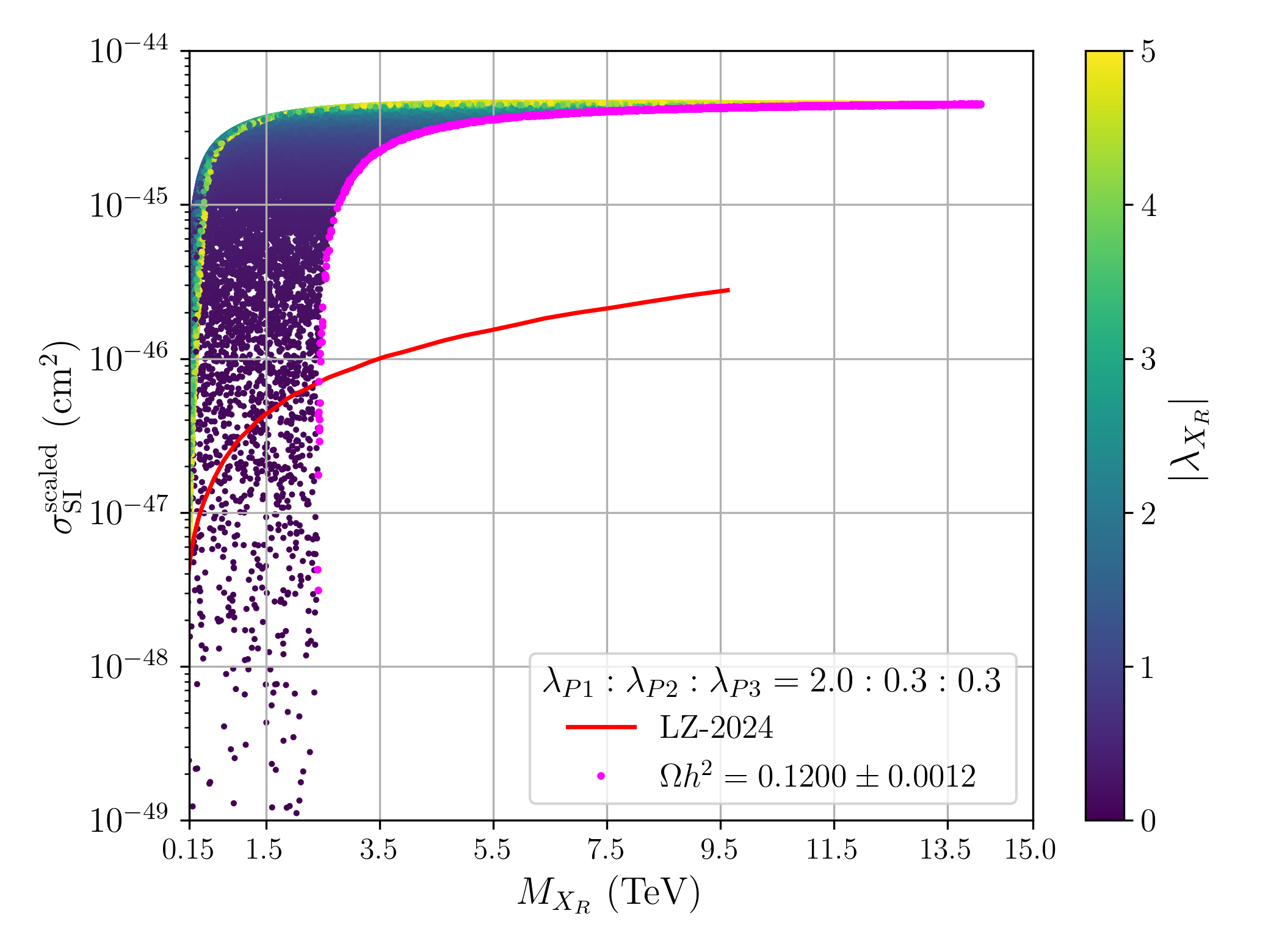}}}
		\mbox{
			\subfigure[]{\includegraphics[height=0.25\textheight,width=0.5\linewidth]{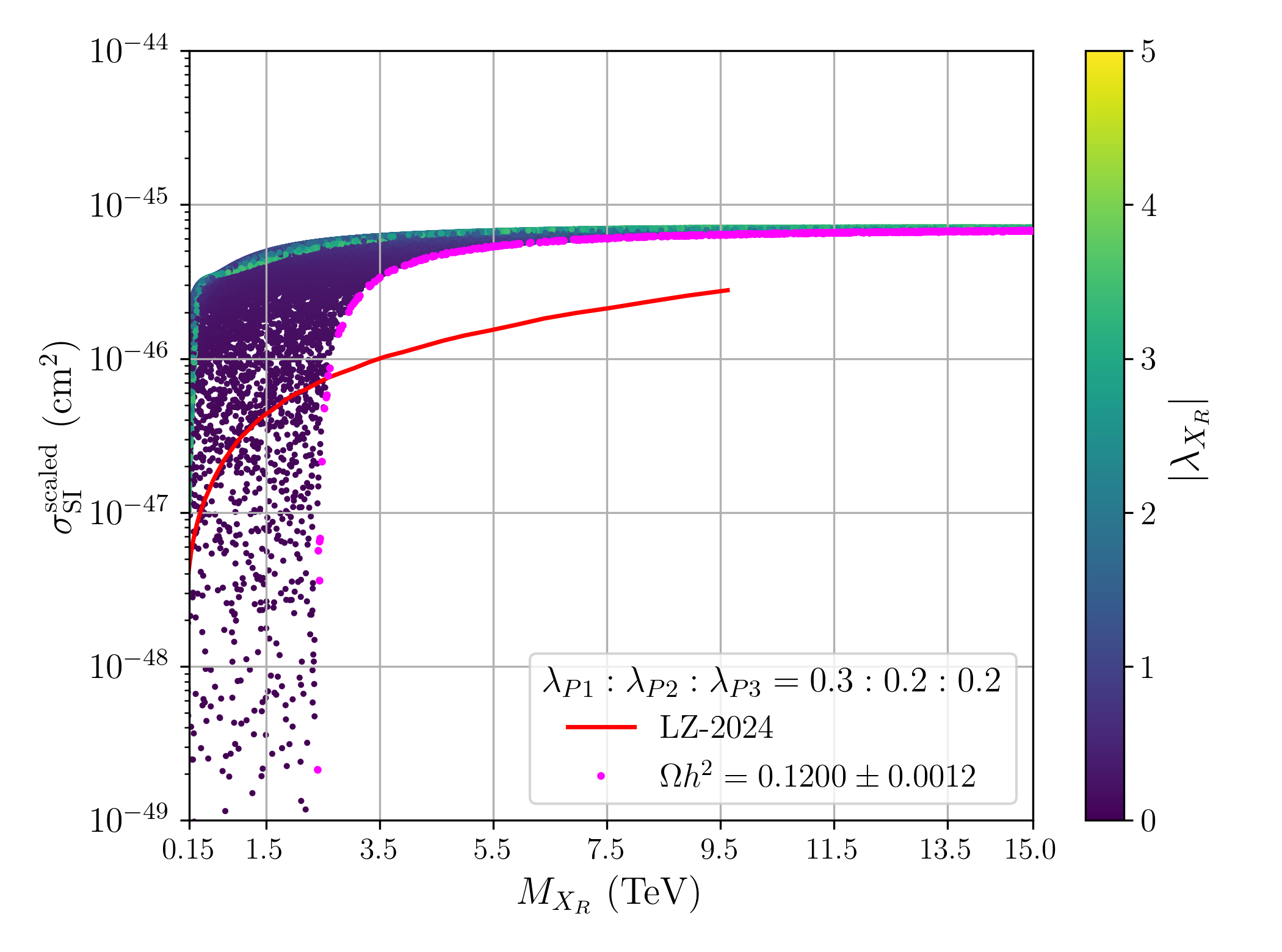}}
			\hspace*{0.25cm}
			\subfigure[]{\includegraphics[height=0.25\textheight,width=0.5\linewidth]{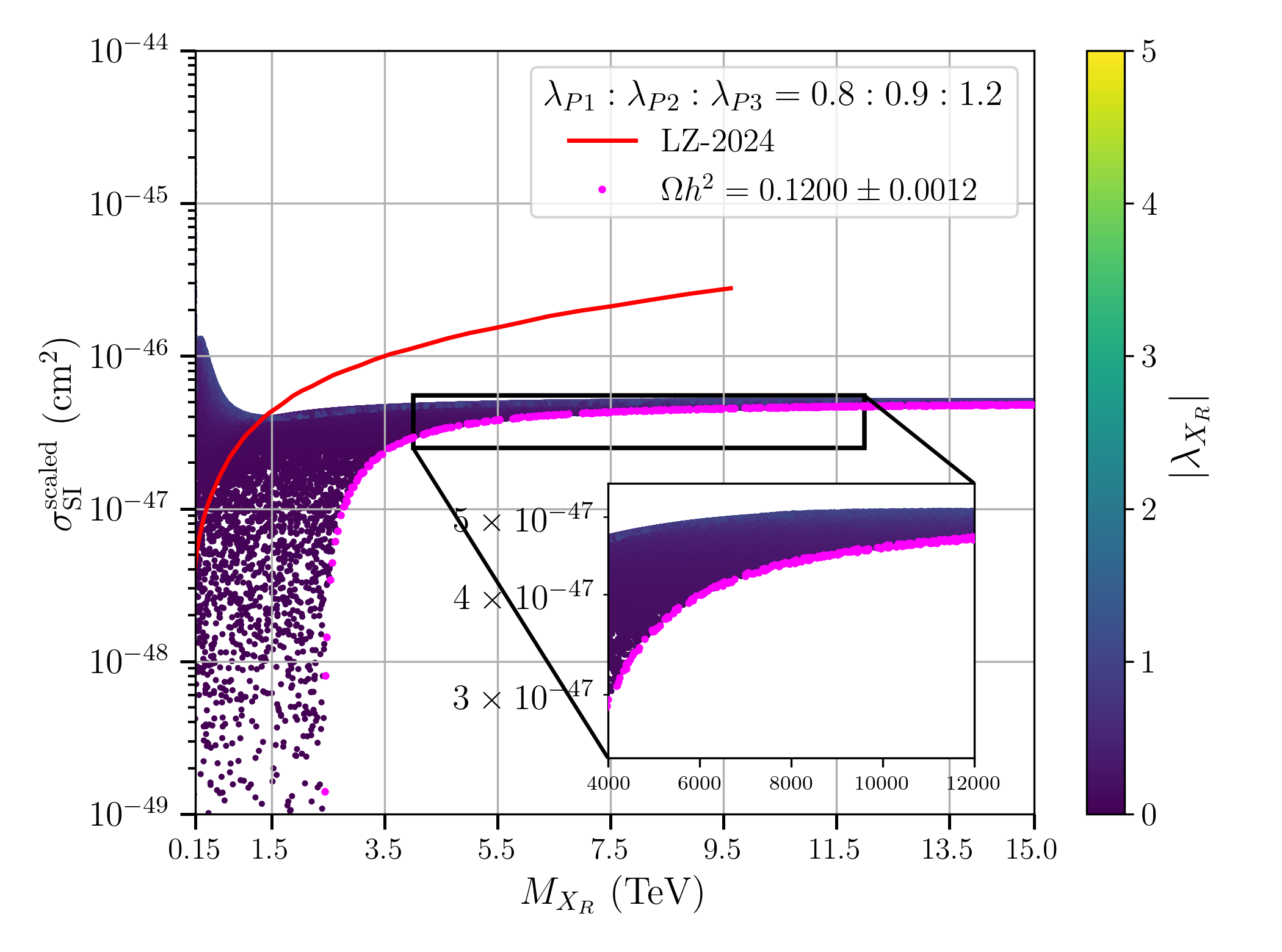}}
		}
	\end{center}
	\caption{Scaled spin-independent cross-section $\sigma_{\mathrm{SI}}^{\mathrm{scaled}}$ verses DM mass $M_{X_R}$ with different values of $\lambda_{X_R}$ for the four ratios of $r_1 : r_2 : r_3=(0.2:.0.4:1.5),\, (2.0:0.3:0.3),\, (0.3:0.2:0.2), (0.8:0.9:1.2) $ described in (a), (b), (c) and (d).}
\label{fig:DM-sigmaSI}
\end{figure}

In \autoref{fig:DM-sigmaSI-fixed-mass} we show the variation of DM relic in (a), unscaled DM-nucleon SI cross-section in (b) and scaled  DM-nucleon SI cross-section of \autoref{sigmanSIscaled} in (c)  verses $\lambda_{X_R}$  for $M_{X_R}=200, \, 1000,\, 2000$ GeV in red, magenta and blue, respectively  for the ratio (0.2 : 0.4 : 1.5). It is evident from \autoref{fig:DM-sigmaSI-fixed-mass}(a) that annihilation and co-annihilation  cross-sections increase as we enhance  $\lambda_{X_R}$ and thus reduce the available DM relic. On the other hand an increment in $\lambda_{X_R}$ proliferates the unscaled  $\sigma_{\text{SI}}$ as can be seen from  \autoref{sigmanSIscaled}(b).  Convoluting these two we get the scaled spin-independent cross-section $\sigma^{\text{scaled}}_{\text{SI}}$ as defined in \autoref{sigmanSIscaled} and  depicted in \autoref{fig:DM-sigmaSI-fixed-mass}(c). It is clear from this figure that $\sigma^{\text{scaled}}_{\text{SI}}$ has a peak at certain values of   $\lambda_{X_R}$ for a given DM mass. For example for $M_{X_R}=200$ GeV it peaks around  $|\lambda_{X_R}|\simeq 0.5$ and after that it drops gradually. For higher mass values such peaks appear for higher $|\lambda_{X_R}|$. For $M_{X_R}=1000$ GeV it appears  around $|\lambda_{X_R}|\simeq 2.5$ and for  $M_{X_R}=2000$ GeV such peaks will appears beyond the plot range around $|\lambda_{X_R}|\simeq 7.0$. This plot is instrumental in understanding the  $\sigma^{\text{scaled}}_{\text{SI}}$  variation with $M_{X_R}$ and $|\lambda_{X_R}|$ in \autoref{fig:DM-sigmaSI}.

\autoref{fig:DM-sigmaSI} shows the correlation between scaled spin-independent scattering cross section per nucleon $\sigma^{\text{scaled}}_{\text{SI}}$ and DM mass $M_{X_R}$ for fixed ratios of the portal couplings $\lambda_{P1}:\lambda_{P2}:\lambda_{P3}$.  Here the DM mass is varied from 150 GeV to 15 TeV and $|\lambda_{X_R}|$ is shown using the colour bar where blue points correspond to lower values and yellow points correspond to the higher values. In all these subplots, we have shown the points that satisfy the observed relic $\relicobs = 0.1200\pm0.0012$ using magenta colour. Also, the red curve shows the bound from the most recent data from the LUX-ZEPLIN (LZ) experiment \cite{LZ:2024zvo} released in 2024. The parameter space above the red curve till $\sim 10$ TeV is ruled out by the LZ experiment for $X_R$ to be a viable DM candidate. The points except the magenta colour correspond to the underabundant points. 

In \autoref{fig:DM-sigmaSI}(a), for the ratio  $\lambda_{P1}:\lambda_{P2}:\lambda_{P3} = 0.2 : 0.4 : 1.5$, we can see that all points with DM masses $M_{X_R} \gsim 2500$ GeV are ruled out as explored by LZ-2024.  Though the LZ bound exists till slightly less than 10 TeV, $M_{X_R} \gsim 2500$ GeV are over abundant regions, thus practically not viable. For $M_{X_R} \lsim 2500$ GeV, the points  are allowed by the PLANCK \cite{Planck:2018vyg} as well as LZ \cite{LZ:2024zvo} are mostly underabundant blue coloured points with smaller  $|\lambda_{X_R}|$.  There are some green coloured points with larger values of $|\lambda_{X_R}|$ which are mostly ruled out by LZ bounds. Similarly, among the points with correct relic as shown in magenta colour very less number are allowed by LZ. The appearance of large $|\lambda_{X_R}|$ with green points  should be in general for large DM-nucleon cross-section, however as we discussed, depending on the mass they can contribute lower  $\sigma^{\text{scaled}}_{\text{SI}}$ as shown in \autoref{fig:DM-sigmaSI-fixed-mass}. The scenario in \autoref{fig:DM-sigmaSI}(b) with the ratio  $(2.0 : 0.3 : 0.3)$, is very similar, where all of the points above $M_{X_R} \gsim 2500$ are ruled out by the LUX-ZEPLIN \cite{LZ:2024zvo}. The observed relic satisfying points that are allowed by LZ has masses $M_{X_R}\lsim 2.5$ TeV and $\lambda_{X_R}\lsim 0.2$.

\autoref{fig:DM-sigmaSI}(c), with $\lambda_{P1}:\lambda_{P2}:\lambda_{P3} = 0.3 : 0.2 : 0.2$, also has situation similar to the previous two cases, where most of the underabundant points allowed by LZ-2024 has $M_{X_R}\lsim 2500$ GeV and smaller $\lambda_{X_R} \lsim 0.2$. with observed DM relic points are with magenta colour. However, in \autoref{fig:DM-sigmaSI}(d), where the fixed ratio is  $\lambda_{P1}:\lambda_{P2}:\lambda_{P3} = 0.8 : 0.9 : 1.2$, we have comparatively different situation. The LZ-2024 \cite{LZ:2024zvo} line is above all the points  with $M_{X_R}>1500$ GeV, and even the observed DM relic points in magenta are also allowed. Here there are no large $\lambda_{X_R}$ points in green colour, as  for$|\lambda_{X_R}| \gsim 1.04$, the individual portal couplings couplings cross perturbativity limit ($\lambda_{P1,P2,P3} \gsim 4 \pi$).  Below we list a few benchmark point to summarize the model and useful to investigate the model further in different experiments.

\subsection{Benchmark points}
Considering the dark matter relic and direct detection bounds  we list a few benchmark points where we also mention their perturbativity limits as present in \autoref{Bps}. Here we focus on the points where the Fixed Point appear at least below the Planck scale ($10^{19}$ GeV). 
	 In the last three rows of this table we also presented if we have a tree-level viable dark matter candidate, the scale at which Fixed Point appears ($\Lambda_{\text{FP}}$) and the scale at which the bounded from below condition of \autoref{BFB} is violated  ($\Lambda_{\text{BFB-break}}$) considering the two-loop RG evolutions for the couplings using the equations given in \autoref{appendix:beta-functions-4plet}. 
	 
	It is interesting to notice that BP1 and BP2  are closer to the correct DM relic to with FPs around GUT scale, i.e. $10^{15, 14}$ GeV, respectively.  The corresponding mass values are of the order of 2.5 and 2.6 TeV, where NLSP and LSP  have compressed mass spectra, which can lead to displaced decays of the charged NLSP, which can lead to displaced charged track or disappearing charged track \cite{SabanciKeceli:2018fsd, Bandyopadhyay:2024plc, Bandyopadhyay:2023joz, Bandyopadhyay:2020otm, Jangid:2020qgo}. 

\begin{table}[H]
	\centering
	\begin{tabular}{lccccc}
		\toprule
		Parameter & BP-1 & BP-2 & BP-3 & BP-4 & BP-5 \\
		\midrule
		$\lambda_{P1}$    & 0.4413   & 0.7712   & 0.1 &4.218& 0.1 \\
		$\lambda_{P2}$    & 0.1638   &-0.2854   & 0.1 &1.931 & 0.07 \\
		$\lambda_{X_R}$   & 0.08845  & 0.1111   & 0.04 & -0.0272& 0.0667 \\
		$M_{X_R}$ [GeV]   & 2540.671 & 2650.000 & 800.00 & 6000.00 & 2000.00 \\
		$M_{X_I}$ [GeV]   & 2545.116 & 2654.092 & 802.01 & 6012.970 & 2000.20 \\
		$M_{X_1^+}$ [GeV] & 2541.296 & 2650.547 & 800.43 &  6000.833 & 2000.26 \\
		$M_{X_2^+}$ [GeV] & 2545.200 & 2654.254 & 802.29 & 6012.846 & 2000.65 \\
		$M_{X^{++}}$ [GeV]& 2543.285 & 2654.173 & 801.42 & 6011.788 & 2000.79 \\
		\midrule
		$\Omega h^2$ & $0.116 $ & $ 0.121 $ & $0.01380 $ & $0.120 $ & $0.08290 $ \\
		$\Omega/\Omega_{\text{obs}}$ & 0.966  & 1.008  & 0.115 & 1.0 & 0.691 \\
		$\sigma^{\text{scaled}}_{\text{SI}} (\times 10^{-46} )$ & $ 0.415 $ & $ 0.6289 $ & $1.272 $ & $0.0146$ & $1.225 $ \\
		Planck & Allowed & Allowed & Allowed & Allowed & Allowed \\
		XENONnT & Allowed & Allowed & Allowed & Allowed & Allowed \\
		\midrule
		DM at tree level & Yes & Yes & Yes & Yes & No \\
		$\Lambda_{\text{FP}} (GeV)$ & $10^{15}$ & $10^{14}$ & $10^{18}$ & $10^{6}$ & $10^{18}$ \\
		$\Lambda_{\text{BFB-break}} (GeV)$ & $>10^{19}$ & $>10^{19}$ & $>10^{19}$ & $10^{18.0}$ & $>10^{19}$ \\
		\bottomrule
	\end{tabular}
	\caption{Bench mark points with the corresponding values of portal couplings ($\lambda_{P1,\,P2,\,P3}$), the mass-spectrum, relic density ($\Omega h^2$), scaled spin-independent nucleon scattering cross-section ($\sigma^{\text{scaled}}_{\text{SI}}$), scale at which FP appears ($\Lambda_{\text{FP}}$), scale at which the tree-level BFB conditions are violated ($\Lambda_{\text{BFB-break}}$)} \label{Bps}
\end{table}
BP3 has relatively lower values of the Higgs portal couplings as well as the lowest mass scale around 800 GeV. This is an underabundant DM relic point where FP appears around $10^{18}$ GeV. BP4 has relatively larger portal couplings $|\lambda_{P1,P2,P3}| >1$, which could be suitable for the first order phase transition \cite{Bandyopadhyay:2025jlg, Espinosa:2011ax}, but one needs to investigate further on this, we leave it to our future work to study this phenomena. This point is also interesting as we have the correct DM relic as well as the FP appearing for a lower scale of $10^6$ GeV.  Finally we come to BP5 with a DM mass around 2 TeV. This BP is interesting as it does not have a viable DM at the tree-level, and one needs to consider one-loop mass correction to achieve the a viable neutral DM. This point gives rise to an underabundant point for DM relic and the FP is achieved around $10^{18}$ GeV.

\section{Discussions and conclusion}\label{conclusion}
In this article we considered an $SU(2)$ quartet with hypercharge $\frac{1}{2}$ with $\mathbb{Z}_2$-odd symmetry and investigated the appearance of the FP points and the interplay of different Higgs portal terms and their possible complex phases. In this respect Higgs portal term $\lambda_{P1}$ and self-coupling term $\lambda_{Q1}$, which are without phases, come out as the complete enhancer of FPs whereas other terms with phases play roles depending on the relative values of enhancers and spoilers of Fixed Points. Interestingly, after the expansion of the potential, we see larger number of terms without phases as compared to the terms with phases enabling a wider region of parameter space with FPs, dissimilar to the cases of ISM/ITM/ITM \cite{Bandyopadhyay:2025ilx}. However,  $\lambda_{P3}$  due to all terms with relative phases behaves like a true spoiler for FP. Nevertheless, the behaviours that are observed here are quite similar to the case of inert 2HDM which also have terms with complex phases that act like a spoiler for FP \cite{Bandyopadhyay:2025ilx}.

After studying the FP behaviour we also look into the one-loop mass spectrum as the tree-level mass spectrum  does not guarantee a neutral scalar as  LSP.  Thus, in spite of being a $\mathbb{Z}_2$ multiplet, it cannot always ensure a viable dark matter candidate at the tree-level. However, since the charged states generally have more quantum corrections as compared to the neutral ones, we  look for mass correction at the quantum level. 
This ensures us with an enhanced parameter space with neutral LSP, i.e. a viable dark matter. Interestingly,  not only the LSP but the NLSP behaviour  also changes depending on the parameter spaces. While the singly charged scalar $X^\pm_1$ is the only possible NLSP for parameter space that has tree-level DM, at one-loop level we see singly charged and doubly charged NLSPs as well as neutral NLSPs. Finally, we performed a dark matter relic calculation, where we studied the behaviour of different portal coupling strengths in enhancing certain Higgs portal annihilation and co-annihilation processes. The parameter space we then compared with the latest LZ  direct detection data \cite{LZ:2024zvo} and a list of benchmark points were given to summarize the scenario with viable dark matter. Unlike the dark matters  coming from the minimal $SU(2)$ multiplet like singlet, doublet and triplet here the allowed mass range can be even above 15 TeV.  One common thing noticed is the compressed spectrum that will make the collider phenomenological study further interesting.  Especially, there are regions where NLSP will only decay to electron or electron/muon and electron/muon/pions, which is very different from ITM \cite{Jangid:2020qgo, Bandyopadhyay:2025jlg}. There is also a possibility to adjust the effective portal coupling such that it is allowed by the LZ experiment, but with relatively larger magnitude of the individual Higgs portal couplings ($\gsim 1$), which can give rise to first order phase transition and gravitational waves. Both of these need separate studies which are on course as future projects.

\acknowledgments

PMPR wants to Chandrima Sen and Disha Ghosh for their help in different stages of the project. PMPR also acknowledges MoE funding for the research work. PB and PMPR acknowledge the Pheonix 2025 for valuable insights.

\appendix
\section{Lagrangian of the Inert Quadruplet Model}\label{Appdx:Lagrangian}
The relevant part of the Lagrangian of our model is given by,
\begin{equation}
	\mathcal{L} = \mathcal{L}_{\text{SM}} + (D^\mu X)^\dagger (D_\mu X) - V_X \label{eq:Lagrangian-of-Model}
\end{equation}
where $\mathcal{L}_{\text{SM}}$ is the Standard Model Lagrangian including the Higgs-potential $V_{\text{Higgs}}$ (\autoref{eq:Higgs-potential}) and the $V_X$ is the scalar potential of the $SU(2)_L$ quartet $X$, which is of the form
\begin{equation}
	V_X = V_{\text{4-plet}} + V_{\text{portal}},
\end{equation}
described by \autoref{eq:4plet-self-potential} and \autoref{eq:4plet-portal-potential}. The covariant derivative acting on the column vector form of the quartet $X$ (\autoref{eq:4plet-definition}) is given by,
\begin{equation}
	D_\mu X = \big(\partial_\mu - i g_2 W^a_\mu T^a - i g_1 Y_X   B_\mu \big)X, \ \ \ \ \ \text{where}\  a = 1,2,3.
\end{equation}
The SU(2) generators $T^a$ in $\mathbf{4}$ representation are given by,
\begin{equation}
	T_1 =
	\begin{pmatrix}
		0 & \frac{\sqrt{3}}{2} & 0 & 0 \\
		\frac{\sqrt{3}}{2} & 0 & 1 & 0 \\
		0 & 1 & 0 & \frac{\sqrt{3}}{2} \\
		0 & 0 & \frac{\sqrt{3}}{2} & 0
	\end{pmatrix}, 
	\ \ \ \ 
	T_2 =
	\begin{pmatrix}
		0 & -\frac{i\sqrt{3}}{2} & 0 & 0 \\
		\frac{i\sqrt{3}}{2} & 0 & -i & 0 \\
		0 & i & 0 & -\frac{i\sqrt{3}}{2} \\
		0 & 0 & \frac{i\sqrt{3}}{2} & 0
	\end{pmatrix},
	\ \ \ \ 
	T_3 =
	\begin{pmatrix}
		\frac{3}{2} & 0 & 0 & 0 \\
		0 & \frac{1}{2} & 0 & 0 \\
		0 & 0 & -\frac{1}{2} & 0 \\
		0 & 0 & 0 & -\frac{3}{2}
	\end{pmatrix}.
\end{equation}

\subsection{Mixing of singly charged scalars in the 4-plet}\label{Appdx:singly-charged-mixing}
Since the 4-plet X is a complex multiplet, singly-charged fields cannot be identified as conjugate $X^+ \ne (X^-)^*$. The mass matrix for them can be obtained from potentials as follows:
\begin{equation}
	\begin{split}
		M^2_{\text{charged}} = \begin{pmatrix}
			-\mu_X^2 + \frac{v_h^2}{6} (3 \lambda_{P1} + \lambda_{P2})  & \frac{\lambda_{P3} v_h^2}{\sqrt{3}} \\
			\frac{\lambda_{P3} v_h^2}{\sqrt{3}} & -\mu_X^2 + \frac{v_h^2}{2} (\lambda_{P1} + \lambda_{P2})
		\end{pmatrix}
	\end{split}
\end{equation}
and hence they are not mass eigenstates. We can go to the singly-charged mass eigenstates $X_1^+$, $X_2^+$ using a orthogonal transformation of the basis given by,
\begin{equation}
	\begin{pmatrix}
		X^+ \\ (X^-)^*
	\end{pmatrix} = \begin{pmatrix}
		\cos\theta_C & \sin\theta_C \\ -\sin\theta_C & \cos\theta_C
	\end{pmatrix}\ \begin{pmatrix}
		X^+_1 \\ X^+_2
	\end{pmatrix}, \label{eq:charge-matrix-rotation}
\end{equation}
where the mixing-angle $\theta_C$ is given by,
\begin{equation}
	\tan\theta_C = \frac{1}{2\sqrt{3}}\,\left(\frac{\sqrt{\lambda_{P2}^2 + 12 \lambda_{P3}^2} - \lambda_{P2}}{\lambda_{P3}}\right). \label{eq:tanthetaC}
\end{equation}
With this rotation described in \autoref{eq:charge-matrix-rotation} and \autoref{eq:tanthetaC}, we can obtain the singly-charged mass eigenstates, $X_1^+ \equiv (X_1^-)^*$, $X_2^+ \equiv (X_1^-)^*$ having masses,
\begin{equation}
	\begin{split}
		M^2_{X_1^+} &=  -\mu_X^2 + \frac{v_h^2}{6} \left(3 \lambda_{P1} + 2 \lambda_{P2} -  \sqrt{\lambda_{P2}^2 + 12 \lambda_{P3}^2}\right), \\
		M^2_{X_2^+} &=  -\mu_X^2 + \frac{v_h^2}{6} \left(3 \lambda_{P1} + 2 \lambda_{P2} +  \sqrt{\lambda_{P2}^2 + 12 \lambda_{P3}^2}\right).
		\label{eq:4plet-singly-charged-masses}
	\end{split} 
\end{equation}

\section{$\beta$-functions the inert quartet model at two-loop level}\label{appendix:beta-functions-4plet}
{\small\begin{align*}
		\beta^{\mathrm{1-loop}}_{g_1} &= \frac{1}{(4\pi)^2}\Big(\frac{43 g_1^3}{10}\Big);\ \ \  \beta^{\mathrm{1-loop}}_{g_2} = \frac{1}{(4\pi)^2}\Big(-\frac{3 g_2^3}{2}\Big);\ \ \ \ \beta^{\mathrm{1-loop}}_{g_3} = \frac{1}{(4\pi)^2}\Big(-7 g_3^3\Big)
\end{align*}}
{\small\begin{align*}
		\beta^{\mathrm{2-loop}}_{g_1} &= \frac{1}{(4\pi)^4}\Big(\frac{217 g_1^5}{50}\,+\,\frac{117 g_1^3 g_2^2}{10}\,+\,\frac{44 g_1^3 g_3^2}{5}\,-\,\frac{17}{10} g_1^3 Y_t^2\Big)\\
		\beta^{\mathrm{2-loop}}_{g_2} &= \frac{1}{(4\pi)^4}\Big(\frac{39 g_1^2 g_2^3}{10}\,+\,\frac{175 g_2^5}{2}\,+\,12 g_2^3 g_3^2\,-\,\frac{3}{2} g_2^3 Y_t^2\Big)\\
		\beta^{\mathrm{2-loop}}_{g_3} &= \frac{1}{(4\pi)^4}\Big(\frac{11 g_1^2 g_3^3}{10}\,+\,\frac{9 g_2^2 g_3^3}{2}\,-\,26 g_3^5\,-\,2 g_3^3 Y_t^2\Big)\\
		\beta^{\mathrm{1-loop}}_{Y_t} &= \frac{1}{(4\pi)^2}\Big(\frac{9 Y_t^3}{2}\,-\,\frac{17 g_1^2 Y_t}{20}\,-\,\frac{9 g_2^2 Y_t}{4}\,-\,8 g_3^2 Y_t\Big)
\end{align*}}

{\small
	\begin{align*}
		&\beta^{\mathrm{2-loop}}_{Y_t} = \frac{1}{(4\pi)^4}\Big(\frac{449 g_1^4 Y_t}{200}\,+\,\frac{19}{15} g_1^2 g_3^2 Y_t\,+\,9 g_2^2 g_3^2 Y_t\,+\,6 \lambda_H^2 Y_t\,+\,\frac{13 \lambda_{P1}^2 Y_t}{8}\,  +\,\frac{3 \lambda_{P1} \lambda_{P2} Y_t}{2}\,+\,\frac{35 \lambda_{P2}^2 Y_t}{36}\,\\ 
		&\ \ \ \ +\,\frac{47 \lambda_{P3}^2 Y_t}{9}\, +\,\frac{393 g_1^2 Y_t^3}{80}\,+\,\frac{225 g_2^2 Y_t^3}{16}\, +\,36 g_3^2 Y_t^3\,-\,\frac{9}{20} g_1^2 g_2^2 Y_t\,-\,\frac{3 g_2^4 Y_t}{4}\,-\,108 g_3^4 Y_t\,-\,12 \lambda_H Y_t^3\, -\,12 Y_t^5\Big)
\end{align*}}

{\small
	\begin{align*}
		\beta^{\mathrm{1-loop}}_{\lambda_H} &= \frac{1}{(4\pi)^2}\Big(\frac{27 g_1^4}{200}\,+\,\frac{9 g_1^2 g_2^2}{20}\,+\,\frac{9 g_2^4}{8}\,+\,24 \lambda_H^2\,+\,\frac{13 \lambda_{P1}^2}{4}\, +\,\frac{7 \lambda_{P1} \lambda_{P2}}{2}\,+\,\frac{53 \lambda_{P2}^2}{36}\,+\,\frac{28 \lambda_{P3}^2}{9}\,\\
		&\ +\,12 \lambda_H Y_t^2\,-\,\frac{9 g_1^2 \lambda_H}{5}\, -\,9 g_2^2 \lambda_H\,-\,6 Y_t^4\Big)
\end{align*}}

{\small
	\begin{align*}
		&\beta^{\mathrm{2-loop}}_{\lambda_H} =\frac{1}{(4\pi)^4}\Big(\frac{165 g_2^6}{16}\,+\,\frac{2019 g_1^4 \lambda_H}{200}\,+\,\frac{117}{20} g_1^2 g_2^2 \lambda_H\,+\,\frac{147 g_2^4 \lambda_H}{8}\,+\,\frac{108 g_1^2 \lambda_H^2}{5}\, +\,108 g_2^2 \lambda_H^2\, +\,32 g_2^2 \lambda_{P3}^2 \sqrt{2}\,
		\\ &\ +\,\frac{63 g_1^4 \lambda_{P1}}{40}  +\,\frac{3}{4} g_1^2 g_2^2 \lambda_{P1}\,+\,\frac{525 g_2^4 \lambda_{P1}}{8}\, +\,\frac{39 g_1^2 \lambda_{P1}^2}{10}\,+\,\frac{369 g_2^2 \lambda_{P1}^2}{4}\,+\,\frac{33 g_1^4 \lambda_{P2}}{40}\,+\,80 g_3^2 \lambda_H Y_t^2\,  +\,\frac{80 g_2^2 \lambda_{P3}^2}{3}\,
		\\ &\  +\,\frac{21}{4} g_1^2 g_2^2 \lambda_{P2}\, +\,\frac{275 g_2^4 \lambda_{P2}}{8}\,+\,\frac{21}{5} g_1^2 \lambda_{P1} \lambda_{P2}\,+\,\frac{201}{2} g_2^2 \lambda_{P1} \lambda_{P2}\,+\,30 Y_t^6\, +\,\frac{53 g_1^2 \lambda_{P2}^2}{30}\, +\,\frac{479 g_2^2 \lambda_{P2}^2}{12}\,  +\,\frac{63}{10} g_1^2 g_2^2 Y_t^2\,
		\\ &+\,\frac{17}{2} g_1^2 \lambda_H Y_t^2\,+\,\frac{45}{2} g_2^2 \lambda_H Y_t^2\, -\,\frac{65 \lambda_H \lambda_{P1}^2}{2}\,  -\,3 \lambda_H Y_t^4-\,\frac{25 \lambda_{P1}^3}{2}\, -\,\frac{8}{5} g_1^2 Y_t^4\, -\,32 g_3^2 Y_t^4\, 
		\\ &\  -\,35 \lambda_H \lambda_{P1} \lambda_{P2}\, -\,\frac{41 \lambda_{P1}^2 \lambda_{P2}}{2}\, -\,\frac{3663 g_1^6}{2000}\, -\,\frac{97 \lambda_H \lambda_{P2}^2}{6}\, -\,20 \lambda_{P1} \lambda_{P2}^2\,  -\,\frac{182 \lambda_{P2}^3}{27}\, -\,\frac{1761}{400} g_1^4 g_2^2\,  -\,144 \lambda_H^2 Y_t^2\,
		\\ &\  -\,\frac{28}{15} g_1^2 \lambda_{P3}^2\, -\,\frac{412 \lambda_H \lambda_{P3}^2}{9}\,  -\,\frac{550 \lambda_{P1} \lambda_{P3}^2}{9}\, -\,\frac{1154 \lambda_{P2} \lambda_{P3}^2}{27}\, -\,\frac{171}{100} g_1^4 Y_t^2\, -\,\frac{9}{4} g_2^4 Y_t^2\, -\,\frac{429}{80} g_1^2 g_2^4\,-\,312 \lambda_H^3 \Big) 
\end{align*}}

{\small
	\begin{align*}
		&\beta^{\mathrm{1-loop}}_{\lambda_{P1}} = \frac{1}{(4\pi)^2}\Big(\frac{27 g_1^4}{100}\,+\,\frac{45 g_2^4}{4}\,+\,12 \lambda_H \lambda_{P1}\,+\,4 \lambda_{P1}^2\,+\,4 \lambda_H \lambda_{P2}\,+\,\frac{\lambda_{P2}^2}{3}\,+\,8 \lambda_{P3}^2\,+\,17 \lambda_{P1} \lambda_{Q1}\,\\
		&\ \ \ +\,7 \lambda_{P2} \lambda_{Q1}\,+\,\frac{31 \lambda_{P1} \lambda_{Q2}}{3}\,+\,\frac{11 \lambda_{P2} \lambda_{Q2}}{9}\,+\,6 \lambda_{P1} Y_t^2\,-\,\frac{27}{10} g_1^2 g_2^2\,-\,\frac{9 g_1^2 \lambda_{P1}}{5}\,-\,27 g_2^2 \lambda_{P1}\Big)
\end{align*}}

{\small
	\begin{align*}
		&\beta^{\mathrm{2-loop}}_{\lambda_{P1}}= \frac{1}{(4\pi)^4}\Big(\frac{2853 g_1^4 g_2^2}{200}\,+\,\frac{1827 g_1^2 g_2^4}{40}\,+\,\frac{27 _1^4 \lambda_H}{10}\,+\,\frac{225 g_2^4 \lambda_H}{2}\,+\,\frac{987 g_1^4 \lambda_{P1}}{100}\,+\,\frac{9}{4} g_1^2 g_2^2 \lambda_{P1}\,+\,\frac{789 g_2^4 \lambda_{P1}}{2}\, 
		\\ &\ \ \ \  +\,\frac{72}{5} g_1^2 \lambda_H \lambda_{P1}\,+\,72 g_2^2 \lambda_H \lambda_{P1}\,+\,\frac{6 g_1^2 \lambda_{P1}^2}{5}\,+\,18 g_2^2 \lambda_{P1}^2\,+\,\frac{51 g_1^4 \lambda_{P2}}{40}\,+\,\frac{1485 g_2^4 \lambda_{P2}}{8}\,+\,\frac{24}{5} g_1^2 \lambda_H \lambda_{P2}\,
		\\ &\ \ \ \  +\,g_2^2 \lambda_{P2}^2\,+\,\frac{48 g_1^2 \lambda_{P3}^2}{5}\,+\,\frac{81 g_1^4 \lambda_{Q1}}{20}\,+\,\frac{675 g_2^4 \lambda_{Q1}}{4}\,+\,\frac{102}{5} g_1^2 \lambda_{P1} \lambda_{Q1}\,+\,471 g_2^2 \lambda_{P1} \lambda_{Q1}\,+\,\frac{42}{5} g_1^2 \lambda_{P2} \lambda_{Q1}\, 
		\\ &\ \ \ \  +\,207 g_2^2 \lambda_{P2} \lambda_{Q1}\,+\,\frac{51 g_1^4 \lambda_{Q2}}{20}\,+\,\frac{425 g_2^4 \lambda_{Q2}}{4}\,+\,\frac{62}{5} g_1^2 \lambda_{P1} \lambda_{Q2}\,+\,287 g_2^2 \lambda_{P1} \lambda_{Q2}\,+\,\frac{22}{15} g_1^2 \lambda_{P2} \lambda_{Q2}\,
		\\ &\ \ \ \  +\,\frac{3 g_1^2 g_2^2 \lambda_{P2}}{5 \sqrt{2}}\,+\,36 g_2^2 \lambda_H \lambda_{P2}\, +\,\frac{17}{4} g_1^2 \lambda_{P1} Y_t^2\,+\,\frac{45}{4} g_2^2 \lambda_{P1} Y_t^2\,+\,\frac{161}{3} g_2^2 \lambda_{P2} \lambda_{Q2}\, +\,40 g_3^2 \lambda_{P1} Y_t^2\, 
		\\ &\ \ \ \  -\,\frac{53 \lambda_{P1}^3}{4}\,-\,\frac{84}{5} g_1^2 g_2^2 \lambda_{P2}\,-\,16 \lambda_H^2 \lambda_{P2}\,-\,32 \lambda_H \lambda_{P1} \lambda_{P2}\,-\,\frac{11 \lambda_{P1}^2 \lambda_{P2}}{2}\,-\,\frac{1}{5} g_1^2 \lambda_{P2}^2\,-\,\frac{34 \lambda_H \lambda_{P2}^2}{3}\, 
		\\ &\ \ \ \  -\,\frac{175 \lambda_{P1} \lambda_{P2}^2}{36}\,-\,\frac{41 \lambda_{P2}^3}{18}\,-\,\frac{304 \lambda_H \lambda_{P3}^2}{3}\,-\,\frac{586 \lambda_{P1} \lambda_{P3}^2}{9}\,-\,\frac{644 \lambda_{P2} \lambda_{P3}^2}{9}\,-\,\frac{15}{2} g_1^2 g_2^2 \lambda_{Q1}\,-\,101 \lambda_{P1}^2 \lambda_{Q1}\, 
		\\ &\ \ \ \  -\,\frac{166}{3} \lambda_{P1} \lambda_{P2} \lambda_{Q1}\,-\,18 \lambda_{P2}^2 \lambda_{Q1}\,-\,\frac{1132 \lambda_{P3}^2 \lambda_{Q1}}{9}\,-\,83 \lambda_{P1} \lambda_{Q1}^2\,-\,\frac{82 \lambda_{P2} \lambda_{Q1}^2}{3}\,-\,\frac{37}{2} g_1^2 g_2^2 \lambda_{Q2}\,-\,61 \lambda_{P1}^2 \lambda_{Q2}\, 
		\\ &\ \ \ \  -\,\frac{82}{9} \lambda_{P1} \lambda_{P2} \lambda_{Q2}\,-\,\frac{158 \lambda_{P2}^2 \lambda_{Q2}}{27}\,-\,\frac{412 \lambda_{P3}^2 \lambda_{Q2}}{9}\,-\,\frac{298}{3} \lambda_{P1} \lambda_{Q1} \lambda_{Q2}\,-\,\frac{76}{9} \lambda_{P2} \lambda_{Q1} \lambda_{Q2}\,-\,47 \lambda_{P1} \lambda_{Q2}^2\, 
		\\ &\ \ \ \  -\,\frac{88 \lambda_{P2} \lambda_{Q2}^2}{27}\,-\,6 g_2^2 \lambda_{P1} \lambda_{P2} \sqrt{2}\, -\,\frac{189}{10} g_1^2 g_2^2 Y_t^2\,-\,\frac{45}{4} g_2^4 Y_t^2\,-\,72 \lambda_H \lambda_{P1} Y_t^2\,\,-\,12 \lambda_{P1}^2 Y_t^2\,-\,24 \lambda_H \lambda_{P2} Y_t^2 
		\\ &\ \ \ \  -\,\frac{171}{100} g_1^4 Y_t^2\,-\,\frac{3663 g_1^6}{1000}\,-\,\frac{525 g_2^6}{8}\,-\,9 g_1^2 g_2^2 \lambda_H\,-\,60 \lambda_H^2 \lambda_{P1}\,-\,72 \lambda_H \lambda_{P1}^2\,-\,\lambda_{P2}^2 Y_t^2\,-\,24 \lambda_{P3}^2 Y_t^2\,-\,\frac{27 \lambda_{P1} Y_t^4}{2}\Big) 
\end{align*}}

{\small\begin{align*}
		&\beta^{\mathrm{1-loop}}_{\lambda_{P2}} = \frac{1}{(4\pi)^2}\Big(\frac{27 g_1^2 g_2^2}{5}+4 \lambda_H \lambda_{P2}+8 \lambda_{P1} \lambda_{P2}+\frac{13 \lambda_{P2}^2}{3}+\frac{16 \lambda_{P3}^2}{3}+4 \lambda_{P2} \lambda_{Q1}+2 \lambda_{P1} \lambda_{Q2}+\frac{86 \lambda_{P2} \lambda_{Q2}}{9}+\\ &\ \ \ 6 \lambda_{P2} Y_t^2-\frac{9 g_1^2 \lambda_{P2}}{5}-27 g_2^2 \lambda_{P2}\Big)
\end{align*}}

{\small
	\begin{align*}
		&\beta^{\mathrm{2-loop}}_{\lambda_{P2}} = \frac{1}{(4\pi)^4}\Big(18 g_1^2 g_2^2 \lambda_H\,+\,\frac{171}{20} g_1^2 g_2^2 \lambda_{P1}\,+\,\frac{1479 g_1^4 \lambda_{P2}}{200}\,+\,\frac{213}{5} g_1^2 g_2^2 \lambda_{P2}\,+\,\frac{279 g_2^4 \lambda_{P2}}{8}\,+\,\frac{24}{5} g_1^2 \lambda_H \lambda_{P2}\,\\
		&\ \ \ +\,\frac{12}{5} g_1^2 \lambda_{P1} \lambda_{P2}\,+\,48 g_2^2 \lambda_{P1} \lambda_{P2}\,+\,\frac{11 g_1^2 \lambda_{P2}^2}{5}\,+\,31 g_2^2 \lambda_{P2}^2\,+\,\frac{32 g_1^2 \lambda_{P3}^2}{5}\,+\,88 g_2^2 \lambda_{P3}^2\,+\,6 g_2^2 \lambda_{P1} \lambda_{P2} \sqrt{2}\,\\
		&\ \ \ +\,\frac{286 \lambda_{P2} \lambda_{P3}^2}{9}\,+\,18 g_1^2 g_2^2 \lambda_{Q1}\,+\,18 g_2^2 \lambda_{P1} \lambda_{Q1}\,+\,\frac{24}{5} g_1^2 \lambda_{P2} \lambda_{Q1}\,+\,102 g_2^2 \lambda_{P2} \lambda_{Q1}\,+\,\frac{3 g_1^4 \lambda_{Q2}}{10}\,+\,48 g_2^2 \lambda_{P3}^2 \sqrt{2}\,\\
		&\ \ \ +\,39 g_1^2 g_2^2 \lambda_{Q2}\,+\,\frac{25 g_2^4 \lambda_{Q2}}{2}\,+\,\frac{12}{5} g_1^2 \lambda_{P1} \lambda_{Q2}\,+\,66 g_2^2 \lambda_{P1} \lambda_{Q2}\,+\,\frac{172}{15} g_1^2 \lambda_{P2} \lambda_{Q2}\,+\,\frac{710}{3} g_2^2 \lambda_{P2} \lambda_{Q2}\,\\
		&\ \ \ +\,\frac{189}{5} g_1^2 g_2^2 Y_t^2\,+\,\frac{17}{4} g_1^2 \lambda_{P2} Y_t^2\,+\,\frac{45}{4} g_2^2 \lambda_{P2} Y_t^2\,+\,40 g_3^2 \lambda_{P2} Y_t^2\,-\,24 \lambda_{P1} \lambda_{P2} Y_t^2\,-\,13 \lambda_{P2}^2 Y_t^2\,-\,16 \lambda_{P3}^2 Y_t^2\,\\
		&\ \ \ -\,\frac{1017}{25} g_1^4 g_2^2\,-\,\frac{1251}{10} g_1^2 g_2^4\,-\,28 \lambda_H^2 \lambda_{P2}\,-\,80 \lambda_H \lambda_{P1} \lambda_{P2}\,-\,\frac{117 \lambda_{P1}^2 \lambda_{P2}}{4}\,-\,\frac{130 \lambda_H \lambda_{P2}^2}{3}\,-\,\frac{27 \lambda_{P2} Y_t^4}{2}\\
		&\ \ \ -\,\frac{193 \lambda_{P1} \lambda_{P2}^2}{6}\,-\,\frac{353 \lambda_{P2}^3}{36}\,-\,\frac{64 \lambda_H \lambda_{P3}^2}{3}\,-\,24 \lambda_{P1} \lambda_{P3}^2\,-\,100 \lambda_{P1} \lambda_{P2} \lambda_{Q1}\,-\,\frac{164 \lambda_{P2}^2 \lambda_{Q1}}{3}\,\\
		&\ \ \ -\,\frac{64 \lambda_{P3}^2 \lambda_{Q1}}{3}\,-\,33 \lambda_{P2} \lambda_{Q1}^2\,-\,\frac{38 \lambda_{P1}^2 \lambda_{Q2}}{3}\,-\,\frac{1136}{9} \lambda_{P1} \lambda_{P2} \lambda_{Q2}\,-\,\frac{1790 \lambda_{P2}^2 \lambda_{Q2}}{27}\,-\,\frac{280 \lambda_{P3}^2 \lambda_{Q2}}{9}\,\\
		&\ \ \ -\,\frac{68}{3} \lambda_{P1} \lambda_{Q1} \lambda_{Q2}\,-\,102 \lambda_{P2} \lambda_{Q1} \lambda_{Q2}\,-\,\frac{448 \lambda_{P1} \lambda_{Q2}^2}{27}\,-\,\frac{1429 \lambda_{P2} \lambda_{Q2}^2}{27}\,-\,\frac{3 g_1^2 g_2^2 \lambda_{P2}}{5 \sqrt{2}}\,-\,24 \lambda_H \lambda_{P2} Y_t^2\Big)
\end{align*}}

{\small
	\begin{align*}
		\beta^{\mathrm{1-loop}}_{\lambda_{P3}} = \frac{1}{(4\pi)^2}\Big(4 \lambda_H \lambda_{P3}\,+\,8 \lambda_{P1} \lambda_{P3}\,+\,6 \lambda_{P2} \lambda_{P3}\,+\,4 \lambda_{P3} \lambda_{Q1}\,+\,6 \lambda_{P3} Y_t^2\,-\,\frac{9 g_1^2 \lambda_{P3}}{5}\,-\,27 g_2^2 \lambda_{P3}\,-\,\frac{2 \lambda_{P3} \lambda_{Q2}}{9}\Big)
\end{align*}}

{\small
	\begin{align*}
		&\beta^{\mathrm{2-loop}}_{\lambda_{P3}} = \frac{1}{(4\pi)^4}\Big(\frac{1479 g_1^4 \lambda_{P3}}{200}\,+\,\frac{69}{20} g_1^2 g_2^2 \lambda_{P3}\,+\,\frac{279 g_2^4 \lambda_{P3}}{8}\,+\,\frac{48}{5} g_1^2 \lambda_{P1} \lambda_{P3}\,+\,48 g_2^2 \lambda_{P1} \lambda_{P3}\,+\,\frac{36}{5} g_1^2 \lambda_{P2} \lambda_{P3}\,
		\\&\ \ \ +\,60 g_2^2 \lambda_{P2} \lambda_{P3}\,+\,\frac{148 \lambda_{P3}^3}{9}\,+\,84 g_2^2 \lambda_{P3} \lambda_{Q1}\,+\,\frac{2}{15} g_1^2 \lambda_{P3} \lambda_{Q2}\,+\,\frac{2}{3} g_2^2 \lambda_{P3} \lambda_{Q2}\,+\,\frac{761 \lambda_{P3} \lambda_{Q2}^2}{81}\,+\,\frac{45}{4} g_2^2 \lambda_{P3} Y_t^2\,
		\\&\ \ \ +\,\frac{29 g_2^2 \lambda_{P2} \lambda_{P3}}{\sqrt{2}}\,+\,\frac{3}{8} g_1^2 g_2^2 \lambda_{P3} \sqrt{2}\,+\,6 g_2^2 \lambda_{P1} \lambda_{P3} \sqrt{2}\,+\,6 g_2^2 \lambda_{P3} \lambda_{Q1} \sqrt{2}\,+\,\frac{513 g_1^4 Y_t^2}{400}\,+\,\frac{135 g_2^4 Y_t^2}{16}\,+\,\frac{17}{4} g_1^2 \lambda_{P3} Y_t^2\,
		\\&\ \ \ +\,40 g_3^2 \lambda_{P3} Y_t^2\,-\,\frac{12}{5} g_1^2 \lambda_H \lambda_{P3}\,-\,28 \lambda_H^2 \lambda_{P3}\,-\,80 \lambda_H \lambda_{P1} \lambda_{P3}\,-\,\frac{12}{5} g_1^2 \lambda_{P3} \lambda_{Q1}\,-\,100 \lambda_{P1} \lambda_{P3} \lambda_{Q1}\,-\,\frac{9 g_1^2 g_2^2 \lambda_{P3}}{20 \sqrt{2}}\,
		\\&\ \ \ -\,\frac{117 \lambda_{P1}^2 \lambda_{P3}}{4}\,-\,52 \lambda_H \lambda_{P2} \lambda_{P3}\,-\,\frac{487}{12} \lambda_{P1} \lambda_{P2} \lambda_{P3}\,-\,\frac{577 \lambda_{P2}^2 \lambda_{P3}}{72}\,-\,\frac{64}{3} \lambda_{P3} \lambda_{Q1} \lambda_{Q2}\,-\,24 \lambda_{P1} \lambda_{P3} Y_t^2\,
		\\&\ \ \ -\,\frac{182}{3} \lambda_{P2} \lambda_{P3} \lambda_{Q1}\,-\,33 \lambda_{P3} \lambda_{Q1}^2\,-\,\frac{394}{9} \lambda_{P1} \lambda_{P3} \lambda_{Q2}\,-\,\frac{253}{9} \lambda_{P2} \lambda_{P3} \lambda_{Q2}\,-\,18 \lambda_{P2} \lambda_{P3} Y_t^2\,-\,\frac{15 \lambda_{P3} Y_t^4}{2}
		\\&\ \ \ -\,\frac{11}{2} g_2^2 \lambda_{P2} \lambda_{P3} \sqrt{2}\,-\,6 g_2^2 \lambda_{P3} \lambda_{Q2} \sqrt{2}\,-\,\frac{189}{40} g_1^2 g_2^2 Y_t^2\,-\,24 \lambda_H \lambda_{P3} Y_t^2\,\Big)
\end{align*}}

{\small
	\begin{align*}
		&\beta^{\mathrm{1-loop}}_{\lambda_{Q1}} = \frac{1}{(4\pi)^2}\Big(\frac{27 g_1^4}{200}\,+\,\frac{297 g_2^4}{8}\,+\,2 \lambda_{P1}^2\,+\,2 \lambda_{P1} \lambda_{P2}\,+\,4 \lambda_{P3}^2\,+\,29 \lambda_{Q1}^2\,+\,\frac{68 \lambda_{Q1} \lambda_{Q2}}{3}\,+\,\frac{17 \lambda_{Q2}^2}{9}\,-\,\frac{81}{20} g_1^2 g_2^2\,\\ & \ \ \ -\,\frac{9 g_1^2 \lambda_{Q1}}{5}\,-\,45 g_2^2 \lambda_{Q1}\Big)
\end{align*}}

{\small
	\begin{align*}
		&\beta^{\mathrm{2-loop}}_{\lambda_{Q1}} = \frac{1}{(4\pi)^4}\Big(\frac{10179 g_1^4 g_2^2}{400}\,+\,\frac{7911 g_1^2 g_2^4}{80}\,+\,\frac{9 g_1^4 \lambda_{P1}}{10}\,+\,\frac{75 g_2^4 \lambda_{P1}}{2}\,+\,\frac{12 g_1^2 \lambda_{P1}^2}{5}\,+\,12 g_2^2 \lambda_{P1}^2\,+\,\frac{9 g_1^4 \lambda_{P2}}{20}\,+\,\frac{136}{5} g_1^2 \lambda_{Q1} \lambda_{Q2}\,
		\\& \ \ \ +\,\frac{1269 g_2^4 \lambda_{Q2}}{2}\,+\,\frac{138 g_1^2 \lambda_{Q1}^2}{5}\,+\,651 g_2^2 \lambda_{Q1}^2\,+\,\frac{8493 g_2^4 \lambda_{Q1}}{8}\,+\,\frac{75 g_2^4 \lambda_{P2}}{4}\,+\,\frac{12}{5} g_1^2 \lambda_{P1} \lambda_{P2}\,+\,12 g_2^2 \lambda_{P1} \lambda_{P2}\,+\,3 g_2^2 \lambda_{P2}^2\,
		\\& \ \ \ +\,\frac{2289 g_1^4 \lambda_{Q1}}{200}\,+\,\frac{27 g_1^4 \lambda_{Q2}}{10}\,+\,658 g_2^2 \lambda_{Q1} \lambda_{Q2}\,+\,\frac{28 g_1^2 \lambda_{Q2}^2}{15}\,+\,77 g_2^2 \lambda_{Q2}^2\,+\,\frac{3 g_1^2 g_2^2 \lambda_{Q2}}{5 \sqrt{2}}\,-\,\frac{3663 g_1^6}{2000}\,-\,\frac{7 \lambda_{P2}^2 \lambda_{Q1}}{3}\,-\,398 \lambda_{Q1}^3\,
		\\& \ \ \ -\,8 \lambda_{P1}^3\,-\,\frac{9}{2} g_1^2 g_2^2 \lambda_{P2}\,-\,12 \lambda_{P1}^2 \lambda_{P2}\,-\,5 \lambda_{P1} \lambda_{P2}^2\,-\,\frac{\lambda_{P2}^3}{2}\,-\,\frac{\lambda_{P2}^2 \lambda_{Q2}}{2}\,-\,\frac{116 \lambda_{P3}^2 \lambda_{Q2}}{3}\,-\,\frac{1161}{20} g_1^2 g_2^2 \lambda_{Q2}\,-\,\frac{16263 g_2^6}{16}\,
		\\& \ \ \ -\,\frac{12}{5} g_1^2 \lambda_{P3}^2\,-\,56 \lambda_{P1} \lambda_{P3}^2\,-\,32 \lambda_{P2} \lambda_{P3}^2\,-\,\frac{108}{5} g_1^2 g_2^2 \lambda_{Q1}\,-\,20 \lambda_{P1}^2 \lambda_{Q1}\,-\,20 \lambda_{P1} \lambda_{P2} \lambda_{Q1}\,-\,\frac{128 \lambda_{P3}^2 \lambda_{Q1}}{3}\,
		\\& \ \ \ -\,\frac{1472 \lambda_{Q1}^2 \lambda_{Q2}}{3}\,-\,\frac{4114 \lambda_{Q1} \lambda_{Q2}^2}{27}\,-\,\frac{416 \lambda_{Q2}^3}{27}\,-\,12 g_2^2 \lambda_{Q1} \lambda_{Q2} \sqrt{2}\,-\,12 \lambda_{P1}^2 Y_t^2\,-\,12 \lambda_{P1} \lambda_{P2} Y_t^2\,-\,24 \lambda_{P3}^2 Y_t^2\Big)
\end{align*}}

{\small
	\begin{align*}
		&\beta^{\mathrm{1-loop}}_{\lambda_{Q2}} = \frac{1}{(4\pi)^2}\Big(\frac{81 g_1^2 g_2^2}{10}\,+\,\lambda_{P2}^2\,+\,26 \lambda_{Q1} \lambda_{Q2}\,+\,\frac{200 \lambda_{Q2}^2}{9}\,-\,4 \lambda_{P3}^2\,-\,\frac{9 g_1^2 \lambda_{Q2}}{5}\,
		-\,45 g_2^2 \lambda_{Q2}\Big)
\end{align*}}

{\small
	\begin{align*}
		&\beta^{\mathrm{2-loop}}_{\lambda_{Q2}} = \frac{1}{(4\pi)^4}\Big(9 g_1^2 g_2^2 \lambda_{P2}\,+\,\frac{6 g_1^2 \lambda_{P2}^2}{5}\,+\,\frac{12 g_1^2 \lambda_{P3}^2}{5}\,+\,\frac{104 \lambda_{P1} \lambda_{P3}^2}{3}\,+\,\frac{56 \lambda_{P2} \lambda_{P3}^2}{3}\,+\,\frac{279}{4} g_1^2 g_2^2 \lambda_{Q1}\,\\ &\ \ \ +\,\frac{104 \lambda_{P3}^2 \lambda_{Q1}}{3}\,+\,18 g_2^2 \lambda_{Q1}^2\,+\,\frac{1509 g_1^4 \lambda_{Q2}}{200}\,+\,\frac{681}{5} g_1^2 g_2^2 \lambda_{Q2}\,+\,\frac{1065 g_2^4 \lambda_{Q2}}{8}\,+\,\frac{436 \lambda_{P3}^2 \lambda_{Q2}}{9}\,\\ &\ \ \ +\,\frac{84}{5} g_1^2 \lambda_{Q1} \lambda_{Q2}\,+\,408 g_2^2 \lambda_{Q1} \lambda_{Q2}\,+\,\frac{298 g_1^2 \lambda_{Q2}^2}{15}\,+\,432 g_2^2 \lambda_{Q2}^2\,+\,12 g_2^2 \lambda_{Q1} \lambda_{Q2} \sqrt{2}\,+\,24 \lambda_{P3}^2 Y_t^2\,\\
		&\ \ \ -\,\frac{3051}{50} g_1^4 g_2^2\,-\,\frac{6183}{20} g_1^2 g_2^4\,-\,\frac{25 \lambda_{P1} \lambda_{P2}^2}{3}\,-\,\frac{25 \lambda_{P2}^3}{6}\,-\,\frac{25 \lambda_{P2}^2 \lambda_{Q1}}{3}\,-\,20 \lambda_{P1}^2 \lambda_{Q2}\,-\,\frac{3 g_1^2 g_2^2 \lambda_{Q2}}{5 \sqrt{2}} \\
		&\ \ \ -\,20 \lambda_{P1} \lambda_{P2} \lambda_{Q2}\,-\,\frac{61 \lambda_{P2}^2 \lambda_{Q2}}{6}\,-\,506 \lambda_{Q1}^2 \lambda_{Q2}\,-\,\frac{21580 \lambda_{Q1} \lambda_{Q2}^2}{27}\,-\,\frac{24494 \lambda_{Q2}^3}{81}\,-\,6 \lambda_{P2}^2 Y_t^2\Big)
\end{align*}}

\section{Residual phases in the expansion of the I4M potential} \label{ResPhase}
The expansion of the portal terms with coefficient $\lambda_{P2}$ and $\lambda_{P3}$ showing the residual phases are given below.
{\small\begin{equation*}
		\begin{split}
			\lambda_{P2} &\sum_{}^{} \Phi_l \  X^*_{lij}\ X_{ijk}\ \Phi^*_{k} = \lambda_{P2}\,\Big(\frac{2}{3} |{\phi^{0}}|^2 | {X^{0}}|^2\, +\, | {\phi^{0}}|^2 | {X^{-}}|^2 \, +\,\frac{1}{3}| {\phi^{+}}|^2 | {X^{0}}|^2 \\  
			&\ \ \ +\frac{2}{3} | {\phi^{+}}|^2 | {X^{+}}|^2 \, +\,|{\phi^{+}}|^2 | X^{++}|^2 \, +\,\frac{1}{3}| {\phi^{0}}|^2 | {X^{+}}|^2 +\frac{2}{\sqrt{3}}  | {\phi^{0}}|  | {\phi^{+}}|  | {X^{0}}|  | {X^{-}}|  \cos ({\theta^{0}}-{\theta^{+}}+{\tilde{\theta}^{0}}-{\tilde{\theta}^{-}}) \\  
			&\ \ \ +\frac{4}{3} | {\phi^{0}}|  | {\phi^{+}}|  | {X^{0}}|  | {X^{+}}|  \cos  ({\theta^{0}}-{\theta^{+}}-{\tilde{\theta}^{0}}+{\tilde{\theta}^{+}}) + \frac{2}{\sqrt{3}} | {\phi^{0}}|  | {\phi^{+}}|  | {X^{+}}|  | {X^{++}}|  \cos  ({\theta^{0}}-{\theta^{+}}-{\tilde{\theta}^{+}}+{\tilde{\theta}^{++}})\Big)
		\end{split}
\end{equation*}}

\begin{equation*}
	\begin{split}
		\Big(\lambda_{P3} &\sum_{}^{} X_{imn}\, X_{jpq}\, \Phi^*_{i}\, \Phi^*_{j}\, \epsilon_{mp}\, \epsilon_{nq}  + \text{h.c.}\Big)  =\lambda_{P3}\,\Big(\frac{4}{\sqrt{3}}  | {\phi^{+}}|^2 | {X^{0}}|  | {X^{++}}|  \cos(2 {\theta^{+}}-{\tilde{\theta}^{0}}-{\tilde{\theta}^{++}})\\  
		&\ \ \  -\frac{4}{3}  |{\phi^{0}}|  |{\phi^{+}}|  |{X^{0}}|  | {X^{+}}|  \cos  ({\theta^{0}}+{\theta^{+}}-{\tilde{\theta}^{0}}-{\tilde{\theta}^{+}})  +4  | {\phi^{0}}|  | {\phi^{+}}|  | {X^{-}}|  | {X^{++}}|  \cos  ({\theta^{0}}+{\theta^{+}}-{\tilde{\theta}^{-}}-{\tilde{\theta}^{++}})\\  
		&\ \ \ -\frac{4}{3}  | {\phi^{0}}|^2 | {X^{0}}|^2 \cos(2 {\theta^{0}}-2 {\tilde{\theta}^{0}}) -\frac{4}{3}  | {\phi^{+}}|^2 | {X^{+}}|^2 \cos(2 {\theta^{+}}-2 {\tilde{\theta}^{+}})  +\frac{4}{\sqrt{3}}  | {\phi^{0}}|^2 | {X^{-}}|  | {X^{+}}|  \cos(2 {\theta^{0}}-{\tilde{\theta}^{-}}-{\tilde{\theta}^{+}})\Big)
	\end{split}
\end{equation*}

\section{Sub-dominant contributions to the relic density $\Omega h^2$}\label{Feyn:DManni}

\begin{figure}[H]
	\begin{tikzpicture}[
		baseline=(current bounding box.center),
		scale=0.8,
		transform shape,
		fermion/.style={dashed, thick},
		boson/.style={decorate, decoration={snake}, thick},
		scalar/.style={dashed, thick}
		]
		\begin{scope}
			\draw[scalar] (0,1) -- (-1.2,1)  node[left] {$X_{1,2}^-$};
			\draw[scalar] (0,-1) -- (-1.2,-1) node[left] {$X^{++}$};
			\draw[scalar] (0,1) -- (0,-1) node[midway, right] {$X_1^+, X_2^+$};
			
			\draw[boson] (0,-1) -- (2,-1) node[right] {$W^+$};
			\draw[scalar] (0,1) -- (2,1) node[right] {$h$};
		\end{scope}
		\hspace*{2.5cm}	
		\begin{scope}[xshift=4cm]
			\draw[scalar] (0,0) -- (-0.6,1)  node[left] {$X_{1,2}^-$} ;
			\draw[scalar] (0,0) --(-0.6,-1)  node[left] {$X^{++}$};
			\draw[boson] (0,0) -- (1.5,0) node[midway, above] {$W^+$};
			\draw[scalar] (1.5,0) -- (2.2,0.9) node[right] {$h$};
			\draw[boson] (1.5,0) -- (2.2,-0.9) node[right]  {$W^+$};
		\end{scope}
		\hspace*{2.5cm}	
		\begin{scope}[xshift=8cm]
			\draw[scalar] (0,1) -- (-1.2,1)  node[left] {$X_{1,2}^-$};
			\draw[scalar] (0,-1) -- (-1.2,-1) node[left] {$X^{++}$};
			\draw[scalar] (0,1) -- (0,-1) node[midway, right] {$X^{++}$};
			
			\draw[boson] (0,1) -- (2,1) node[right] {$W^+$};
			\draw[scalar] (0,-1) -- (2,-1) node[right] {$h$};
		\end{scope}
	\end{tikzpicture}
	\caption{Contributions to $\Omega^{-1}$ from the modes $X_{1,2}^- \, X_{++}  \rightarrow h W^+$}
\end{figure}
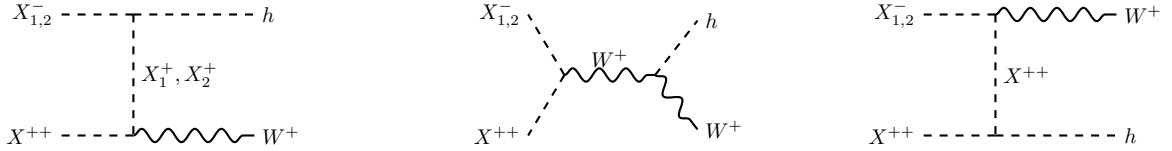
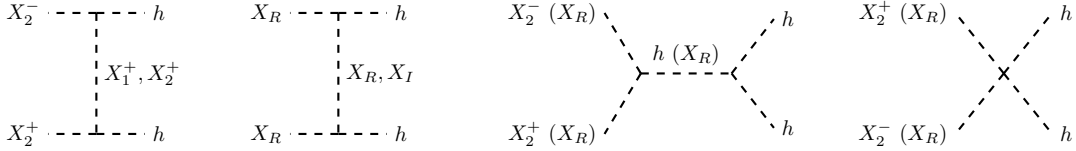
\begin{figure}[H]
	\begin{tikzpicture}[
		baseline=(current bounding box.center),
		scale=0.8,
		transform shape,
		fermion/.style={dashed, thick},
		boson/.style={decorate, decoration={snake}, thick},
		scalar/.style={dashed, thick}
		]
		\begin{scope}
			\draw[scalar] (0,1) -- (-0.8,1)  node[left] {$X_2^-$};
			\draw[scalar] (0,-1) -- (-0.8,-1) node[left] {$X_2^+$};
			\draw[scalar] (0,1) -- (0,-1) node[midway, right] {$X_1^+, X_2^+$};
			
			\draw[scalar] (0,1) -- (0.8,1) node[right] {$h$};
			\draw[scalar] (0,-1) -- (0.8,-1) node[right] {$h$};
		\end{scope}
		
		\begin{scope}[xshift=4cm]
			\draw[scalar] (0,1) -- (-0.8,1)  node[left] {$X_R$};
			\draw[scalar] (0,-1) -- (-0.8,-1) node[left] {$X_R$};
			\draw[scalar] (0,1) -- (0,-1) node[midway, right] {$X_R, X_I$};
			
			\draw[scalar] (0,1) -- (0.8,1) node[right] {$h$};
			\draw[scalar] (0,-1) -- (0.8,-1) node[right] {$h$};
		\end{scope}
		
		\begin{scope}[xshift=9cm]	
			\draw[scalar] (0,0) -- (-0.6,1)  node[left]  {$X_2^- \ (X_R)$ };
			\draw[scalar] (0,0) --(-0.6,-1)  node[left]  {$X_2^+\ (X_R)$};
			\draw[scalar] (0,0) -- (1.5,0) node[midway, above] {$h\ (X_R)$};
			\draw[scalar] (1.5,0) -- (2.2,0.9) node[right] {$h$};
			\draw[scalar] (1.5,0) -- (2.2,-0.9) node[right] {$h$};
		\end{scope}
		\begin{scope}[xshift=15cm]
			\draw[scalar] (0,0) -- (-0.8,1) node[left] {$X_2^+ \ (X_R)$};
			\draw[scalar]  (0,0) -- (-0.8,-1) node[left] {$X_2^- \ (X_R)$};
			\draw[scalar] (0,0) -- (0.8,1) node[right] {$h$};
			\draw[scalar] (0,0) -- (0.8,-1) node[right] {$h$};
		\end{scope}		
	\end{tikzpicture}
	\caption{Contributions to $\Omega^{-1}$ from the modes $X_2^+ \, X_2^-  \rightarrow h h$}
\end{figure}

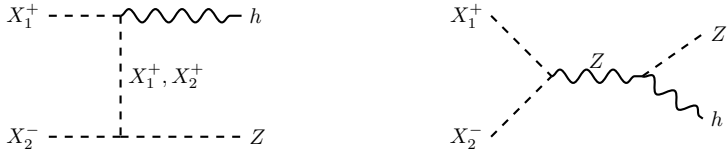
\begin{figure}[H]
	\begin{tikzpicture}[
		baseline=(current bounding box.center),
		scale=0.8,
		transform shape,
		fermion/.style={dashed, thick},
		boson/.style={decorate, decoration={snake}, thick},
		scalar/.style={dashed, thick}
		]
		\begin{scope}
			\draw[scalar] (0,1) -- (-1.2,1)  node[left] {$X_1^+$};
			\draw[scalar] (0,-1) -- (-1.2,-1) node[left] {$X_2^-$};
			\draw[scalar] (0,1) -- (0,-1) node[midway, right] {$X_1^+, X_2^+$};
			
			\draw[boson] (0,1) -- (2,1) node[right] {$h$};
			\draw[scalar] (0,-1) -- (2,-1) node[right] {$Z$};
		\end{scope}
		\hspace*{2.5cm}	
		\begin{scope}[xshift=4cm]
			\draw[scalar] (-1,1) -- (0,0);
			\draw[scalar] (-1,-1) -- (0,0);
			\draw[boson] (0,0) -- (1.5,0) node[midway, above] {$Z$};
			\draw[scalar] (1.5,0) -- (2.5,0.7) node[right] {$Z$};
			\draw[boson] (1.5,0) -- (2.5,-0.7) node[right] {$h$};
			
			\node[left] at (-1,-1) {$X_2^-$};
			\node[left] at (-1,1) {$X_1^+$};
		\end{scope}
	\end{tikzpicture}
	\caption{Contributions to $\Omega^{-1}$ from the modes $X_1^+ \, X_2^-  \rightarrow h Z$}
\end{figure}

Three other sub-dominant processes that have diagrams similar to some dominant processes.

$X_2^+ X_2^- \longrightarrow W^+ W^-$ has similar diagrams like $X_1^+ X_1^- \longrightarrow W^+ W^-$

$X^{++} X_I \rightarrow W^+ W^-$ has similar diagrams with $X^{++} X_R \rightarrow W^+ W^-$

$X_I X_I \rightarrow W^+ W^-$ has similar diagram to $X_R X_R \longrightarrow W^+ W^-$

\bibliographystyle{unsrt}
\bibliography{references}

\begin{thebibliography}{10}

\bibitem{ATLAS:2012yve}
Georges Aad et~al.
\newblock {Observation of a new particle in the search for the Standard Model
  Higgs boson with the ATLAS detector at the LHC}.
\newblock {\em Phys. Lett. B}, 716:1--29, 2012.

\bibitem{CMS:2012qbp}
Serguei Chatrchyan et~al.
\newblock {Observation of a New Boson at a Mass of 125 GeV with the CMS
  Experiment at the LHC}.
\newblock {\em Phys. Lett. B}, 716:30--61, 2012.

\bibitem{Jangid:2020qgo}
Shilpa Jangid and Priyotosh Bandyopadhyay.
\newblock {Distinguishing Inert Higgs Doublet and Inert Triplet Scenarios}.
\newblock {\em Eur. Phys. J. C}, 80(8):715, 2020.

\bibitem{Bandyopadhyay:2024plc}
Priyotosh Bandyopadhyay, Snehashis Parashar, Chandrima Sen, and Jeonghyeon
  Song.
\newblock {Probing Inert Triplet Model at a multi-TeV muon collider via vector
  boson fusion with forward muon tagging}.
\newblock {\em JHEP}, 07:253, 2024.

\bibitem{Bandyopadhyay:2020djh}
Priyotosh Bandyopadhyay, Shilpa Jangid, and Manimala Mitra.
\newblock {Scrutinizing Vacuum Stability in IDM with Type-III Inverse seesaw}.
\newblock {\em JHEP}, 02:075, 2021.

\bibitem{Bandyopadhyay:2020otm}
Priyotosh Bandyopadhyay and Antonio Costantini.
\newblock {Obscure Higgs boson at Colliders}.
\newblock {\em Phys. Rev. D}, 103(1):015025, 2021.

\bibitem{Chiang:2020rcv}
Cheng-Wei Chiang, Giovanna Cottin, Yong Du, Kaori Fuyuto, and Michael~J.
  Ramsey-Musolf.
\newblock {Collider Probes of Real Triplet Scalar Dark Matter}.
\newblock {\em JHEP}, 01:198, 2021.

\bibitem{Chakrabarty:2015yia}
Nabarun Chakrabarty, Dilip~Kumar Ghosh, Biswarup Mukhopadhyaya, and Ipsita
  Saha.
\newblock {Dark matter, neutrino masses and high scale validity of an inert
  Higgs doublet model}.
\newblock {\em Phys. Rev. D}, 92(1):015002, 2015.

\bibitem{Bandyopadhyay:2021ipw}
Priyotosh Bandyopadhyay and Shilpa Jangid.
\newblock {Discerning singlet and triplet scalars at the electroweak phase
  transition and gravitational wave}.
\newblock {\em Phys. Rev. D}, 107(5):055032, 2023.

\bibitem{Jangid:2020dqh}
Shilpa Jangid, Priyotosh Bandyopadhyay, P.~S. Bhupal~Dev, and Arjun Kumar.
\newblock {Vacuum stability in inert higgs doublet model with right-handed
  neutrinos}.
\newblock {\em JHEP}, 08:154, 2020.

\bibitem{Khan:2016sxm}
Najimuddin Khan.
\newblock {Exploring the hyperchargeless Higgs triplet model up to the Planck
  scale}.
\newblock {\em Eur. Phys. J. C}, 78(4):341, 2018.

\bibitem{Kanemura:2023wap}
Shinya Kanemura and Yushi Mura.
\newblock {Criterion of perturbativity with the mass-dependent beta function in
  extended Higgs models}.
\newblock {\em Phys. Rev. D}, 110(7):075016, 2024.

\bibitem{Gonderinger:2009jp}
Matthew Gonderinger, Yingchuan Li, Hiren Patel, and Michael~J. Ramsey-Musolf.
\newblock {Vacuum Stability, Perturbativity, and Scalar Singlet Dark Matter}.
\newblock {\em JHEP}, 01:053, 2010.

\bibitem{Athron:2018ipf}
Peter Athron, Jonathan~M. Cornell, Felix Kahlhoefer, James Mckay, Pat Scott,
  and Sebastian Wild.
\newblock {Impact of vacuum stability, perturbativity and XENON1T on global
  fits of $\mathbb {Z}_2$ and $\mathbb {Z}_3$ scalar singlet dark matter}.
\newblock {\em Eur. Phys. J. C}, 78(10):830, 2018.

\bibitem{Bandyopadhyay:2023joz}
Priyotosh Bandyopadhyay, Mariana Frank, Snehashis Parashar, and Chandrima Sen.
\newblock {Interplay of inert doublet and vector-like lepton triplet with
  displaced vertices at the LHC/FCC and MATHUSLA}.
\newblock {\em JHEP}, 03:109, 2024.

\bibitem{Benincasa:2022elt}
Nico Benincasa, Luigi Delle~Rose, Kristjan Kannike, and Luca Marzola.
\newblock {Multi-step phase transitions and gravitational waves in the inert
  doublet model}.
\newblock {\em JCAP}, 12:025, 2022.

\bibitem{Banerjee:2019luv}
Shankha Banerjee, Fawzi Boudjema, Nabarun Chakrabarty, Guillaume Chalons, and
  Hao Sun.
\newblock {Relic density of dark matter in the inert doublet model beyond
  leading order: The heavy mass case}.
\newblock {\em Phys. Rev. D}, 100(9):095024, 2019.

\bibitem{Branco:2011iw}
G.~C. Branco, P.~M. Ferreira, L.~Lavoura, M.~N. Rebelo, Marc Sher, and Joao~P.
  Silva.
\newblock {Theory and phenomenology of two-Higgs-doublet models}.
\newblock {\em Phys. Rept.}, 516:1--102, 2012.

\bibitem{Ferreira:2004yd}
P.~M. Ferreira, R.~Santos, and A.~Barroso.
\newblock {Stability of the tree-level vacuum in two Higgs doublet models
  against charge or CP spontaneous violation}.
\newblock {\em Phys. Lett. B}, 603:219--229, 2004.
\newblock [Erratum: Phys.Lett.B 629, 114--114 (2005)].

\bibitem{Datta:2016nfz}
Amitava Datta, Nabanita Ganguly, Najimuddin Khan, and Subhendu Rakshit.
\newblock {Exploring collider signatures of the inert Higgs doublet model}.
\newblock {\em Phys. Rev. D}, 95(1):015017, 2017.

\bibitem{Parashar:2024cqq}
Snehashis Parashar, Priyotosh Bandyopadhyay, Chandrima Sen, and Jeonghyeon
  Song.
\newblock {Signatures of the Inert Triplet Model from Vector Boson Fusion at a
  Muon Collider}.
\newblock {\em PoS}, ICHEP2024:305, 2025.

\bibitem{Kanemura:2025xkf}
Shinya Kanemura, Yushi Mura, and Tetsuo Shindou.
\newblock {Classification of Higgs sectors from group theoretical properties of
  UV gauge theories}.
\newblock {\em Phys. Rev. D}, 113(5):055020, 2026.

\bibitem{Jangid:2023jya}
Shilpa Jangid and Hiroshi Okada.
\newblock {Exploring CP-violation in Y=0 inert triplet with real singlet}.
\newblock {\em Phys. Rev. D}, 108(5):055025, 2023.

\bibitem{Chowdhury:2015yja}
Debtosh Chowdhury and Otto Eberhardt.
\newblock {Global fits of the two-loop renormalized Two-Higgs-Doublet model
  with soft Z$_{2}$ breaking}.
\newblock {\em JHEP}, 11:052, 2015.

\bibitem{Bandyopadhyay:2025jlg}
Priyotosh Bandyopadhyay and Snehashis Parashar.
\newblock {Dark clouds to silver linings over the hyperchargeless scalar
  triplets}.
\newblock {\em Phys. Rev. D}, 112(5):055032, 2025.

\bibitem{Bandyopadhyay:2024gyg}
Priyotosh Bandyopadhyay and Snehashis Parashar.
\newblock {Probing a scalar singlet-triplet extension of the standard model via
  vector boson fusion at a muon collider}.
\newblock {\em Phys. Rev. D}, 110(11):115032, 2024.

\bibitem{Zeng:2019tlw}
Yu-Pan Zeng, Chengfeng Cai, Dan-Yang Liu, Zhao-Huan Yu, and Hong-Hao Zhang.
\newblock {Probing quadruplet scalar dark matter at current and future $pp$
  colliders}.
\newblock {\em Phys. Rev. D}, 101(11):115033, 2020.

\bibitem{AbdusSalam:2013eya}
Shehu~S. AbdusSalam and Talal~Ahmed Chowdhury.
\newblock {Scalar Representations in the Light of Electroweak Phase Transition
  and Cold Dark Matter Phenomenology}.
\newblock {\em JCAP}, 05:026, 2014.

\bibitem{Chowdhury:2023uyd}
Talal~Ahmed Chowdhury, Kareem Ezzat, Shaaban Khalil, Ernest Ma, and Dibyendu
  Nanda.
\newblock {Higgs quadruplet impact on W mass shift, dark matter, and LHC
  signatures}.
\newblock {\em Phys. Rev. D}, 109(7):075039, 2024.

\bibitem{Jurciukonis:2024bzx}
Darius Jur{\v{c}}iukonis and Lu{\'\i}s Lavoura.
\newblock {On the extension of the SM through a scalar quadruplet}.
\newblock 6 2024.

\bibitem{Chakraborty:2025kcl}
Amit Chakraborty, Shreecheta Chowdhury, Nilanjana Kumar, and Vandana Sahdev.
\newblock {Search for quadruplet scalars using boosted decision trees at the
  LHC}.
\newblock {\em Phys. Rev. D}, 113(9):095017, 2026.

\bibitem{Bandyopadhyay:2016oif}
Priyotosh Bandyopadhyay and Rusa Mandal.
\newblock {Vacuum stability in an extended standard model with a leptoquark}.
\newblock {\em Phys. Rev. D}, 95(3):035007, 2017.

\bibitem{Bandyopadhyay:2017tlq}
Priyotosh Bandyopadhyay, Eung~Jin Chun, and Rusa Mandal.
\newblock {Scalar Dark Matter in Leptophilic Two-Higgs-Doublet Model}.
\newblock {\em Phys. Lett. B}, 779:201--205, 2018.

\bibitem{Bandyopadhyay:2021kue}
Priyotosh Bandyopadhyay, Shilpa Jangid, and Anirban Karan.
\newblock {Constraining scalar doublet and triplet leptoquarks with vacuum
  stability and perturbativity}.
\newblock {\em Eur. Phys. J. C}, 82(6):516, 2022.

\bibitem{Bandyopadhyay:2025ilx}
Priyotosh Bandyopadhyay and Pram Milan~P. Robin.
\newblock {Fate of the scalar quartic couplings in the inert models}.
\newblock {\em Phys. Rev. D}, 112(1):015013, 2025.

\bibitem{Shrock:2016hqn}
Robert Shrock.
\newblock {Study of the six-loop beta function of the $\lambda\phi^4_4$
  theory}.
\newblock {\em Phys. Rev. D}, 94(12):125026, 2016.

\bibitem{Shrock:2023xcu}
Robert Shrock.
\newblock {Search for an ultraviolet zero in the seven-loop beta function of
  the {\ensuremath{\lambda}}{\ensuremath{\phi}}44 theory}.
\newblock {\em Phys. Rev. D}, 107(5):056018, 2023.

\bibitem{Chun:2009mh}
E.~J. Chun.
\newblock {Minimal dark matter in type III seesaw}.
\newblock {\em JHEP}, 12:055, 2009.

\bibitem{Cirelli:2005uq}
Marco Cirelli, Nicolao Fornengo, and Alessandro Strumia.
\newblock {Minimal dark matter}.
\newblock {\em Nucl. Phys. B}, 753:178--194, 2006.

\bibitem{Bandyopadhyay:2010wp}
Priyotosh Bandyopadhyay and Eung~Jin Chun.
\newblock {Displaced Higgs production in type III seesaw}.
\newblock {\em JHEP}, 11:006, 2010.

\bibitem{Ahriche:2007jp}
Amine Ahriche.
\newblock {What is the criterion for a strong first order electroweak phase
  transition in singlet models?}
\newblock {\em Phys. Rev. D}, 75:083522, 2007.

\bibitem{Espinosa:2011ax}
Jose~R. Espinosa, Thomas Konstandin, and Francesco Riva.
\newblock {Strong Electroweak Phase Transitions in the Standard Model with a
  Singlet}.
\newblock {\em Nucl. Phys. B}, 854:592--630, 2012.

\bibitem{Kannike:2023bfh}
Kristjan Kannike.
\newblock {Constraining the Higgs trilinear coupling from an SU(2) quadruplet
  with bounded-from-below conditions}.
\newblock {\em JHEP}, 01:176, 2024.

\bibitem{Staub:2013tta}
Florian Staub.
\newblock {SARAH 4 : A tool for (not only SUSY) model builders}.
\newblock {\em Comput. Phys. Commun.}, 185:1773--1790, 2014.

\bibitem{SabanciKeceli:2018fsd}
Asl{\i} Sabanc{\i}~Keceli, Priyotosh Bandyopadhyay, and Katri Huitu.
\newblock {Long-lived triplinos and displaced lepton signals at the LHC}.
\newblock {\em Eur. Phys. J. C}, 79(4):345, 2019.

\bibitem{LZ:2024zvo}
J.~Aalbers et~al.
\newblock {Dark Matter Search Results from 4.2{\,}{\,}Tonne-Years of Exposure
  of the LUX-ZEPLIN (LZ) Experiment}.
\newblock {\em Phys. Rev. Lett.}, 135(1):011802, 2025.

\bibitem{Belanger:2013oya}
G.~Belanger, F.~Boudjema, A.~Pukhov, and A.~Semenov.
\newblock {micrOMEGAs$\_$3: A program for calculating dark matter observables}.
\newblock {\em Comput. Phys. Commun.}, 185:960--985, 2014.

\bibitem{Belyaev:2012qa}
Alexander Belyaev, Neil~D. Christensen, and Alexander Pukhov.
\newblock {CalcHEP 3.4 for collider physics within and beyond the Standard
  Model}.
\newblock {\em Comput. Phys. Commun.}, 184:1729--1769, 2013.

\bibitem{Planck:2018vyg}
N.~Aghanim et~al.
\newblock {Planck 2018 results. VI. Cosmological parameters}.
\newblock {\em Astron. Astrophys.}, 641:A6, 2020.
\newblock [Erratum: Astron.Astrophys. 652, C4 (2021)].

\bibitem{XENON:2023iku}
E.~Aprile et~al.
\newblock {Searching for Heavy Dark Matter near the Planck Mass with XENON1T}.
\newblock {\em Phys. Rev. Lett.}, 130(26):261002, 2023.

\bibitem{Cline:2013gha}
James~M. Cline, Kimmo Kainulainen, Pat Scott, and Christoph Weniger.
\newblock {Update on scalar singlet dark matter}.
\newblock {\em Phys. Rev. D}, 88:055025, 2013.
\newblock [Erratum: Phys.Rev.D 92, 039906 (2015)].

\end{thebibliography}
\end{document}